\documentclass[prd,preprint,superscriptaddress,preprintnumbers,eqsecnum,showpacs,nofootinbib,
nobibnotes]{revtex4-1}
\usepackage[utf8]{inputenc}
\usepackage{wrapfig}
\usepackage{ulem}

\usepackage{diagbox}

\usepackage{mathtools}
\usepackage{amsfonts}
\usepackage{mathrsfs}
\usepackage{bbm}
\usepackage{slashed}

\usepackage{graphicx}
\usepackage{color}
\usepackage{array}

\usepackage{placeins}
\usepackage{booktabs}
\usepackage{makecell}
\usepackage{epstopdf}
\usepackage{subcaption}

\usepackage{xspace}
\usepackage{siunitx}
\usepackage{hyperref}
\usepackage[nameinlink]{cleveref}
\usepackage{appendix}

\usepackage{xcolor}
\hypersetup{
	colorlinks,
	linkcolor={red!75!black},
	citecolor={blue!75!black},
	urlcolor={blue!75!black}
}

\setkeys{Gin}{width=0.48\textwidth}

\graphicspath{
	{./figures/}
	{../figures/}
}

\def\qslash{q\llap{/}}

\def\id{1\!\mbox{l}}
\def\s0#1#2{\mbox{\small{$ \frac{#1}{#2} $}}}
\def\0#1#2{\frac{#1}{#2}}

\def\colT#1{T^{#1}_{\rm c}}

\newcommand{\Tr}{{\text{Tr}}}

\newcommand{\identityflavour}{\mathbbm{1}_{\rm f}}

\newcommand{\fatg}{{\rm{I}}\!\!{\rm{I}}\!\!\Gamma}

\newcommand*{\eg}{e.g.\@\xspace}
\newcommand*{\ie}{i.e.\@\xspace}

\newcommand{\bal}{\begin{align}}
\newcommand{\eal}{\end{align}}

\newcommand{\be}{\begin{equation}}
\newcommand{\ee}{\end{equation}}
\newcommand{\bea}{\begin{eqnarray}}
\newcommand{\eea}{\end{eqnarray}}

\newcommand{\overbar}[1]{\mkern 2mu\overline{\mkern-3mu#1\mkern-1.7mu}\mkern 1.7mu}

\def\ie{{\it i.e.}, }
\def\eg{{\it e.g.}, }
\def\pslash{p\hspace{-0.18cm}\slash}

\def\s#1{{\scriptscriptstyle #1}}

\def\fig#1{Fig.~\ref{#1}}
\def\Fig#1{Fig.~\ref{#1}}

\def\Sec#1{\Cref{#1}}
\def\sec#1{\Cref{#1}}
\def\App#1{\Cref{#1}}
\def\app#1{\Cref{#1}}
\def\Eq#1{Eq.~(\ref{#1})}
\def\eq#1{(\ref{#1})}

\newcommand{\gettitle}{Fully coupled functional equations for the quark sector of QCD}

\hypersetup{
	pdftitle={\gettitle},
	pdfauthor={Gao, Pawlowski,Papavassiliou},
	pdfkeywords={functional methods}
		{correlations functions} {phase structure}
		{functional renormalization group} {Dyson-Schwinger equation},
	bookmarksopen=true,
	bookmarksopenlevel=2,
	bookmarksnumbered=true
}

\begin{document}

\title{\gettitle}

\author{Fei Gao}
\affiliation{Institut f{\"u}r Theoretische Physik,
	Universit{\"a}t Heidelberg, Philosophenweg 16,
	69120 Heidelberg, Germany
}
\author{Joannis Papavassiliou}
\affiliation{
Department of Theoretical Physics and IFIC, University of Valencia and CSIC, E-46100 Valencia, Spain
}

\author{Jan M. Pawlowski}
\affiliation{Institut f{\"u}r Theoretische Physik,
	Universit{\"a}t Heidelberg, Philosophenweg 16,
	69120 Heidelberg, Germany
}
\affiliation{ExtreMe Matter Institute EMMI,
	GSI, Planckstr.1,
	64291 Darmstadt, Germany
}

\begin{abstract}

We present a comprehensive study  of the quark sector of $2+1$ flavour QCD, based on a self-consistent treatment of the coupled system of Schwinger-Dyson equations for the quark propagator and the full quark-gluon vertex in the one-loop dressed approximation. The individual form factors of the quark-gluon vertex are expressed in a special tensor basis obtained from a set of gauge-invariant operators. The sole external ingredient used as input to our equations is the Landau gauge gluon propagator with $2+1$ dynamical quark flavours, obtained from studies with Schwinger-Dyson equations, the functional renormalization group approach,  and large volume lattice simulations. The appropriate renormalization procedure required in order to self-consistently accommodate external inputs stemming from other functional approaches or the lattice is discussed in detail, and the value of the gauge coupling is accurately determined at two vastly separated renormalization group scales.

Our analysis establishes a clear hierarchy among the vertex form factors. We identify only three dominant ones, in agreement with previous results. The components of the quark propagator obtained from our approach are in excellent agreement with the results from Schwinger-Dyson equations, the functional renormalization group, and lattice QCD simulation, a simple benchmark observable being the chiral condensate in the chiral limit, which is computed as $(244\,\textrm{MeV})^3$. The present approach has a wide range of applications, including the self-consistent computation of bound-state properties and finite temperature and density physics, which are briefly discussed.

\end{abstract}

\maketitle

\section{Introduction}\label{sec:Introduction}

In functional approaches to QCD, the task of computing quark-, gluon-, and hadron correlation functions is formulated in terms of closed coupled  diagrammatic relations between them, which must then be solved numerically.
In all these approaches, such as Schwinger-Dyson equations (SDEs), functional renormalization group (fRG), $n$-particle irreducible methods (nPI),
and bound state methods (Bethe-Salpeter (BS), Faddeev- and higher-order equations),
the diagrammatic relations are built out of the propagators of the fundamental and composite QCD fields.
For reviews on functional methods in QCD, see,
\eg~\cite{Roberts:1994dr, Alkofer:2000wg, Maris:2003vk, Fischer:2006ub, Binosi:2009qm, Maas:2011se,  Huber:2018ned} (SDEs),~\cite{Pawlowski:2005xe, Gies:2006wv, Rosten:2010vm, Braun:2011pp, Pawlowski:2014aha, Dupuis:2020fhh} (fRG), and ~\cite{Cloet:2013jya, Eichmann:2016yit, Sanchis-Alepuz:2017jjd} (bound-states).

Functional approaches allow for an attractively simple and versatile access to the dynamical mechanisms that drive numerous fundamental QCD phenomena. Moreover, their flexibility in using as external inputs correlation functions stemming from distinct non-perturbative setups, \eg lattice~\cite{Cucchieri:2007md, Bogolubsky:2007ud, Bowman:2007du, Bogolubsky:2009dc, Oliveira:2009eh, Skullerud:2002ge, Skullerud:2003qu, Kizilersu:2006et, Rojas:2013tza, Oliveira:2016muq, Sternbeck:2017ntv, Oliveira:2018fkj, Athenodorou:2016oyh, Duarte:2016ieu, Aguilar:2019uob, Aguilar:2021lke}), is a particularly welcome feature, which increases their quantitative reliability and their range of applicability. However, such inputs are not always available, prominent and important examples being QCD at finite temperature and density, as well as the hadron spectrum. Hence, in the past two decades, functional methods have evolved into a self-contained quantitative approach to QCD, allowing for quantitative predictions within a ``first principle'' framework, without the need of external inputs.

This ongoing progress requires quantitative computations involving the full tensor structure of correlation functions, and in particular that of the three- and four-point functions, that dominantly drive the dynamics of QCD. Specifically, the quark-gluon vertex is the pivotal ingredient of the matter dynamics of QCD, being intimately connected with fundamental phenomena such as chiral symmetry breaking and quark mass generation, bound state formation, \eg~\cite{Bender:2002as, Alkofer:2008tt, Chang:2009zb, Aguilar:2010cn, Eichmann:2011vu, Binosi:2014aea, Gomez-Rocha:2014vsa, Gomez-Rocha:2015qga, Eichmann:2016hgl, Binosi:2016rxz}, and the QCD phase structure at finite temperature and chemical potential, \eg~\cite{Braun:2007bx,  Braun:2009gm, Qin:2010nq, Fischer:2011mz, LUO:2013hea, Fister:2013bh, Fischer:2013eca,  Fischer:2014ata, Christiansen:2014ypa, Shi:2014zpa,Cui:2015xta, Eichmann:2015kfa, Cyrol:2017qkl, Contant:2018zpi, Maelger:2019cbk, Fu:2019hdw, Braun:2019aow, Gao:2020qsj, Gao:2020fbl, Braun:2020ada}.

To date, the quark-gluon vertices employed in most SDE studies are not based on a full solution of the corresponding dynamical equations, but are rather put together from quark and ghost dressing functions with the aid of the Slavnov-Taylor identities (STIs); see, \eg~\cite{Ball:1980ay, Curtis:1990zs, Davydychev:2000rt, Fischer:2003rp, Aguilar:2014lha, Aguilar:2016lbe, Bermudez:2017bpx, Oliveira:2018fkj, Aguilar:2018epe, Chang:2021vvx} or rely on perturbative expansion schemes; see, \eg~\cite{Pelaez:2017bhh, Pelaez:2020ups, Barrios:2021cks}. These are operationally simple and suggestive treatments, with an impressive array of very successful applications, ranging from the properties of hadrons to the phase structure of QCD. However, within the STI constructions, the strength associated with the classical tensor structure requires a phenomenological \textit{infrared} enhancement, whose size is adjusted by means of the constituent quark masses. The latter, including their momentum dependence, are equivalent to the physical amount of chiral symmetry breaking, and hence, in such an approach, the quantitative strength of chiral symmetry breaking is a phenomenological input rather than a prediction. To be sure, the need for such an enhancement may be attributed to the insufficient knowledge of some of the ingredients comprising these  STIs (\eg quark-ghost kernel~\cite{Aguilar:2016lbe}). Nonetheless, in view of the results in the present work, as well as of previous considerations within functional approaches~\cite{Alkofer:2008tt, Fischer:2009jm, Chang:2010hb, Williams:2014iea, Mitter:2014wpa, Blum:2015lsa, Williams:2015cvx, Binosi:2016wcx, Cyrol:2017ewj, Tang:2019zbk, Gao:2020qsj, Gao:2020fbl}, it seems to originate mainly from the omission of important tensor structures that are simply not accessible through the standard STI construction.

This situation calls for a self-consistent treatment of the full quark-gluon vertex within the SDE formalism in the Landau gauge. The determination of the eight relevant form factors from their dynamical equations requires the solution of the  coupled system of gluon, ghost and quark propagators, the quark-gluon vertex, as well as additional vertices, for unqenched SDE works see \cite{Williams:2015cvx, Tang:2019zbk}. The most complete results in this direction have been obtained within functional methods for two-flavour QCD, see~\cite{Williams:2014iea, Mitter:2014wpa, Mitter:2014wpa} (quenched), and~\cite{Williams:2015cvx, Cyrol:2017ewj} (unquenched). Recently, the fRG results of~\cite{Cyrol:2017ewj} have been used as input for a 2+1--flavour analysis within the SDE approach, both in the vacuum and at finite temperature and density~\cite{Gao:2020qsj, Gao:2020fbl}. This endeavour requires a well-defined calculational SDE scheme, where one could unambiguously identify and reliably compute the dominant components of this vertex, either self-consistently or with the aid of a given input.

In the present work we put forward a systematic approximation scheme for the set of functional equations governing the quark sector of QCD, by studying in detail the coupled system of SDEs for the quark propagator (quark gap equation)  and the quark-gluon vertex. Our SDE analysis reveals that the quark dynamics is dominated by {\it three} specific tensor structures of the quark-gluon vertex, in agreement with earlier considerations~\cite{Williams:2014iea, Mitter:2014wpa, Williams:2015cvx, Cyrol:2017ewj, Gao:2020qsj, Gao:2020fbl}. It is important to stress that, apart from the dressing associated with the classical tensor, the other two dominant dressings are not accessible by means of an STI-based construction. In fact, the numerical impact of these latter dressings at the level of the gap equations is crucial, furnishing directly the required amount of chiral symmetry breaking without the need to resort to artificial enhancing factors. In our opinion, this demonstrates conclusively that no artificial enhancement is required once
the contributions from the appropriate tensorial structures have been properly taken into account.
Importantly, we also find that certain tensor structures, which in previous STI treatments seemed dominant precisely due to the use of such enhancing factors, turn out to be clearly subleading.
Consequently, the present detailed analysis enables us to restrict our considerations to the three most relevant tensors, thus arriving at a reduced set of fully coupled SDEs, which are solved iteratively together with the quark gap equation.

A central ingredient of the  system of equations considered in this work is the gluon propagator, entering both in the gap equation and the SDE for the quark-gluon vertex. The gluon propagator obeys its own SDE~\cite{Roberts:1994dr, Alkofer:2000wg, Maris:2003vk, Fischer:2006ub, Binosi:2009qm, Huber:2018ned}, which depends on the quark propagator and further correlation functions, a fact that leads to a proliferation of coupled equations. Even though the complete treatment of such as extended system has already been implemented for $N_f=2$ flavour QCD~\cite{Mitter:2014wpa, Cyrol:2017ewj}, in the present work we prefer to maintain the focus on the novel features of our approach rather than be sidetracked by a technically exhaustive analysis. To that end, we treat the gluon propagator as an external ingredient: within our most elaborate and trustworthy approximation, we consider a renormalization point at large, perturbative, momenta with $\mu=40$\,GeV, and use the SDE data for the gluon propagator from~\cite{Gao:2020qsj, Gao:2020fbl} as external input. These SDE data are based on the fRG two-flavour computation of~\cite{Cyrol:2017qkl}, as are the gluon data of~\cite{Fu:2019hdw}, which are also used as input, for the purpose of estimating our systematic error. Finally, we also consider gluon data from $N_f=2+1$ lattice simulations~\cite{Boucaud:2018xup, Zafeiropoulos:2019flq, Aguilar:2019uob}, and a renormalization point of $\mu=4.3$\,GeV for comparison.  While the lattice data offer the smallest systematic error, their  momentum range is only $p\lesssim 5$\,GeV. As we will see, all the different inputs lead to quantitatively compatible results.

Note also that the gluon propagator is rather insensitive to the details of the quark dynamics, within the range of pion and current quark masses considered here; for a detailed evaluation in the two-flavour case and pion masses in the range $m_\pi \approx 0 - 300$\,MeV~\cite{Cyrol:2017ewj}. To be sure, this property does not persist when additional families of active quarks are added to the theory, since, in this case, one implements effectively a transition from infinite to finite quark masses. In fact, as has been clearly established in the analysis of~\cite{Ayala:2012pb}, the sequential inclusion of quark families affects the quantitative behaviour of the gluon propagator, markedly suppressing its infrared support.

A further important ingredient of the current approach is the MOM-type renormalization scheme used here: In this scheme, both the full dressings of primitively divergent vertices as well as the respective bare renormalization constants approach unity
at the symmetric point $\bar p=\mu$, for asymptotically large $\mu\to\infty$.
For this reason we shall call it MOM${}^2$. The  MOM${}^2$ is the natural scheme employed
in the fRG-approach to QCD, and hence underlies our input data. It has been also used in \cite{Gao:2020qsj, Gao:2020fbl}, and is explained for the first time in detail in \App{app:RG-schemes}. The respective results are particularly stable  under vast changes in the value of the renormalization point $\mu$. Finally, a chief advantage of this scheme is its relative operational simplicity and low computational cost, combined with quantitative reliability and systematic error control.

The article is organised as follows. In \Sec{sec:notation} we review some general features of the SDE and fRG approaches, and introduce the notation that will be used in this work. In \sec{sec:quarkgap} we set up the gap equation and discuss its renormalization. Then, in \sec{sec:QuarkGluon} we focus on the quark-gluon vertex, present the tensorial basis that will be employed, and derive the system of integral equations satisfied by its form factors. In \sec{sec:gencon} we present a detailed discussion of how to implement self-consistently the renormalization of the SDEs when an external input is employed. In \sec{sec:CurrentQuark} we discuss the procedure that fixes the values of the current quark masses, and introduce the light chiral condensate as our benchmark observable. In \sec{sec:results} we present and discuss the central results of our analysis, with special emphasis on the quark mass and the eight form factors of the quark-gluon vertex, evaluated at the symmetric point. Then, in \sec{sec:Stability} we confirm the stability of our results under variations of the ultraviolet (UV) cutoff, the renormalization point, and the inputs used for the gluon propagator. In \sec{sec:QuantApprox} we capitalise on the hierarchy displayed among the vertex form factors, and propose a simplified treatment that reduces the numerical cost without compromising the accuracy of the results. In \sec{sec:summary} we summarise our approach and present our conclusions. Finally, in \App{app:RG-schemes} we offer numerous technical details of the present MOM-type scheme (MOM${}^2$). We also explain how to map our running vertex coupling in the MOM${}^2$ scheme to the MOM coupling as well as comparing Taylor couplings in the present approach and the lattice. Our results are in quantitative agreement with two-loop perturbation theory at large momenta, and match the respective lattice results for small momenta.  Finally, in \App{app:cof} we present the kernels of the vertex SDE.

\section{General considerations}\label{sec:notation}

In this section we briefly comment on certain important aspects of functional approaches that are relevant for the ensuing analysis, and introduce the notation that will be employed in this work.

\subsection{The action}\label{sec:action}

The starting point is the classical action of QCD in covariant gauges, given by
\begin{align}
	S[\phi]= \int d^4 x\, \left[ \frac{1}{4}  ( F^a_{\mu\nu})^2+\bar q\left(\slashed{D} +m_q\right) q
	+\frac{1}{2 \xi} \left( \partial_\mu A^a_\mu\right)^2 - \bar c^a \,\partial_\mu D^{ab}_\mu\, c^b\right]\,,
	\label{eq:S_QCD}\end{align}
where the ghost has a positive dispersion, typically used in fRG applications to QCD; for a recent review see \cite{Dupuis:2020fhh}. The covariant derivative, $D_\mu$, and the field strength tensor, $F_{\mu\nu}$, are given by

\begin{align}\label{eq:F+D}
	F_{\mu\nu}^a = \partial_\mu A^a_\nu-\partial_\nu A^a_\mu + g_s f^{abc} A_\mu^b A_\nu^c\,,\qquad \textrm{and}\qquad
	D_\mu = \partial_\mu- i g_s A_\mu^a t^a\,, \qquad [t^a , t^b ] =i \, f^{abc} t^c\,.
\end{align}
The first two terms in \eq{eq:S_QCD} are the Yang-Mills and Dirac actions, respectively; in the latter we have suppressed the summation over group indices in the fundamental representation, as well as Dirac and flavour indices. The remaining terms in \eq{eq:S_QCD} encode the gauge fixing and ghost sector. In \eq{eq:F+D}, the covariant derivative in the fundamental representation reads \mbox{$\partial_\mu- i g_s A_\mu^a T^a$}, where $T^a$ are the corresponding generators, while that of the adjoint representation is given by $ \partial_\mu\delta^{ab}-   g_s  f^{abc} A_\mu^c$. The computations in the present work are carried out in the Landau gauge, $\xi=0$.

\subsection{SDE setup and renormalization}\label{sec:SDE-setup}
In contradistinction to the flow equations of the fRG approach, the SDEs depend also on derivatives of
the classical QCD action in \eq{eq:S_QCD}. More specifically, we need the bare action, whose parameters absorb the UV infinities of the diagrams. The mapping from bare fields, $\phi^{(0)}$, to renormalized finite fields, $\phi$, is given by
\begin{subequations}\label{eq:Bare-Ren}
	\begin{align}\label{eq:bare-renFields}
		A^{(0)}_\mu = Z_3^{1/2} A_\mu\,,
		\qquad c^{(0)} = \tilde Z_3^{1/2} c\,,
		\qquad \bar c^{(0)} = \tilde Z_3^{1/2} \bar c \,,\qquad q^{(0)} = Z_2^{1/2} q\,,\qquad \bar q^{(0)} = Z_2^{1/2} \bar q\,,
	\end{align}
	while for the strong coupling, masses, and gauge fixing parameters we have, correspondingly,
	\begin{align}\label{eq:bare-renCoup}
		g^{(0)}_s = Z_g g\,, \qquad m_q^{(0)}  = Z_{m_q} m_q\,,
		\qquad \xi^{(0)} = Z_\xi \,\xi\,.
	\end{align}
	Then, the bare QCD action, $S_{\textrm{bare}}$, reads in terms of the renormalized fields and coupling parameters,
	\begin{align}
		\label{eq:barSQCD}
		S_\textrm{bare}[\phi^{(0)}; g_s^{(0)}, m_q^{(0)}]= S[Z^{1/2}_3 A_\mu, \tilde Z^{1/2}_3 c, \tilde Z^{1/2}_3 \bar c, Z_2^{1/2} q, Z_2^{1/2} \bar q, Z_g g_s, Z_{m_q} m_q]\,.
	\end{align}
From \eq{eq:barSQCD} we may define the renormalization constants of the three-gluon vertex, $Z_1$, the four-gluon vertex, $Z_4$, the ghost-gluon vertex, $\tilde Z_1$, and the quark-gluon vertex, $Z^{f}_1$, and relate them as
	\begin{align}
		\label{eq:bare-renVert}
		Z_1= Z_g Z_3^{3/2}\,,\quad Z_4= Z_g^2 Z_3^{2}\,,\quad  \tilde Z_1= Z_g Z_3^{1/2}\tilde Z_3^{1/2}\,,
		\qquad Z^{f}_1 = Z_g Z_3^{1/2} Z_2\,.
	\end{align}
\end{subequations}

\subsection{fRG setup}\label{sec:fRG setup}
The central object of functional approaches to QCD is the one-particle irreducible (1PI) effective action,  $\Gamma[\phi]$, where $\phi$ is a ``superfield'', whose components are the fundamental  renormalized fields of QCD,
including the auxiliary ghost field introduced through the gauge-fixing,
\begin{align}
	\phi=(A_\mu\,,\, c\,,\, \bar c\,,\, q\,,\, \bar q)\,.
\end{align}
While this is typically rather implicit in most SDE applications, it is commonly the starting point in fRG studies. Derivatives of the effective action $\Gamma[\phi]$ w.r.t.\ the fields are the 1PI $n$-point correlation functions, denoted by
\begin{subequations}
	\label{eq:npointfunc}
	\begin{align}
		\Gamma_{\phi_1\cdots\phi_n}^{(n)}(p_1,...,p_n) = \frac{\delta^n
			\Gamma}{\delta \phi_1(p_1)\cdots \phi_n(p_n)}\,,
	\end{align}
	where all momenta are considered as incoming. Vertices $\Gamma^{(n)}$ are expanded in a complete tensor basis $\{ {\cal T}^{(i)}_{\phi_{i_1}\cdots\phi_{i_n}} \}$, the standard fRG notation in QCD being
	\begin{align}\label{eq:VertDressings}
		\Gamma_{\phi_{i_1}\cdots\phi_{i_n}}^{(n)}(p_1,...,p_n) = \sum_{i} \lambda_{\phi_{i_1}\cdots\phi_{i_n}}(p_1,...,p_n)\,
		{\cal T}^{(i)}_{\phi_{i_1}\cdots\phi_{i_n}}(p_1,...,p_n)\,,
	\end{align}
\end{subequations}
with $\lambda_{\phi_{i_1}\cdots\phi_{i_n}}$ denoting the scalar form factors (dressings).

Note that the renormalization factors defined in \eq{eq:Bare-Ren} have a natural relation to the full dressings of the primitively divergent $n$-point functions in the fRG-approach, $Z_{\phi_i,k}(p)$, $M_{q,k}(p)$, defined in \eq{eq:PropDressings}, and $\lambda^{(1)}_{ \phi_{i_1}\cdots \phi_{i_n},k}$, defined in \eq{eq:VertDressings}; for a detailed account see \cite{Pawlowski:2005xe, Dupuis:2020fhh}.

\subsection{Running couplings}\label{sec:effcharges}

We next consider the different ``avatars'' of the strong running coupling
$\alpha_s(\bar p)=g^2_s(\bar p)/4 \pi $, which can be deduced from the form factors
$\lambda^{(1)}$ associated with the  classical tensor structures of the four fundamental QCD vertices.
In particular, in the present analysis we will employ the running couplings obtained from
the ghost-gluon and quark-gluon vertices, given by
\begin{align}
\label{eq:avatars}
\alpha_{c\bar c A}(\bar p) =\frac{1}{4 \pi} \frac{[\lambda^{(1)}_{c\bar c A}(\bar p)]^2}{Z_A(\bar p)Z_c^2(\bar p) }\,,\qquad\qquad
\alpha_{q\bar q A}(\bar p) = \frac{1}{4 \pi} \frac{[\lambda^{(1)}_{q\bar q A}(\bar p)]^2}{Z_A(\bar p)Z_q^2(\bar p)}	\,,
\end{align}
where $\bar p$ is a symmetric-point configuration, and $Z_A$, $Z_c$ and $Z_q$ are the dressings of the two-point functions (suppressing color),
\begin{align}
\Gamma^{(2)}_{AA\,\mu\nu}(p) =&\, Z_A(p) p^2\, P_{\mu\nu}(p) +\frac{1}{\xi}\, p_\mu p_\nu \,, \nonumber\\ \Gamma^{(2)}_{c\bar c}(p) =&\, Z_c(p) p^2\,,\nonumber\\[1ex]
\Gamma^{(2)}_{q\bar q}(p) =&\, Z_q(p)\left[ i\,\pslash +M_q(p)\right]\,,
\label{eq:PropDressings}
\end{align}
where we have introduced the transverse projection operator
\begin{align}\label{eq:TransProject}
{P}_{\mu\nu}(p) = \delta_{\mu\nu}- \frac{p_\mu p_\nu}{p^2}\,,
\end{align}
usually denoted by $\Pi^\bot_{\mu\nu}(p)$ in the fRG literature. Note that the above two-point functions are the {\it inverses} of the
gluon, ghost, and quark propagators, respectively.

By virtue of the fundamental STIs of the theory, all QCD couplings coincide for large values of $\bar p$,
\begin{align}\label{eq:STIcouplings}
	\alpha_i(\bar p)=\alpha_s(\bar p)\,,\quad i= (c\bar c A\,,\, q\bar q A, A^3, A^4)\qquad \textrm{for perturbative}\quad \bar p =\bar p_\textrm{pert}\,.
\end{align}
As the momentum $\bar p$ gets smaller, the various $\alpha_i(\bar p)$ start deviating from each other, due to differences induced by non-trivial contributions from the scattering kernels appearing in the STIs. As has been pointed out in~\cite{Cyrol:2017ewj}, the amount of chiral symmetry breaking obtained from the gap equation appears to be particularly sensitive to the UV coincidence of the couplings described by (\ref{eq:STIcouplings}). The preservation of (\ref{eq:STIcouplings}) is a indispensable feature of any quantitatively reliable framework; in particular, special truncation schemes such as the PT-BFM~\cite{Binosi:2009qm} are tailor-made for this task.

\section{The quark gap equation}\label{sec:quarkgap}

\begin{figure}[t]
	\includegraphics[width=0.8\columnwidth]{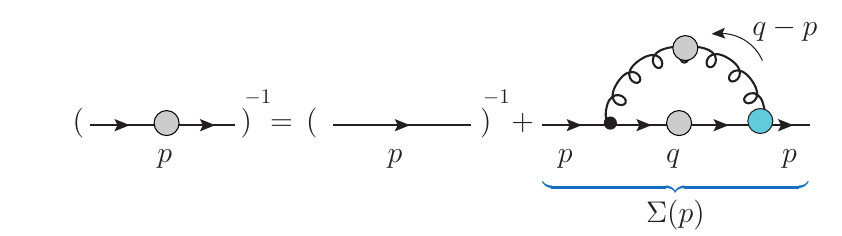}
	\caption{Diagrammatic representation of the quark gap equation. Gray (blue)
          circles denote full propagators (vertices), black dots denote classical vertices.\hspace*{\fill}}\label{fig:QuarkDSE}
\end{figure}

The quark gap equation~\cite{Roberts:1994dr, Alkofer:2000wg, Maris:2003vk, Fischer:2006ub, Binosi:2009qm, Maas:2011se, Huber:2018ned} relates the inverse quark propagator, $\Gamma^{(2)}_{q\bar q}(p)$, to its classical counter part, $S^{(2)}_{q\bar q}(p)$, the quark and gluon propagators, and the classical and full quark-gluon vertices, see \fig{fig:QuarkDSE}. Schematically it reads
\begin{align}
\Gamma^{(2)}_{q \bar q}(p)  = S^{(2)}_{q\bar q} + Z^{f}_{1} g_s \int_q\;   G_{\!AA} (q-p)\,( -i  \gamma)\, G_{q \bar q}(q) \,\Gamma^{(3)}_{ q\bar qA} (q,-p)\, ,
\label{eq:QuarkGap}
\end{align}
where we suppress all Lorentz and color indices, and $g_s$ stands for the gauge coupling.
The four-dimensional momentum integration has been abbreviated by
\begin{align}\label{eq:MomReg}
\int_q := \int_\textrm{reg} \frac{d^4{q}}{(2\pi)^4} \,,
\end{align}
where the subscript $\textrm{``reg''}$ indicates a suitable regularization of the momentum integral; common choices include the dimensional regularization or an appropriately implemented momentum cutoff.
The respective cutoff parameter (\eg $\epsilon$ or $\Lambda^2$) appears also in all
renormalization constants, and in particular the quark-gluon vertex renormalization, $Z_1^f$, as well as the wave function renormalization, $Z_2$, and the mass renormalization, $Z_{m_q}$, of the quark. The last two factors enter into
\eq{eq:QuarkGap} through $S^{(2)}_{q\bar q}(p)$, the second derivative of the bare QCD action, \eq{eq:barSQCD}, with respect to the renormalized quark and anti-quark fields,  see \eq{eq:bare-renFields},
\begin{align}
	S^{(2)}_{q\bar q}(p) =  i Z_{2}\, \pslash +Z_{m_q}\, m_q\,,
	\label{eq:bareS2}
\end{align}
where $m_q$ denotes the bare current quark mass.

The full gluon propagator, $G_{\!AA\,\mu\nu}^{ab}(p)$, in the Landau gauge, and the quark propagator, $G_{q \bar q}^{ab}(p)$, are given by
\begin{align}
	G_{\!AA\,\mu\nu}^{ab}(p) =\delta^{ab}\, {P}^{\mu\nu}(p)G_{\!A}(p)    \,,\qquad  \qquad G_{q \bar q}^{ab}(p)= \delta^{ab} G_{q}(p)\,.
	\label{eq:props}
\end{align}
In \eq{eq:props}\,, $G_{\!A}(p)$ is the scalar part of the gluon propagator, and $G_q(p)$ carries only the Dirac structure but not the trivial color structure. Both $G_{\!A}(p$ and $G_q(p)$ can be described in terms of the scalar dressings introduced in (\ref{eq:PropDressings}), to wit,
\begin{align}
	G_{\!A}(p) = \frac{1}{Z_A(p) p^2}    \,,\qquad  G_{q}(p)= \frac{1}{Z_q(p)\left[ i \,  \pslash + M_q(p) \right] }\,,
	\label{eq:ncprops}
\end{align}
where $M_q(p)$ is the momentum-dependent mass function. Note that in the fRG-approach, for large cutoff scales,
the functions $Z_A(p)$ and $Z_q(p)$ tend towards the corresponding (finite) wave function renormalizations, while $M_q(p)$ tends to the bare quark mass.

Finally, $\left[\Gamma^{(3)}_{\bar q q A}\right]^a_{\nu}(q,-p)$ denotes the quark-gluon vertex,
in accordance with the general definition of (\ref{eq:npointfunc}), with all momenta considered as incoming.

The presence of the transverse projection operator ${P}_{\mu\nu}$ in (\ref{eq:QuarkGap}) makes natural the use of the {\it transversely projected} version of the quark-gluon vertex. Specifically, for the purposes of the present work we introduce the transversely projected vertex $\fatg_\mu(q,-p)$, defined
through
\begin{align}\label{eq:TransverseQG}
{P}_{\mu\nu}(p-q)\left[\Gamma^{(3)}_{\bar q q A}\right]^a_{\nu}(q,-p) = \identityflavour\,  \colT{a} \, \fatg_\mu(q,-p)\,,
\end{align}
where $\identityflavour$ denotes the identity matrix in flavour space. Note that while $\left[\Gamma^{(3)}_{\bar q q A}\right]^a_{\nu}(q,-p)$ requires twelve tensors for its full decomposition, $\fatg_\mu(q,-p)$ is comprised by a subset of only eight; for more details see, \eg \cite{Mitter:2014wpa, Cyrol:2017ewj}.

With the above definitions, the color contractions in \eq{eq:QuarkGap} can be easily carried out, and we arrive at the standard form of the gap equation,
\begin{align}
	Z_q(p)\left[ i \,  \pslash +M_q(p) \right]   =  Z_{2}\, i \pslash +Z_{m_q}\,m_q  +\Sigma(p)\,,
	\label{eq:QuarkGapDetails}
\end{align}
with the renormalized self-energy
\begin{align}\label{eq:Sigma}
\Sigma(p) = Z^{f}_{1} g_s C_{\!f}  \int_q\frac{1}{Z_A(q-p) (q-p)^2}\,
\gamma_{\mu}\, \frac{1}{Z_q(q) \left[i \,  \qslash + M_q(q)\right]} \, \fatg_\mu(q,-p) \,,
\end{align}
where $C_{\!f}$ denotes the Casimir eigenvalue of the fundamental representation, with $C_{\!f} =4/3$ for $SU(3)$.

Note that \Eq{eq:QuarkGapDetails} is finite due to the regularization of the loop integral, as indicated in \eq{eq:MomReg}.
As mentioned there, the cutoff-dependences of the loop integral and of $Z_2$, $Z_{m_q}$, and $Z^{f}_{1}$, cancel against each other,
giving finally rise to cutoff-independent functions $Z_q(p)$ and $M_q(p)$.
As we discuss in the next section, an analogous renormalization procedure renders the vertex $\fatg_\mu(q,-p)$ cutoff-independent.

The gap equation in \eq{eq:QuarkGapDetails} can be projected on its Dirac vector and scalar parts by multiplying it
with either $\id$ or $\pslash$ and performing the corresponding traces. This leads us to the standard set of coupled SDEs for $Z_q(p)$ and $M_q(p)$,
\begin{align}
Z_q(p) p^2= Z_{2} p^2 - Z^{f}_1 {\rm tr}[i\pslash\Sigma(p)]\,,
\qquad
M_q(p) =  Z^{-1}_q(p)(Z_{m_q} m_q + Z^{f}_1 \,{\rm tr}[ \Sigma(p)])\,.
\label{eq:system}
\end{align}
The different parts in \eq{eq:system}  depend manifestly on the UV-cutoff $\Lambda$, and even diverge for  $\Lambda\to\infty$, while the \textit{finite} gap equation \eq{eq:system} is cutoff independent, but $\mu$-dependent.

We next specify the renormalization conditions at a given renormalization scale $\mu$. We employ a variant (MOM${}^2$) of the non-perturbative version of the momentum subtraction (MOM) scheme. Within the MOM${}^2$ scheme, the renormalized quantum corrections of all primitively divergent vertices with momenta $p_1,...,p_n$ vanish at a symmetric point $\bar p^2=\mu^2$, when
\begin{align}\label{eq:SymPoint}
p^2_i=\bar p^2\,,\qquad \qquad \forall i=1,...,n\,,
\end{align}
exactly as happens in the standard MOM case. In particular, the dressings of the two-point functions reduce to unity,
\begin{subequations}\label{eq:RG-conditions}
\begin{align}\label{eq:RG2}
Z_A(\mu)=1\,, \qquad Z_c(\mu)=1\,,\qquad Z_q(\mu)=1\,, \quad M_q(\mu) =m_q\,,
\end{align}
where $Z_c(p)$ is the dressing associated with the ghost propagator. Similarly, in the case of the vertices,
the symmetric point dressings $\lambda^{(1)}_{\phi_1\cdots\phi_n}(\bar p):=\left. \lambda^{(1)}_{\phi_1\cdots\phi_n}(p_1,...,p_n))\right|_{p_i^2=\bar p^2}$ of the classical tensor structures satisfy
\begin{align}\label{eq:RGvert}
\lambda^{(1)}_{A^3}(\mu)=g_s\,,\qquad \lambda^{(1)}_{A^4}(\mu)=g_s^2\,,\qquad\lambda^{(1)}_{c\bar c A}(\mu)=g_s\,,\qquad\lambda^{(1)}_{q\bar q A}(\mu)=g_s\,.
\end{align}
\end{subequations}
Evidently, all renormalization constants also depend on the subtraction point $\mu$.

Within the renormalization scheme defined above, we have that $\Gamma_{q\bar q}(p^2=\mu^2) = i \pslash + m_q$, and in the standard MOM scheme the respective renormalization factors would be given by
\begin{align}\label{eq:ZMOM}
Z_{2} = 1+ \left.\frac{{\rm tr}[i\pslash Z^{f}_1\Sigma(p)]}{p^2}\right |_{p^2=\mu^2}\,,\qquad Z_{m_q} = 1- \left.\frac{{\rm tr}[ Z^{f}_1 \Sigma(p)]}{m_q}\right|_{p^2=\mu^2} \,.
\end{align}
However, this is no longer the case within the  MOM${}^2$ scheme, where, instead, we use \eq{eq:RGvert} and a modification of \eq{eq:ZMOM}, implemented by a rescaling of the field and triggered by the fRG input data. This is discussed further in \sec{sec:gencon} and the technical details are provided in \app{app:RG-schemes} and in particular in \app{app:MOM2}. Roughly speaking, for \eq{eq:system} and \eq{eq:ZMOM} it amounts to splitting the $Z_2$ and $\Sigma$ into a loop part with momenta $q^2\leq \mu^2$ and one with $q^2\geq \mu^2$, thus emulating the Wilsonian momentum split typically implemented within the fRG approach. Then, the contributions from the region with $q^2\geq \mu^2$ are absorbed into a rescaling of the quark fields. This removes the $\Lambda$-dependence from the different parts of \eq{eq:system}, insuring explicitly the multiplicative nature of renormalization.

The solution of the quark gap equation requires the knowledge of the gluon propagator and the quark-gluon vertex, which, in turn, depend on the quark propagator and further correlation functions, thus leading to an extended system of coupled integral equations, which must be solved simultaneously. Such a complete, fully back-coupled analysis, subject to certain simplifying approximations, is indeed feasible, and has been presented within functional approaches for $N_f=2$ flavour QCD in~\cite{Mitter:2014wpa, Cyrol:2017ewj}. However, the main purpose of the present work is the detailed analysis of the system of quark propagator and quark-gluon vertex, as well as the discussion of quantitative approximation schemes. For this reason we opt for a simpler treatment, which permits us to maintain our focus on the novel aspects of our approach. In particular, the gluon propagator entering into both the gap equation and the vertex SDE will be treated as an external ingredient. Thus, rather than solving its own dynamical equation, we will employ the results obtained in the unquenched lattice simulations of~\cite{Boucaud:2018xup, Zafeiropoulos:2019flq, Aguilar:2019uob} and the functional analysis of~\cite{Cyrol:2017qkl, Fu:2019hdw}.

\section{SDE of the quark-gluon vertex}\label{sec:QuarkGluon}

\begin{figure}[t]
	\vspace{.3cm}
	\includegraphics[width=1\columnwidth]{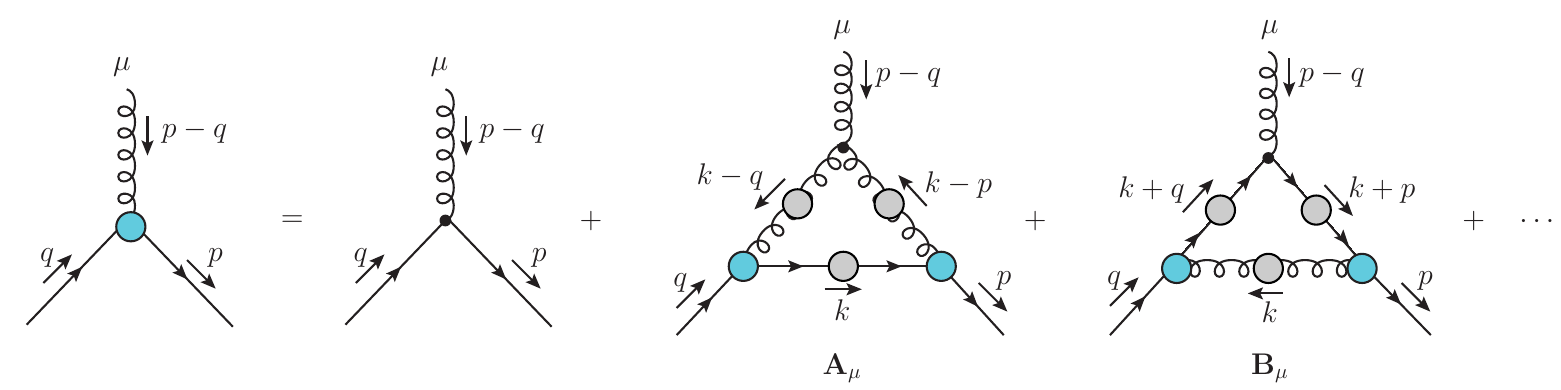}
	\caption{Diagrammatic representation of the quark-gluon SDE. Gray (blue)
		circles denote full propagators (vertices), black dots denote classical vertices. The ellipses denote higher order contributions: diagrams without perturbative one-loop counterparts .\hspace*{\fill}
	}\label{fig:SDEVertex}
\end{figure}

In this section we set up and discuss the SDE for the $\fatg_{\!\!\mu}$ defined in \eq{eq:TransverseQG}, which enters in the quark gap equation.
In the present work we consider the ``one-loop dressed'' approximation of this SDE, which is diagrammatically depicted in Fig.~\ref{fig:SDEVertex}. The terms omitted from this SDE correspond to terms that do not lead to perturbative one-loop contributions. All such graphs may be systematically accounted for by carrying out the so-called ``skeleton expansion'' of the relevant kernels.
In particular, the two graphs depicted in Fig.~\ref{fig:SDEVertex} correspond to the
lowest order terms in the skeleton expansion of the kernels ${\bar q} q A A$ and ${\bar q} q {\bar q} q$.
This functional equation will be projected on its different tensorial components,
thus furnishing a set of dynamical equations governing the respective form factors.

The SDE for the vertex $\fatg_{\!\!\mu}$ is expressed as
\begin{align}
	\label{eq:DSEv}
	\fatg_{\!\!\mu}(q,-p) = Z^{f}_1 g_s {P}_{\mu\nu}(p-q) \,(-i \gamma_\nu) + {\bold A}_\mu(q,-p) + {\bold B}_\mu(q,-p)\,,
\end{align}
with the contributions of the graphs ${\bold A}_\mu(q,-p)$ and ${\bold B}_\mu(q,-p)$ in Fig.~\ref{fig:SDEVertex} given by
\begin{align}
	{\bold A}_\mu(q,-p) =&\,
	\frac{Z_1 N_c}{2} {P}_{\mu\nu}(p-q)
	\!\int_k\Gamma^{(0)}_{\nu\alpha\beta}\, G_{\!A}(k-q)\,G_{\!A}(k-p)\,\fatg_{\!\alpha}(k,-p) \,G_{q}(k) \,\fatg_{\!\beta}(q,-k) \,,
	\nonumber\\[1ex]\nonumber
	{\bold B}_\mu(q,-p) =&\, -\frac{Z^{f}_1}{2N_c} {P}_{\mu\nu}(p-q)\\[1ex]
	&\hspace{.9cm}\times  \int_k G_{\!A}(k)\,
	\fatg_{\!\alpha}(k+p,-p) G_{q}(k+p)\, (-i\gamma_\nu )\,G_{q}(k+q) \fatg_{\!\alpha}(q,-k-q) \,.
	\label{eq:AandB}
\end{align}
In the above formulas, $N_c=3$ for $SU(3)$, the vertex renormalization constants
$Z_1$ and $Z^{f}_1$ were defined after \eq{eq:barSQCD}, and
$\Gamma^{(0)}_{\nu\alpha\beta}$ denotes the classical three-gluon vertex,
\begin{align}
	\Gamma^{(0)}_{\nu\alpha\beta} = g_s \bigl[(2k-p-q)_\nu g_{\alpha\beta} + (2q-p-k)_\alpha g_{\nu\beta} + (2p-q-k)_\beta g_{\alpha\nu}\bigr]\,,
\end{align}
where we have factored out the color factor $f^{abc}$.

The vertex $\fatg_{\!\!\mu}$ may be decomposed in a basis formed by the transverse projections ${P}_{\mu\nu} {\cal T}_{i}^\mu$ of eight independent tensorial structures, denoted by ${\cal T}_{i}^\mu$, which can be derived from gauge-invariant quark-gluon operators~\cite{Mitter:2014wpa, Cyrol:2017ewj}, according to
\begin{align}
\label{eq:barqAq_tensors}
\bar q \slashed{D} q \to {\cal T}_1^{\mu},\qquad \bar q \slashed{D}^2 q \to {\cal T}_2^{\mu}, {\cal T}_3^{\mu},{\cal T}_4^{\mu},\qquad
\bar q \slashed{D}^3 q \to {\cal T}_5^{\mu}, {\cal T}_6^{\mu}, {\cal T}_7^{\mu},\qquad \bar q \slashed{D}^4 q \to {\cal T}_8^{\mu} \,.
\end{align}
The full tensor basis with 12 elements is then given in terms of transverse and longitudinal projections
of the tensors \eq{eq:barqAq_tensors}.
Specifically, introducing  ${P}^\textrm{L}_{\mu\nu} := \delta_{\mu\nu}-{P}_{\mu\nu}$, a concrete choice is given by~\cite{Mitter:2014wpa, Cyrol:2017ewj},
\begin{align}\label{eq:fullbasis}
\Bigl( \{ {P}_{\mu\nu} {\cal T}_{i}^\mu \}\,,\,  {P}^\textrm{L}_{\mu\nu} {\cal T}_{1,2,6,8}^\mu\,\Bigr) \,,
\end{align}
where the projection operators ${P}_{\mu\nu}$ and ${P}^\textrm{L}_{\mu\nu}$
carry the gluon momentum.

In particular, for the transversally projected quark gluon vertex we have
\begin{align}\label{eq:Gamma-mu}
	\fatg^{\mu}(q, -p)= \sum_{i=1}^8  \lambda_{i}(q,-p) P^{\mu\nu}(q-p) {\cal T}_{i}^\nu (q,-p)\,,
\end{align}
where the short-hand notation $\lambda_{i}:=\lambda^{(i)}_{q\bar q A}$ was introduced.

With the aid of \eq{eq:Gamma-mu}, and through appropriate tensor contractions,
the starting SDE of \eq{eq:DSEv} may be converted into a system of coupled integral equations for the $\lambda_i(p,q)$.
Specifically, one obtains
\begin{align}
\lambda_i(q,-p)= Z^{f}_1 g_s \,\delta_{i {\s 1}} + \, a_i(q,-p) + \, b_i(q,-p)\,,  \qquad i=1,...,8
\label{eq:forlambdapre}
\end{align}
with
\begin{eqnarray}
a_i(q,-p) &=& \frac{Z_1 N_c}{2}\,\! \int \frac{d^4{k}}{(2\pi)^4}\lambda_j(k,-p)\lambda_k(q,-k) G_{\!A}(k-q)\,G_{\!A}(k-p) K_{ijk}(p,q,k) \,,
\nonumber\\
b_i(q,-p) &=& -\frac{Z^{f}_1}{2N_c} \!\int \frac{d^4{k}}{(2\pi)^4} \lambda_j(k+p,-p)\lambda_k(q,-k-q) G_{\!A}(k) \widetilde{K}_{ijk}(p,q,k)\,,
\label{eq:aandb}
\end{eqnarray}
where the kernels $K_{ijk}(p,q,k)$ and $\widetilde{K}_{ijk}(p,q,k)$ contain combinations of $Z_q$, $M_q$, and the various momenta; further information on their precise structure is provided in \app{app:cof}.

The renormalization condition corresponding to \eq{eq:RG-conditions} dictates that, at the symmetric point $\bar p^2=\mu^2$, we must impose
\begin{align}\label{eq:Zf1-renorm}
	Z^f_1 g_s = g_s -  \bigl[a_1(q,-p) + b_1(q,-p)\bigr]_{p^2=q^2=\mu^2} \,.
\end{align}
This leads us to the final, explicitly renormalized coupled integral equations for the $\lambda_i(p,q)$,
\begin{align}
\lambda_i(p,q)=  a_i(p,q)+b_i(p,q) +\left( g_s - \bigl[a_i(q,-p) + b_i(q,-p)\bigr]_{p^2=q^2=\mu^2} \right) \delta_{i {\s 1}}  \,,
\label{eq:forlambda}
\end{align}
which satisfies manifestly \eq{eq:RGvert}.

So far we have described the standard non-perturbative MOM scheme. Now we implement the MOM${}^2$ scheme, described in detail in \app{app:RG-schemes}. Roughly speaking, as in the case of the gap equation, we split the $Z$'s and the loop contributions into those with $k^2\leq \mu^2 $ and those with $k^2\geq \mu^2$. After appropriate rescalings, as was done with the gap equation \eq{eq:system}, all parts in the vertex SDE are manifestly independent of the ultraviolet cutoff $\Lambda$.

As we will see in detail in \sec{sec:results}, the numerical treatment of the system of coupled integral equations given by Eqs.(\ref{eq:system}), (\ref{eq:forlambda}), and (\ref{eq:aandb}), reveals a clear hierarchy among the dressings $\lambda_i$. In particular, depending on their numerical impact, the $\lambda_i$ may be naturally separated  into ``{\it dominant}'', ``{\it subleading}'', and ``{\it negligible}''.

Specifically, the three dominant components of the quark gluon vertex are $\lambda_{1,4,7}$, associated with the tensor structures
\begin{subequations} \label{eq:1-8}
\begin{align}\label{eq:147}
	{\cal T}_{1}^\mu(p,q) =-i \gamma^{\mu}\,, \qquad {\cal T}_{4}^\mu(p,q) = (\pslash+\qslash)\gamma^\mu\,,
	\qquad {\cal T}_{7}^\mu(p,q) =\frac{i}{2} [\pslash,\qslash]\gamma^\mu\,.
\end{align}
As we will see in \sec{sec:results} , keeping only these three form factors in the coupled SDE analysis, \ie the terms corresponding to $i=1,4,7$ in \eq{eq:aandb}, already furnishes quantitatively accurate results for our benchmark observable, the RG-invariant chiral condensate. It is important to emphasise that out of these three dominant structures, only $\lambda_{1}$ is accessible to an STI-based derivation of the quark gluon vertex, in the spirit of the original BC construction.

The three subleading components, $\lambda_{2,5,6}$, are associated with the basis elements
\begin{align}\label{eq:256}
	{\cal T}_{2}^\mu(p,q) =(q-p)^\mu\,,
	\qquad	{\cal T}_{5}^\mu(p,q) =i(\pslash+\qslash) (p-q)^\mu\,,\qquad
	{\cal T}_{6}^\mu(p,q)  = i(\pslash-\qslash)(p-q)^\mu\,.
\end{align}
These three dressings may be obtained from the STI-based constructions, implemented only in the vacuum.
Therefore, in view of the numerous applications to QCD at finite temperature and density,
SDE-based computations of these subleading tensor structures, such as the one put forth here, are clearly preferable.

Finally, the form factors associated with the tensors
\begin{align}\label{eq:38}
	{\cal T}_{3}^\mu(p,q) = (\pslash-\qslash)\gamma^\mu\,, \qquad
	{\cal T}_{8}^\mu(p,q) = -\frac12[\pslash,\qslash] (p-q)^\mu\,,
\end{align}
\end{subequations}
are negligible, having no appreciable numerical impact on our benchmark observable or any other relevant quantity (see also \cite{Gao:2020qsj, Gao:2020fbl}).

This concludes the description of our SDE setup.

\section{External input and self-consistent renormalization}\label{sec:gencon}

In this section we discuss self-consistent renormalization schemes for the SDE with a given external input. This
issue is addressed both in general and for the given input data for the gluon propagator used here.
In addition, we detail the origin and characteristics of these data.

In \sec{sec:MultRenorm} we elaborate on the implementation of multiplicative renormalization in the present MOM${}^2$ scheme in the present non-perturbative approach; there, and in \app{app:RG-schemes}, we also emphasise the differences to the standard MOM scheme. In \sec{sec:InputGluon} we provide an overview on the gluon propagator data used as input, in \sec{sec:SC-alphas} we discuss the general self-consistent determination of the value of the renormalized coupling $\alpha_s(\mu)$ at the renormalization scale $\mu$, and in \sec{sec:SC-alphasDet} we determine $\alpha_s(\mu)$ for the gluon input data specified in \sec{sec:SC-alphas}.

\subsection{Multiplicative renormalization in the MOM${}^2$ scheme} \label{sec:MultRenorm}

The self-consistent implementation of multiplicative renormalization at the level of the non-perturbative SDEs constitutes a yet unresolved problem, which has been treated only approximately within numerical applications,
see, \eg~\cite{Bloch:2001wz, Bloch:2002eq, Fischer:2003rp, Aguilar:2010cn, Aguilar:2018epe, Huber:2018ned, Huber:2020keu}.
In the present context, the complications stemming from this issue manifest themselves at the level of the gap equation by the presence of the factor $Z_1^f$ in the definition of the quark self-energy $\Sigma (p)$, and at the level of the SDE for $\fatg^{\mu}$ through the factors $Z_1$ and $Z_1^f$ entering in the expressions for ${\bold A}_\mu$ and ${\bold B}_\mu$, respectively.

Evidently, the renormalization constants $Z_1, Z_f$ display a non-trivial (``marginal'') dependence on the UV cutoff, which is required for rendering the diagrams finite. However, the order-by-order  cancellation known from perturbation theory does not translate straightforwardly to the non-perturbative setup of the SDEs. In this work we adopt the MOM${}^2$ scheme, which is a modification of the standard MOM scheme and its approximation used in the SDEs. In fact, the MOM${}^2$ scheme is commonly employed in fRG applications to QCD~\cite{Cyrol:2017ewj, Fu:2019hdw}, and has been also used in recent SDE applications~\cite{Gao:2020fbl, Gao:2020qsj}. Our gluon input data are taken from these sources, and hence, the MOM${}^2$ scheme is the natural one for their implementation. In \app{app:RG-schemes}, for the first time, we present a technical derivation from a Wilsonian approach to the path integral, as well as discussion of the mapping of the MOM${}^2$ quark-gluon coupling, $\alpha_{s,\textrm{MOM}^2}$, to the standard MOM coupling, $\alpha_{s,\textrm{MOM}}$. The full setup will be explained elsewhere. Its spirit is entailed in the following consideration, already mentioned below \eq{eq:ZMOM} and \eq{eq:forlambda} : for a given RG-scale $\mu$, we split the loop contribution to the $Z$'s and the diagrams into those with loop momenta $q^2 \leq \mu^2$ and $q^2 \geq \mu^2$. The latter contributions are absorbed into a respective rescaling of the fields. This leaves us with a  (unique by virtue of the STIs) ratio of renormalization functions in front of all diagrams, always occuring together with the coupling. Hence, we simply absorb this ratio in the definition of the coupling. The remaining contributions to the MOM${}^2$ $Z's$ are finite and, at the RG-point $p^2=\mu^2$, they only carry powers of the running coupling $\alpha_s(\mu)$, and no logarithms. Accordingly, these contributions vanish for $\mu\to \infty$ and hence can be dropped for sufficiently large RG-scale $\mu$, invoking asymptotic freedom: $\alpha_s(\mu\to\infty)\to 0$. This leads to $Z_{\textrm{MOM}^2}\to 1$ for all renormalization functions at the level of the SDEs. Now, in the MOM${}^2$ scheme, the MOM condition \eq{eq:RG-conditions} is implemented at the level of the correlation functions of the rescaled fields. Note, that consequently correlation functions of the standard renormalized fields do not satisfy \eq{eq:RG-conditions}. Finally, the MOM${}^2$ scheme can be easily implemented at the technical level, by means of a standard BHPZ subtraction of the diagrams at the renormalization point, setting $Z_i=1$ with $Z_i\in (Z_3,\tilde Z_3, Z_2, Z_g)$, leading to unity vertex renormalizations via the STIs, \ie$Z_1=\tilde Z_1=Z_3=Z_4=Z_1^f=1$. The conditions on the $Z$'s reflect the rescaling of fields and vertices. For these trivial $Z$'s, the BHPZ subtraction at the RG scale $\mu$ evidently implements the MOM conditions. Moreover, all parts of the SDEs are finite, and the loop integration can be extended to infinity.

For more details, including the relations of the running couplings as well as $\Lambda_\textrm{QCD}$ in the standard MOM scheme and present MOM${}^2$ scheme, we refer the reader to \app{app:RG-schemes}. As already mentioned above, the use of the fRG input data for the gluon propagator is one of the main reasons for resorting to this modification of the standard MOM scheme, as then the RG condition on the input data and the SDEs coincide. Another reason is its operational simplicity. In particular, in the MOM${}^2$ scheme the renormalization functions need not to be tuned for achieving cutoff-independence. Nonetheless, while the MOM${}^2$ scheme is the natural and fully consistent RG scheme within the fRG approach, the considerations here and in \app{app:RG-schemes} do not constitute a proof of the full self-consistency of the present operational procedure in the SDE, which is the subject of ongoing work.

In summary, for the numerical treatment of the system of integral equations presented here, we simply implement the substitution
\begin{align}
	\Sigma(p)\to \Sigma(p)\!\!\mid_{Z_1^f=1} \,,  \qquad\qquad 	\bigl[a_i(p,q), b_i(p,q)\bigr] \to \bigl[a_i(p,q), b_i(p,q)\bigr]_{Z_1^f=1=Z_1}\,,
	\label{mulren}
\end{align}
as well as a BHPZ-subtraction. It is evident from the discussion above and in \app{app:RG-schemes}, that the simplifications implemented by (\ref{mulren}) are bound to induce a residual small cutoff- and  $\mu$-dependence to the results obtained, which are discussed in \sec{sec:Stability}.

\begin{figure}[t]
	\hspace{-1cm}
	\includegraphics[width=0.45\textwidth]{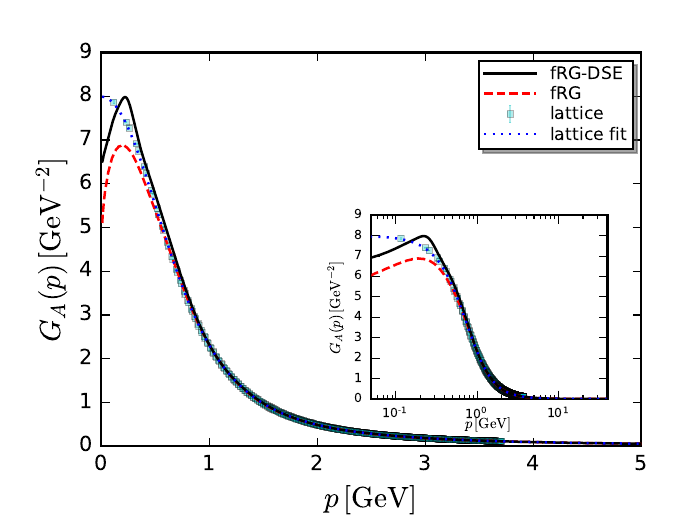}\hspace{1cm}
	\includegraphics[width=0.45\textwidth]{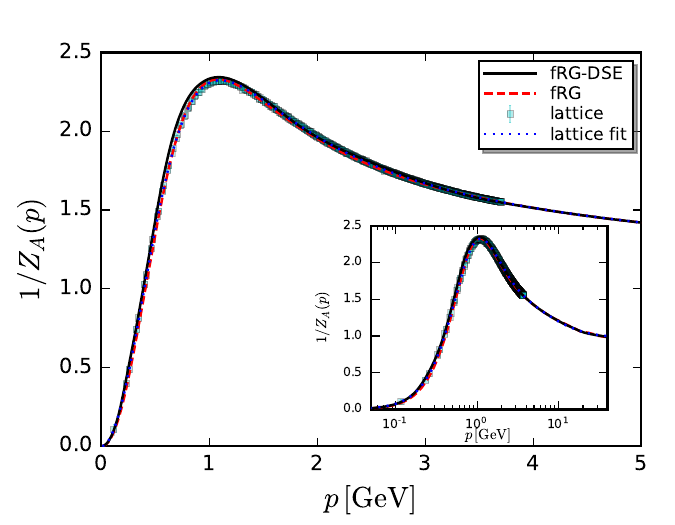}
	\caption{ 2+1--flavour gluon propagator, $G_{\!A}(p^2)$, and  dressing function $1/Z_A(p)= p^2\,G_{\!A}(p)$.
		Lattice simulations: \cite{Boucaud:2018xup, Zafeiropoulos:2019flq, Aguilar:2019uob}, fRG-DSE approach: \cite{Gao:2020fbl, Gao:2020qsj},  fRG approach: \cite{Fu:2019hdw}. The computations in~\cite{Gao:2020fbl, Gao:2020qsj, Fu:2019hdw} are based on the 2-flavour input fRG data from~\cite{Cyrol:2017ewj}.\hspace*{\fill}} \label{fig:Ginput}
\end{figure}
%

\subsection{Gluon propagator} \label{sec:InputGluon}

The gluon propagator can be computed from its own SDE; for the most recent results in Yang-Mills theory, see~\cite{Cyrol:2014kca, Huber:2016tvc, Huber:2017txg, Huber:2020keu},
while for 2+1--flavour solutions of the fRG-assisted SDE, see~\cite{Gao:2020fbl, Gao:2020qsj}.
Consequently, we could extend the current system to a fully self-coupled one, the only input being the strong coupling and the current quark masses. However, in this work we concentrate rather on the novel key ingredient, namely the computation of the full transversally projected quark-gluon vertex and its properties. Therefore, we simply take quantitative input data from either the lattice~\cite{Boucaud:2018xup, Zafeiropoulos:2019flq, Aguilar:2019uob}, the SDE~\cite{Aguilar:2008xm, Aguilar:2012rz, Gao:2020fbl, Gao:2020qsj}, or the fRG~\cite{Fu:2019hdw}.

For the present computation, the input data have to cover the momentum region \mbox{$p^2 \in [0,\Lambda^2]$}, where $\Lambda$ is the UV cutoff of the loop integrals in the SDEs. We emphasise that the loops are finite in the limit $\Lambda\to\infty$, and the contributions of loop momenta $q^2 \gg \mu^2$ decay rapidly thanks to the BPHZ subtractions. The introduction of a UV cutoff in the numerical implementation is done only for operational speed-up and stability. Accordingly, the checks of UV cutoff-independence in the MOM${}^2$ scheme are only tests of the numerical procedure. This is to be contrasted with the checks of UV cutoff-independence in the standard MOM scheme. There, the validity of the determination of the correct $\Lambda$-dependence, and hence of the values of the renormalization functions, is indeed probed.

In the present work, we consider UV cutoffs in the range $\Lambda=50-5000$\,GeV for testing the cutoff-independence of the results, see \sec{sec:Lambda-Indep}. While the functional input data cover the full momentum regime,  lattice input data are restricted within $p\lesssim 4$\,GeV. Consequently, they have to be extrapolated towards the UV; the best extrapolation is provided by the functional input data, which agree quantitatively with the lattice data for $p\gtrsim 1$\,GeV.

Next, we provide a physically motivated fit of the resulting
``functional-lattice'' propagator, valid within the regime $p\in [0, 40]$\,GeV, which incorporates explicitly
the one-loop resummed running of $G_{\!A}(p)$. In particular,
\begin{align}
G_{\!A}(p)=\frac{(a^2+p^2)/(b^2+p^2)}{M^2(p^2)+p^2\left[1+ c\,\ln \left(d^2\, p^2 + e^2\, M^2(p^2)\right)\right]^{\gamma}}\,,\qquad \textrm{with}\qquad
M^2(p)=\frac{f^4}{g^2+p^2}\,,
\label{eq:propfit}
\end{align}
where $\gamma=(13-4/3 N_f)/(22-4/3 N_f)$ denotes the one-loop anomalous dimension of the gluon propagator in the Landau gauge. The optimized
values of the fitting parameters are given by
\mbox{$\{a,b,c,d,e\}=\{1\, {\rm GeV}, 0.735 \,{\rm GeV},\, 0.12,\, 0.0257\, {\rm GeV}^{-1}, 0.081 \,{\rm GeV}^{-1}\}$}, together with
\mbox{$\{f,g\}=\{0.65 \,{\rm GeV}, 0.87 \, {\rm GeV}\}$}.
As can be seen in Fig. \ref{fig:Ginput}, this fit matches very accurately the input points in the physically relevant region of momenta.

Note also that, as can be seen in Fig. \ref{fig:Ginput}, the input data for $G_{\!A}(p)$ differ in the infrared, \ie for $p^2 \lesssim 1$ GeV (\textit{``scaling''}~\cite{Gao:2020fbl, Gao:2020qsj, Fu:2019hdw}, vs \textit{``decoupling''} or \textit{``massive''}~\cite{Boucaud:2018xup, Zafeiropoulos:2019flq, Aguilar:2019uob}; for related discussions, see, \eg~\cite{Aguilar:2008xm, Boucaud:2008ky, Fischer:2008uz}). Nonetheless, the quark propagator obtained using either of them, as well as the computed physical observables, agree within our systematic error bars. The reason for this is related to the fact that, inside the quantum loops considered here, the gluon propagator $G_{\!A}(p)$ is eventually multiplied by $p^2$; as a result, the infrared differences are largely washed out, and the relevant quantity, $Z^{-1}_A(p) = p^2 G_{\!A}(p)$, is practically identical for both.

\subsection{Self-consistent determination of $\alpha_s(\mu)$} \label{sec:SC-alphas}

A necessary ingredient for our analysis is the value of the dressing $\lambda_1$ at the symmetric point, which,
for sufficiently large values of $\mu$ is equal to the (unique) perturbative $g_s$, or the $\alpha_s$ defined in \eq{eq:avatars}, see also (\ref{eq:RGvert}). In our study we use as external input the $N_f=2+1$ gluon propagator from lattice and functional  methods, the renormalization procedures adopted in those earlier computations need be incorporated into the present SDE treatment, such that a self-consistent value for $\alpha_s(\mu)$ may be obtained. This naturally leads us to the MOM${}^2$ renormalization scheme used here,  as well as within the fRG and SDE  computations in \cite{Mitter:2014wpa, Cyrol:2016tym, Cyrol:2017ewj, Gao:2020fbl, Gao:2020qsj}.

The self-consistent calibration of $\alpha_s(\mu)$, in any scheme, may be implemented according to two different procedures, \textit{(i)} and \textit{(ii)}, detailed below. In \textit{(i)}, one compares correlation functions computed within the present SDE setup, whose form depends on the
value of $\alpha_s$ used, with data for them originating from the same framework that
provides the required external input.
In \textit{(ii)}, one invokes self-consistency conditions
between the results of the current SDE approach and those derived from the STIs.

We emphasise that both procedures are optimised when implemented
in the perturbative and semi-perturbative regime with,
\begin{align}\label{eq:ppert}
p\gtrsim p_\textrm{pert}\,,\qquad \textrm{with} \qquad p_\textrm{pert}\approx 4\,\textrm{GeV}\,,
\end{align}
where the truncation errors are small and under control; instead, their extension
to the non-perturbative infrared regime is bound to  worsen the calibration.
In fact, while in the regime of (\ref{eq:ppert})
STI- as well as vertex-couplings agree at least up to two loops,
they deviate markedly from each other as $p \to 0$, see~\cite{Cyrol:2017ewj}. Moreover,
the regularity assumption that is implicit in the direct use of the STIs for the determination of transverse couplings
[as in \eq{eq:STI-coupling}] may fail in the infrared; for more details, see~\cite{Aguilar:2011xe,Aguilar:2016vin,Cyrol:2016tym, Cyrol:2017ewj}.

The concrete implementation of \textit{(i)} and \textit{(ii)} is presented in detail below.

\textit{(i)} The setup in the present work only requires the data for the $2+1$-flavour gluon propagator as external input from either distinct SDE approaches, the fRG, or lattice simulations. Within all these frameworks, one has access to
data sets not only for this specific input but also for additional
correlation functions, as well as derived couplings, \eg via \eq{eq:avatars}.
While the latter are not needed as explicit inputs for the SDE, they can be used as a means of calibrating the calculation, because they
can be recomputed from their own dynamical equation within the present SDE setup. In doing so, it is clear
that their momentum dependence changes as the value of $\alpha_s(\mu)$ is varied.
Thus, the external data sets may be reproduced
for a unique self-consistent choice of $\alpha_s(\mu)$, which calibrates our approach (for $p\gtrsim p_\textrm{pert}$).

If a propagator is chosen for the purpose of calibration, the ghost propagator is clearly the best choice, as it is governed by a rather simple SDE, whose only other ingredient is the ghost-gluon vertex, which is protected by Taylor's non-renormalization theorem. For example, if one were to use as external
input the gluon propagator from the lattice, the calibration proceeds by computing the ghost dressing function $Z_c(p^2)$ within our SDE setup,
adjusting the  $\alpha_s(\mu)$ such that the lattice data for $Z_c(p^2)$ will be best reproduced.

If one of the running couplings, $\alpha_i(\bar p)$, is employed for the calibration,
we compute its shape within both the approach that furnishes the external input
and within the SDE setup. Then, self-consistency requires that the SDE $\alpha_s(\mu)$ is chosen such that the
difference between the two $\alpha_i(\bar p)$'s is minimised, for $\bar p\gtrsim p_\textrm{pert}$.

We close with the remark that, in the present setup, all procedures mentioned above lead to $\alpha_s(\mu)$ that agree within the small numerical and systematic errors. We consider this an important self-consistency check of the SDE approach put forth here.

\textit{(ii)} If the additional results needed for the implementation of \textit{(i)} are unavailable, one can use the STI satisfied by the quark-gluon vertex in order to fix $\alpha_s(\mu)$. Specifically, for $\bar p\gtrsim p_\textrm{pert}$ one uses the relation
\be\label{eq:STI-coupling}
\lambda_1(\bar p)  = g_s(\mu) L_1 (\bar p)\,,
\ee
where $L_1$ is the solution of the STI for the longitudinally projected classical tensor structure, see~\cite{Aguilar:2010cn, Cyrol:2017ewj}. As in \textit{(i)},
minimising the difference between the two sides of \eq{eq:STI-coupling} singles out a unique $\alpha_s(\mu)$.

This concludes our general discussion of the self-consistent determination of $\alpha_s(\mu)$.

\subsection{Value of  $\alpha_s(\mu)$} \label{sec:SC-alphasDet}

In this work  we use two classes of data for fixing $\alpha_s(\mu)$, and employ both procedures, \textit{(i)} and \textit{(ii)}, described above;
both procedures \textit{(i)} and \textit{(ii)} are used when $G_{\!A}(p)$ is obtained from functional methods, while \textit{(ii)} is applied when $G_{\!A}(p)$ is taken from the lattice. For both classes of data, and for very different renormalization scales [$\mu=4.3,\, 40$\,GeV], we will produce results for the quark propagator and the pion decay constant that agree within our estimated systematic error; this coincidence, in turn, constitutes a non-trivial check of the  systematic errors. We next describe the determination of $\alpha_s(\mu)$ for both cases:

\textit{Functional data sets:} Here we employ procedure \textit{(i)}. The functional input for $G_{\!A}(p)$ is provided by fRG~\cite{Fu:2019hdw} and SDE~\cite{Gao:2020fbl, Gao:2020qsj} data, renormalized at $\mu= 40$\,GeV. Note that the respective SDE and fRG relations for correlation functions are expanded about their $N_f=2$ counterparts, computed within the fRG~\cite{Cyrol:2017ewj}. The respective data sets also include $\alpha_{q \bar q A}(\bar p)$, thus providing directly $\alpha_s(\mu)$ at $\mu=40$\,GeV; this allows us to minimise the difference between input and output $\alpha_{q \bar q A}(\bar p)$, for $\bar p\gtrsim p_\textrm{pert}$.

\textit{Functional and lattice data sets:} For both, functional and lattice input data, we employ procedure \textit{(ii)}. We use the lattice data for $G_{\!A}(p)$  from~\cite{Boucaud:2018xup, Zafeiropoulos:2019flq, Aguilar:2019uob}, and, in line with our arguments, we chose the maximal lattice momentum available for our renormalization scale, namely $\mu=4.3$\,GeV. The STI function $L_1$ in \eq{eq:STI-coupling} is computed based on the quenched computation in~\cite{Aguilar:2016lbe} (Fig.17, fourth panel), properly accounting for unquenching effects. Then, we minimise the difference between the left- and right-hand sides in \eq{eq:STI-coupling}. The same procedure is applied to the functional data set at both $\mu=4.3$\,GeV and $\mu=40$\,GeV.

Both procedures  are now applied at two rather disparate renormalization scales $\mu$, namely $\mu_1= 4.3$\,GeV and $\mu_2= 40$\,GeV, for which we obtain the MOM${}^2$ scheme values,
\begin{align}
 \alpha_{s}(4.3\,\textrm{GeV}) = 0.433\,,  \qquad \alpha_{s}(40\,\textrm{GeV}) = 0.166\,,
\label{eq:twomu}
\end{align}
which are fully compatible with earlier SDE and fRG considerations. The coincidence of the couplings for both procedures \textit{(i)} and \textit{(ii)} is a further non-trivial consistency check of the present RG scheme, and corroborates the correct implementation of the MOM${}^2$ scheme in the present SDE approach, for more details see \app{app:RG-schemes}. The value of the coupling at the scale $M_Z$ obtained in this setup is $\alpha_s(M_Z)=0.14$. The deviation from the standard value $\alpha_{s,\textrm{phys}}(M_Z) \approx 0.118$ is due to the presence of only three active flavours in our analysis and the modifications in the present scheme in comparison to the standard MOM scheme.

In \app{app:MOM2-MOM} we discuss in detail the mapping between the present MOM${}^2$ results to MOM results, in particular for the running quark-gluon coupling and $\Lambda_{\textrm{QCD}}$, see \eq{eq:MOM2toMOM}, \eq{eq:MOMLambda+zs}, \eq{eq:alphaMOMp} and \fig{fig:Couplingcomp}. For the details, we refer the reader to this Appendix. Here we only quote results for the running coupling in the MOM scheme at selected momenta, $p=4.3, 40$\,GeV, together with the respective $\Lambda_{\textrm{QCD}}$,
\begin{align}
	\alpha_s(4.3\,\textrm{GeV}) = 0.365\,,  \qquad \alpha_s(40\,\textrm{GeV}) =  0.140\,,\qquad  \Lambda_\textrm{QCD}=  708(3)\, \textrm{MeV}\,,
	\label{eq:twomuMOM}
\end{align}
We also obtain $\alpha_s(M_Z) = 0.119$, to be compared with $\alpha_s(M_Z)\approx 0.118$ (six flavour QCD) and $\Lambda_{\textrm{QCD}}=710$\,MeV \cite{Deur:2016tte}. For comparison see also the recent lattice estimate $\Lambda_{\textrm{QCD}}=664$\,MeV in \cite{Zafeiropoulos:2019flq, Cui:2019dwv}. Note that the definition of $\Lambda_{\textrm{QCD}}$ in \cite{Deur:2016tte, Zafeiropoulos:2019flq, Cui:2019dwv}	involve a rescaling relative to our definition, which has been taken into account in the values given here.

\section{Current quark masses and benchmark predictions }\label{sec:CurrentQuark}

In this section we determine the  fundamental parameters of QCD, the current-quark masses $m_q=(m_l, m_s)$, where we have assumed isospin symmetry with identical up and down quark current masses: $m_{u/d}=m_l$. In addition,
we provide results for a benchmark observable, namely
the light chiral condensate, $\Delta_l=- \langle \bar l(x) l(x)\rangle$, which allows us to
evaluate the veracity of the present approximations. In particular, we find that our
$\Delta_l$ is in excellent quantitative agreement with the most recent lattice estimates reported in~\cite{Aoki:2019cca}.

The current quark masses $m_q(\mu)$ at a given $\mu$ are fixed from the physical pion mass, $m_{\pi}$, and the ratio of strange and light current quark masses, $m_s(\mu)/m_l(\mu)$. This procedure has been used both in~\cite{Mitter:2014wpa, Cyrol:2017ewj, Fu:2019hdw, Gao:2020qsj, Gao:2020fbl} (see also the reviews~\cite{Fischer:2006ub, Eichmann:2016yit, Fischer:2018sdj, Dupuis:2020fhh}), and in lattice simulations (see, \eg the compilation in \cite{Aoki:2019cca}).

Note that,  due to the identical (one-loop) RG-running of all $m_q(\mu)$,
the mass ratio $m_s(\mu)/m_l(\mu)$ tends to a constant for asymptotically large $\mu$,
\begin{align}
	\label{eq:ratio}
        \lim_{\mu\to \infty} \frac{m_s(\mu)}{m_l(\mu)}=\frac{m_s}{m_l}\,,\qquad \qquad m_q(\mu)
        \to \frac{m_q}{[\ln (\mu/\Lambda_{\textrm{QCD}})]^{{\gamma_m}} }\,, \quad \textrm{with}\quad \gamma_m=\frac{12}{33- 2 N_f}\,.
\end{align}
In the present work we use $\mu=40$\,GeV, and compare the results to those obtained with a considerably lower $\mu=4.3$\,GeV, whose choice was dictated by the restricted momentum range of the lattice input.

For $\mu=40$\,GeV, the $m_q(\mu)$ will be determined using the values
\begin{align}
	\label{eq:PhysPar}
	\Bigl( m_q(\mu), \mu=40\,\textrm{GeV}\Bigr) :\ m_\pi = 138\,\textrm{MeV}\,,\qquad \textrm{and} \qquad \frac{m_s}{m_l}=27\,.
\end{align}

\subsection{Pagels-Stokar formula and Gell-Mann--Oakes--Renner relation} \label{sec:PS+GMOR}

Ideally, the pion mass, $m_\pi$, and decay constant, $f_\pi$, should be determined from the on-shell properties of the BS wave function of the pion. We emphasise, that in the present work we do not aim at a precise determination of the pion decay constant, and the approximate values quoted below are only given here for the benefit of the reader. Importantly, this approximate value does not enter the computation of correlation functions with the SDE, nor is it used fo the determination of absolute scales. The latter are set by our quantitative gluon input, and the accuracy of the present approximation is solely tested by our benchmark observable, the chiral condensate in the chiral limit.

Accordingly, we simply employ a standard Euclidean approximations for the pion decay constant, given by the Pagels-Stokar (PS)  formula for $f_\pi$, \eq{eq:PS}. For the pion mass we use the Gell-Mann--Oakes--Renner (GMOR) relation for $m_\pi$ in terms of the chiral condensate, \eq{eq:fpi+mpi}. The GMOR relation is correct up to order ${\cal O}(m_l^2)$ within an expansion about the chiral limit, while the PS formula is known to underestimate $f_\pi$ by  $\lesssim 10\% $ (see, \eg  \cite{Bender:1997jf,Gao:2017gvf} and the reviews~\cite{Fischer:2006ub, Bashir:2012fs, Eichmann:2016yit}). We emphasise that this low value does not undermine the precision of our analysis, given that $f_\pi$ is a derived quantity that does not feed back into the SDEs.

The PS formula reads
\begin{align}
	f^{(\textrm{PS})}_\pi =\,\frac{4N_c}{N_\pi}\int_p \frac{Z_2}{Z_q(p)}\frac{{\overbar M}_q(p)}{\left[p^2+M_q^2(p)\right]^2}\left[M_l(p)  -\frac{p^2}{2}M_q^{\prime}(p)\right]\,,
	\label{eq:PS}
\end{align}
with $M_q'(p)= \partial_{p^2} M_q(p)$, and the subtracted mass function ${\overbar M}_{\!q}(p)$,
\begin{align}\label{eq:barM}
	{\overbar M}_{\!q}(p) := M_q(p) - m_q\,\frac{\partial M_q(p)}{\partial m_q} \,, \qquad \textrm{with}\qquad  \lim_{p\to\infty} p^2\,{\overbar M}_{\!q}(p) = 0\,.
\end{align}
Note that \Eq{eq:barM} applies to both $m_l$ or $m_s$. We use the standard PS formula, and in the present MOM${}^2$ scheme this amounts to $Z_2\approx 1$ at $\mu=40$\,GeV, with a $\lesssim 1\%$ error. Note that, in contradistinction to the standard MOM scheme, $Z_2$ does not depend on the UV cutoff, which may be safely sent to infinity. Moreover, we have $Z_2(\mu\to\infty)=1$, for more details see \app{app:RG-schemes}.

The normalization $N_\pi$ of the pion wave function is given by
\begin{align}
	N_\pi = \frac{1}{2} \left[ f_\pi + \sqrt{f^2_\pi + 8 N_c \,I_l} \, \right] \,, \label{eq:Npi}
\end{align}
with
\begin{align}
	I_q := \int_p \frac{{\overbar M}_q^2(p) \,\left[ p^2\, Z_q(p) Z_q''(p)+2 Z_q(p)\,Z_q'(p)-p^2 Z_q''(p) \right]}
	{Z^2_q(p)\,\left[p^2+M^2_q(p)\right]}\,,
        \label{eq:Iq}
\end{align}
where the abbreviations $Z'(p)= \partial_{p^2} Z(p)$, and $Z''(p)= \partial_{p^2} Z'(p)$ have been used. Note that, due to the asymptotic behavior of ${\overbar M}_{\!q}(p)$ stated
in \eq{eq:barM}, the integral is finite.
The above expressions for $N_\pi$ are approximate; a more complete treatment requires all components of the pion wave function, and will be considered elsewhere.

It is well-known that, in the chiral limit, we have $I_{q,\chi}=0$ and $N_\pi=f_\pi$~\cite{Maris:1997hd}.
However, our approximations deviate slightly from this result, furnishing a  $I_{q,\chi}$ which fails to vanish
by an amount that induces a $3\%$ discrepancy between  $N_\pi$ and $f_\pi$.
We effectively account for this small error by setting
\begin{align}\label{eq:DeltaI}
	I_q \to I_q -I_{q,\chi}\,,
\end{align}
thus compensating, in a simple way, for the contributions of the omitted form factors.
The numerical impact of this adjustment will be discussed in \Sec{sec:DetCurQuark}, see \eq{eq:fpiwo}.

In the chiral limit we arrive at
\begin{align} \label{eq:fpiChiral}
	f^{(\textrm{PS})}_{\pi,\chi}= 84.9\,\textrm{MeV} \,,
\end{align}
in comparison to the FLAG estimate of $	f^\textrm{lat}_{\pi,\chi}=86.7$\,MeV and $f^\textrm{lat}_{\pi}=92.1 (0.6)\,$MeV for physical quark masses.

Turning to the GMOR relation, we have
\begin{align}
	m^2_\pi=\frac{2 \, m_l}{ f^2_\pi}\Delta_l + {\cal O}(m_l^2)\,,\qquad \textrm{with} \qquad \Delta_l  = -\langle \bar u u\rangle=-\langle \bar d d\rangle\,,
	\label{eq:fpi+mpi}
\end{align}
where $m_l$ is a $\mu$-independent current quark mass,
and $\Delta_l$ denotes the finite and $\mu$-independent light quark condensate; for a
concise discussion of its RG properties, see ~\cite{Miransky:1994vk}.

Since in our calculations enter the $\mu$-dependent current quark masses, $m_l(\mu)$, rather than $m_l$,
for the purposes of the present work we find it advantageous to capitalise on the property
\begin{align}
m_l \,\Delta_l  =  m_l(\mu) \,\Delta_l(\mu)\,,
\label{eq:mDRGI}
\end{align}
and recast \eq{eq:fpi+mpi} in the form
\begin{align}
	m^2_\pi=\frac{2 \, m_l(\mu)}{ f^2_\pi}\Delta_l(\mu)\,.
	\label{eq:GMOR}
\end{align}
Evidently, in order to extract from \eq{eq:GMOR} the value of $m_l(\mu)$, one requires knowledge of $f_\pi$ and $\Delta_l(\mu)$. Given the aforementioned shortcomings of the PS formula, we will simply use the physical value of $f_\pi$ as input in \eq{eq:fpi+mpi}. Consequently, the systematic error associated with the determination of the pion mass is a combination of the size of the subleading corrections in the $m_l(\mu)$-expansion of \eq{eq:GMOR}, and the error in the computation of the chiral condensate $\Delta_l(\mu)$. We therefore use a very conservative systematic error estimate of 10\% percent, whose derivation details are discussed separately below.

\subsection{Light chiral condensate} \label{sec:ChiralCond}

The benchmark observable for our computation is the $\mu$-independent $\Delta_l$,
whose {\it chiral limit} value, $\Delta_{l,\chi}$,
can be extracted from the UV behaviour of the corresponding constituent quark mass $M_{l,\chi}(p)$~\cite{Miransky:1994vk},
\begin{align}\label{eq:UVlimDeltal}
  \lim_{p\to\infty}  M_{l,\chi}(p) =\frac{2 \pi^2\gamma_m}{3}
  \frac{\Delta_{l,\chi}}{p^2  [\ln (p/\Lambda_{\textrm{QCD}})]^{1-\gamma_m}}\,.
\end{align}
with $\gamma_m$ defined in \eq{eq:ratio}.

Of course, as mentioned above, the quantity that we will employ in \eq{eq:GMOR} is rather $\Delta_l(\mu)$. The latter
is usually computed in lattice simulations at a given scale $\mu_\textrm{lat}$, which is typically far lower than the one used in the present SDE approach; for more details and an overview of the respective lattice results
see~\cite{Aoki:2019cca}. For sufficiently large $\mu$,  $\Delta_l$ and $\Delta_l(\mu)$
are related by
\begin{align}\label{eq:Deltalmu}
	\Delta_l(\mu) = \Delta_l \,[\ln (\mu/\Lambda_{\textrm{QCD}})]^{{\gamma_m}} \,.
	\end{align}
With  $m_l(\mu)$ defined in \eq{eq:ratio} and \eq{eq:Deltalmu}, the combination $m_l(\mu) \Delta_l(\mu)$ is
RG-invariant, as stated in \eq{eq:mDRGI}.

\begin{figure}[t]
	\includegraphics[width=0.45\textwidth]{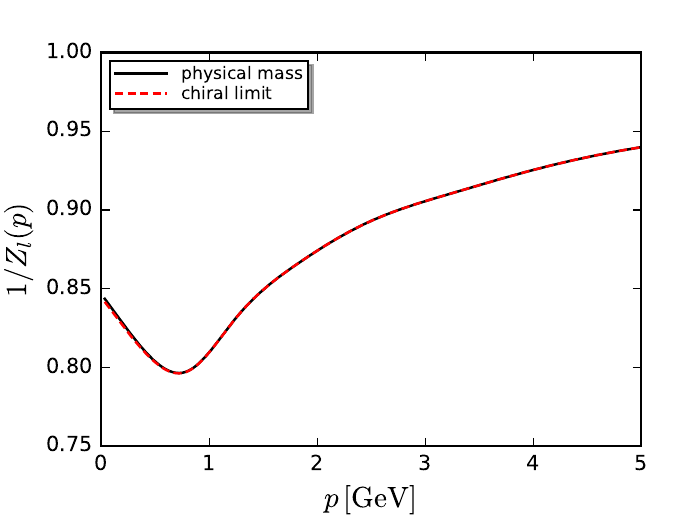}\hspace{1cm}
	\includegraphics[width=0.45\textwidth]{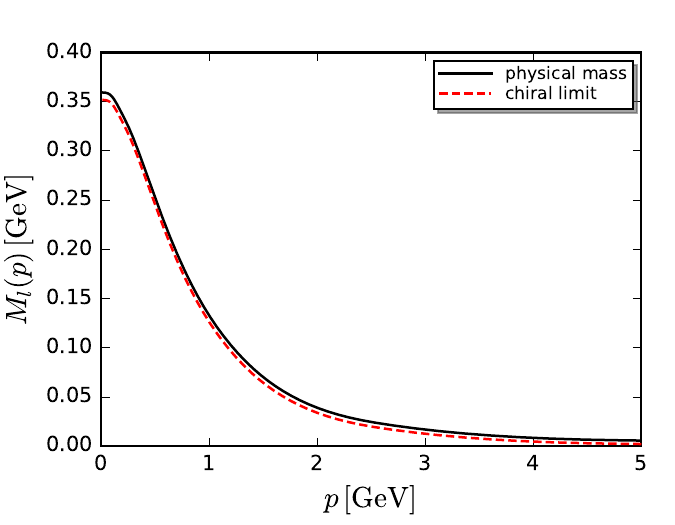}
	\caption{ Quark dressing, $1/Z_q(p^2)$, and mass function, $M_q(p^2)$,  in the chiral limit and for physical quark masses. \hspace*{\fill} }\label{fig:MassFun}
\end{figure}

Combining the above relations with our SDE results for the quark propagator we arrive at our first non-trivial prediction. We use our results for $M_l(p)$ and $1/Z_l(p)$ in the chiral limit, shown in \fig{fig:MassFun}, which are necessary for our analysis; a complete discussion is provided in \sec{sec:results}. Specifically, the value of $\Delta_{l,\chi}$ is given by
\begin{align}\label{eq:DeltaChiral}
	\Delta_{l,\chi} = (244.3(6)\,\textrm{MeV})^3\,,
\end{align}
where $m_s$ is kept fixed and $m_l\to 0$.

To fully appreciate the above prediction, we emphasise that, as can be shown \textit{analytically} by identifying the diagrams included in each perturbative order, the current approach encodes the full two-loop running of $M_l(p)$, and is numerically consistent with the full three-loop running; for a more detailed discussion, see \sec{sec:results}.

We next combine \eq{eq:DeltaChiral} with \eq{eq:Deltalmu} to obtain our prediction for $\Delta_{l,\chi}(\mu)$, and compare it with the FLAG result of~\cite{Aoki:2019cca}, obtained  from different lattice groups. There, the renormalization scale is $\mu_\textrm{lat}=2$\,GeV, and the estimate for the QCD scale is $\Lambda^\textrm{lat}_\textrm{QCD}=343(12)$\,MeV in the $\overbar{\textrm{MS}}$ scheme. Instead, in the present SDE computation we have $\Lambda_\textrm{QCD}=295(2)$\,MeV in the MOM${}^2$ scheme, see \eq{eq:LQCDMOM21loop} and \fig{fig:LambdaQCD1loop} in \app{app:MOM2-MOM}. With \eq{eq:Deltalmu} and \eq{eq:DeltaChiral}, this leads us to
\begin{align}\label{eq:DeltalmuRes}
\Delta_{l,\chi}(\mu_\textrm{lat})= (269.3(7)\,\textrm{MeV})^3\,,	\qquad \qquad \left[\Delta_{l,\chi}(\mu_\textrm{lat})\right]_\textrm{FLAG}= (272(5)\,\textrm{MeV})^3\,.
\end{align}
Taking into account the subtleties in the conversion and application of RG scales, the agreement between these two values is rather impressive,
providing non-trivial support for the quantitative reliability of the present approximation. We emphasise that the prediction \eq{eq:DeltaChiral}, and hence \eq{eq:DeltalmuRes}, have been obtained without any phenomenological input on the strength of chiral symmetry breaking.

Finally, $\Delta_{l,\chi}$ can be used for the determination of the physical ($m_q\neq 0$) chiral condensate $\Delta_{l}$ through
\begin{align} \label{eq:DiffDeltal}
	\Delta_{l}= \Delta_{l,\chi} + (\Delta_{l}- \Delta_{l,\chi})  = \Delta_{l,\chi} +
	\int_p \left( \frac{Z_2}{Z_q(p)} \frac{{\overbar M}_l(p)}{p^2 +M_l^2(p)} -   \frac{Z_2}{Z_{q,\chi}(p)} \frac{{M}_{l,\chi}(p)}{p^2 +M_{l,\chi}^2(p)}\right) \,,
\end{align}
where $Z_2\approx 1$ in the MOM${}^2$ scheme, for more details see \app{app:RG-schemes}. The integral in \eq{eq:DiffDeltal} is simply the difference of the loop expressions for the chiral condensates, and is finite. In \eq{eq:DiffDeltal} we have already set the multiplicative renormalization factors to unity, according to the procedure described in \sec{sec:MultRenorm}. The calculated value for $\Delta_{l}$ is reported in the following subsection.
%

\subsection{Determination of the current quark masses} \label{sec:DetCurQuark}

With the groundwork laid in \sec{sec:PS+GMOR} and \sec{sec:ChiralCond}, we now determine the current quark masses. Using  \eq{eq:PhysPar}, \eq{eq:PS}, \eq{eq:GMOR}, \eq{eq:DiffDeltal}, and a pion decay constant $f_\pi=92.1$\,MeV, we arrive at the current quark masses $m_q$ at \mbox{$\mu= 40$}\,GeV,
\begin{subequations}
		\label{eq:mpi-mq}
\begin{align}
	m_q(\mu) = 2.7 \,\textrm{MeV}\,,  \qquad m_s = 73\,\textrm{MeV}\,, \qquad \textrm{with}\qquad m_\pi=138\,\textrm{MeV} \,,
\end{align}
and the predictions
\begin{align}\label{eq:DeltaPhys}
\Delta_{l} = (300\,\textrm{MeV})^3\,, \qquad M_q(0)=351\, \textrm{MeV}\,, 	\qquad \qquad  f^{(\textrm{PS})}_\pi=87.6\,\textrm{MeV}\,.
\end{align}
\end{subequations}
From \eq{eq:DeltaPhys} we deduce that $f^{(\textrm{PS})}_\pi$ has an error of 5\%, which is well within our conservative estimate for the systematic error in the range of 10\%. While $f^{(\textrm{PS})}_\pi$ does not enter in the determination of the pion mass, and we are safely within the regime of chiral perturbation theory at physical pion masses, we assign a conservative 10\% systematic error to our current quark masses. We emphasise that this error is not inherent to our SDE-computation, but affects deduced observables.  Indeed, our benchmark result \eq{eq:DeltalmuRes} for the chiral condensate in the chiral limit agrees with the lattice results within the statistical error of the latter (less than 1\% deviation).

For completeness we also report the result for $f^{(\textrm{PS})}_\pi$ without the correction to $ N_\pi$ implemented by \eq{eq:DeltaI}; we have
\begin{align}
	\label{eq:fpiwo}
[f^{(\textrm{PS})}_\pi]_{I_{q,\chi} \neq 0}= 86.9\,\textrm{MeV}\,,
\end{align}
in very good agreement with the $f^{(\textrm{PS})}_\pi$ in \eq{eq:DeltaPhys}.

All results presented thus far have been obtained using
the fRG data for the gluon propagator, renormalized at $\mu=40$\,GeV, as input
in the SDEs.
In order to illustrate that our predictions are essentially independent of the input propagator,
we also report the results obtained when a fit to the gluon lattice data is employed,
which has the fRG  perturbative behavior built in it.
Note also that the fRG and lattice data differ in the deep infrared [see \fig{fig:Ginput}], as commented in \sec{sec:InputGluon}.
With this particular input we arrive at
\begin{subequations}
	\label{eq:mpi-mq-lattice}
	\begin{align}
		m_q(\mu) = 2.7\,\textrm{MeV}\,,  \qquad m_s =73\,\textrm{MeV}\,, \qquad \textrm{with}\qquad m_\pi=138\,\textrm{MeV} \,,
	\end{align}
	and the predictions
	\begin{align}\label{eq:DeltaPhyslattice}
		\Delta_{l} = (301\,\textrm{MeV})^3\,, \qquad M_q(0)=350\, \textrm{MeV}\,, 	\qquad \qquad  f^{(\textrm{PS})}_\pi=88.1\,\textrm{MeV}\,.
	\end{align}
\end{subequations}
In addition, the results for chiral condensate in the chiral limit, $\Delta_{l,\chi} = (246\,\textrm{MeV})^3$ and $\Delta_{l,\chi}(\mu_\textrm{lat})= (271\,\textrm{MeV})^3$ are in quantitative agreement with \eq{eq:DeltaChiral} and \eq{eq:DeltalmuRes}, obtained for $\mu=40$\,GeV. This coincidence, and the comparison of \eq{eq:mpi-mq-lattice} with \eq{eq:mpi-mq} shows explicitly that, within our estimated systematic error, the behaviour of the gluon propagator in the deep infrared does not affect the observables considered here.

This concludes the discussion of the determination of the current quark masses. In summary, we have shown that the present SDE approach allows for a quantitatively reliable computation of the constituent quark mass function $M_q(p)$, and hence the chiral condensate in \eq{eq:DeltalmuRes}, without any phenomenological input.

\section{ Numerical results}\label{sec:results}

In this section we present and discuss the central results of our numerical analysis, focusing mainly on the general features displayed by $Z_q$, $M_q$, and the $\lambda_i$. The stability of these  results under variations of the UV cutoff, the gluon inputs, and the RG scale will be addressed in \Sec{sec:Stability}, while the implementation of reliable simplifications will be analysed in \Sec{sec:QuantApprox}.

Within the setup described in Secs.~\ref{sec:QuarkGluon}, \ref{sec:gencon} and \ref{sec:CurrentQuark}, the SDE system of quark propagator and quark-gluon vertex is solved iteratively. To that end, it is convenient to use as the starting point of the iteration input data for quark and gluon propagators and the respective BC vertex,
or, when the fRG-DSE gluon is utilized~\cite{Gao:2020qsj, Gao:2020fbl},
the respective quark-gluon vertex data. Moreover, the dressings of the three dominant tensor structures, $\lambda_{1,4,7}$, are fed back into their own coupled SDEs and those of the other vertex dressings; instead, the remaining $\lambda_{2,3,5,6,8}$  are only used in the quark gap equation.
We have checked that this procedure affects the key quantities only very marginally:
when the sub-dominant dressings are fed back into the quark-gluon SDE, the numerical changes induced lie comfortably within the estimated error bars of our method.

The algorithm we employ for the numerical integrations is the standard Gauss–Legendre quadrature.
To that end, we pass to spherical coordinates, with the loop measure given by
\begin{equation}
d^4q=\frac{1}{2}q^2\, \sin^2\theta \sin\phi \, dq^2\,d\theta \,d\phi \,d\psi \,,
\end{equation}
where $q^2\in[0,\infty)$, $\theta\in[0,\pi]$, $\phi\in[0,\pi]$, and $\psi\in[0, 2\pi]$.
The azimuthal angle $\psi$ may be integrated trivially.

We next introduce the parametrization $q^2=\Lambda\kappa (\Lambda/\kappa)^{y}$, where $\kappa$ and $\Lambda$ denote
the infrared and UV cutoffs, respectively.
The new integration variable $y$ is related to $q^2$ by \mbox{$y =  \ln(q^2/\Lambda\kappa)/\ln(\Lambda/\kappa)$}; evidently, $y\in[-1, 1]$.
In addition, the angular integrals are written in terms of the
new integration variables $z=\cos\theta$ and $z^\prime=\cos\phi$, again with $z,z'\in[-1, 1]$.
These changes of variables facilitate the use of the Gauss–Legendre quadrature, according to which, an integral of a function $f(x)$, with $x\in[-1, 1]$,
is written as
\begin{align}
\int_{-1}^1 dx f(x) \approx \sum_{i = 1}^n w_i f(x_i) \,, \label{GaussLeg}
\end{align}
where the nodes $x_i$ and weights $w_i$ are uniquely determined by requiring that \eq{GaussLeg} becomes exact for all polynomials of degree less than $2n$. Specifically, one may show that the $x_i$ are the roots of the $n$-th Legendre polynomial, $P_n(x)$, and the weights are given by the formula
\begin{align}
w_i = \frac{2}{(1 - x_i^2)[P^\prime_n(x_i)]^2} \,.
\end{align}
For the actual calculation, we set $n=40$ for the integration over $y$, and $n=20$ for the integrations over $z$ and $z'$.

Finally, once the integrations have been carried out, one is left with a sizable system of coupled non-linear algebraic equations,
which is solved by means of Broyden's method, with the numerical precision set at $10^{-5}$.

As discussed in \Sec{sec:Stability} and \app{app:RG-schemes}, the RG scale $\mu$ should be taken as large (\ie as ``perturbative'') as possible; therefore, we choose $\mu=40\,$GeV. However, the lattice data for the gluon propagator are limited by  $p \lesssim 5$\,GeV, and have to be extended by a perturbative fit; instead, the functional data for the gluon propagator cover the full momentum range of interest. Therefore, we employ the functional data from~\cite{Fu:2019hdw, Gao:2020fbl, Gao:2020qsj} as our input for the present computation, while the lattice data from~\cite{Boucaud:2018xup, Zafeiropoulos:2019flq, Aguilar:2019uob} are used
as benchmark results for the low momentum regime, see \fig{fig:Ginput}.

The above setup represents our best approximation, and allows us to compute the quark-gluon vertex as well as the quark propagator without any phenomenological input; in particular, the only parameters to be fixed are the fundamental parameters of QCD, namely the current quark masses at the RG scale $\mu$, see \sec{sec:CurrentQuark}.

\begin{figure}[t]
	\centering
	\begin{subfigure}[t]{.48\textwidth}
		\includegraphics[width=1\textwidth]{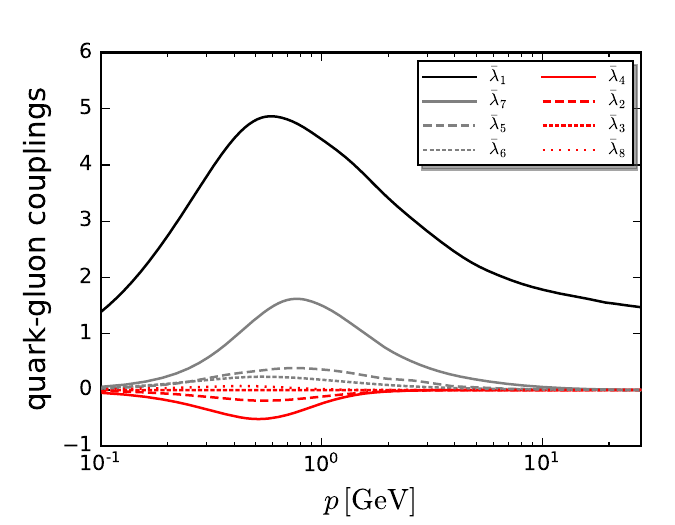}
		\caption{Quark-gluon couplings $\bar\lambda_i(p)$, \eq{eq:barlambda}. \emph{Black:} classical tensor structure,  \eq{eq:avatars}. \emph{Grey:} chirally symmetric non-classical tensor structures. \emph{Red:}  chiral-symmetry breaking tensor structures.\hspace*{\fill} }
		\label{fig:VertexDressings}			
	\end{subfigure}
	\hfill
	\begin{subfigure}[t]{0.48\textwidth}
		\includegraphics[width=1\textwidth]{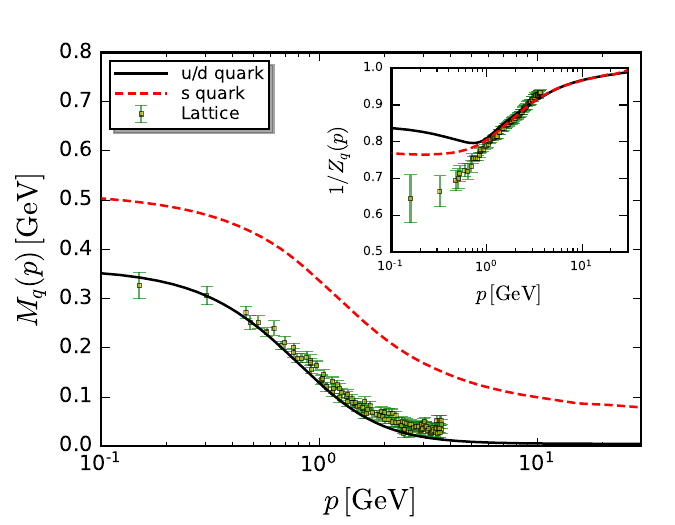}
		\caption{Quark mass functions $M_q(p)$ and propagator dressings $1/Z_q(p)$ (inset) for $q=l,s$, see \eq{eq:PropDressings}. Lattice results from \cite{Bowman:2005vx}: light quark mass function $M_l(p)$ and dressing $1/Z_l(p)$.\hspace*{\fill} }		
		\label{fig:QuarkMass}
	\end{subfigure}
	\caption{Numerical results for the coupled system of SDEs for quark-gluon vertex (light quark, \emph{left panel}), and the light and strange quark propagators (\emph{right panel}).
		\hspace*{\fill}}
	\label{fig:qg-couplings+quark-prop}
\end{figure}
Our main results are summarised in \fig{fig:qg-couplings+quark-prop}. In particular, in \fig{fig:VertexDressings} we show all dressings of the quark-gluon vertex, while in \fig{fig:QuarkMass} we display the quark mass functions, $M_q(p)$, and the quark wave function renormalizations $1/Z_q(p)$ (inset) for $q=l,s$. Note that, in order to best expose the physical relevance of the different tensor structures and the corresponding dressings, we introduce dimensionless couplings, $\bar\lambda_i(\bar p)$, with $i=1,...,8$, see, \eg \cite{Mitter:2014wpa}; specifically, we concentrate on the symmetric point $\bar p$ and define
\begin{align}\label{eq:barlambda}
	\bar\lambda_i(\bar p) =  \frac{\bar p^{\,n_i}\lambda_i(\bar p) }{ Z_q(\bar p) Z_A^{1/2}(\bar p ) } \,,\qquad \textrm{with}\qquad n_1=0, \, \ n_{2,3,4}=2 \,,\  n_{5,6,7} = 2\,, \ n_{8} = 3\,.
\end{align}
In \eq{eq:barlambda}, the multiplication of $\lambda_i$ by $\bar p^{\,n_i}$ renders the $\bar\lambda_i$ dimensionless, and the division by the wave function renormalizations leaves us with the respective eight running couplings.

\begin{figure}[t]
	\centering
	\begin{subfigure}[t]{0.49\textwidth}
		\includegraphics[width=1\textwidth]{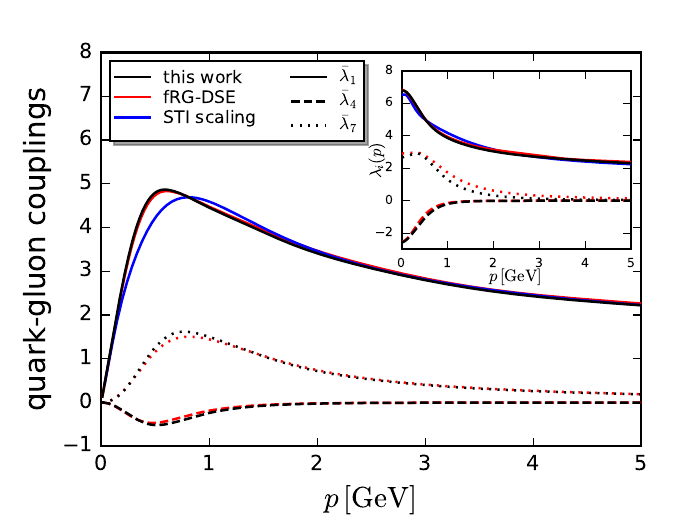}
		\caption{Dominant quark gluon couplings $\lambda_{1,4,7}$: full solution SDE (this work, black), full solution \cite{Gao:2020fbl} (fRG-DSE, red), couplings with STI, $\bar\lambda_1$, and scaling relations, $\bar\lambda_{4,7}$ (STI-scaling, blue).\hspace*{\fill}  }
		\label{fig:ResultsCompareLitqg}
	\end{subfigure}
	\hfill
	\begin{subfigure}[t]{0.49\textwidth}
		\includegraphics[width=1\textwidth]{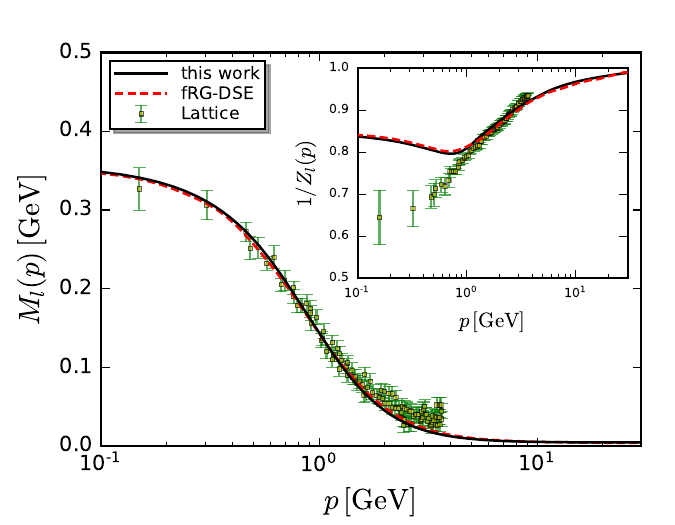}
		\caption{Light quark mass function $M_l(p)$ and dressing function $1/Z_l(p)$ (inset):  full solution (this work, full black), full solution \cite{Gao:2020fbl} (fRG-DSE, dashed red),  lattice results \cite{Bowman:2005vx} (Lattice, green).
			\hspace*{\fill}}
		\label{fig:ResultsCompareLitGap}
	\end{subfigure}
	\caption{Full results for dominant quark-gluon couplings $\bar\lambda_{1,4,7}$ (this work, black) from the present work in comparison to full results from \cite{Gao:2020fbl} (fRG-DSE, red), lattice results (Lattice, green), and the approximation \textit{STI-scaling}: $\lambda_{1,2,6}$ from STI, and $\lambda_{4,5,6,7}$ from scaling relations,  see discussion at the end of \sec{sec:QuantApprox} and \cite{Cyrol:2017ewj, Gao:2020qsj}.  The $\lambda_i$ are dressings of tensor structures in \Eq{eq:147} and  measured in  $\lambda_1$,  $\lambda_4[(\rm GeV)^{-1}]$ and  $\lambda_7[(\rm GeV)^{-2}]$.  \hspace*{\fill}}
	\label{fig:ResultsCompareLit}
\end{figure}
%

For all quarks $q=l,s$, the products $1/(4 \pi) \bar\lambda_i(\bar p)\bar\lambda_j(\bar p)$ can be interpreted as the interaction strength of a one-gluon exchange between corresponding quark currents. For example, \mbox{$\alpha_{q \bar q A}(\bar p) = 1/(4 \pi)  [\bar\lambda_1(\bar p)]^2$}, see \eq{eq:avatars}, simply measures the interaction strength or running coupling of a one-gluon exchange between the two quark currents $\bar q(t)\gamma_{\mu} q(p-t)$. Similarly, the combinations $1/(4 \pi) \bar\lambda_i(\bar p)\bar\lambda_j(\bar p)$ can be understood as the interaction strength of a one-gluon exchange between the respective tensor currents $\bar q\,\overbar {\cal T}_i\, q$ and $\bar q\,\overbar {\cal T}_j\, q$, with the dimensionless tensor structures $  \overbar {\cal T}_i= {\cal T}_i/(\bar p^2)^{n_i}$, with $i,j=1,...,8$. As in the case of the gluon dressing, one sees the prominent enhancement around $\bar p \approx 0.5 - 1$\,GeV, which is crucial for obtaining from the gap equation the correct amount of chiral symmetry breaking. Moreover, we note the clear suppression of all $\bar\lambda_i$ for $\bar p\to 0$.

In \fig{fig:ResultsCompareLit}, we compare our results to those  obtained through a combined setup~\cite{Gao:2020fbl, Gao:2020qsj} (fRG-DSE), where the relevant SDEs are expanded about the two-flavour QCD correlation functions of~\cite{Cyrol:2017ewj}. As we may infer from \fig{fig:ResultsCompareLit}, the present SDE results are quantitatively consistent with those of~\cite{Gao:2020fbl, Gao:2020qsj}. This is an additional, highly relevant reliability check for the \textit{respective} results of different but similar functional approaches. In our opinion, the confirmation of this quantitative agreement, and further successful comparisons of this type, provide important information about the respective systematic error. Together with \textit{apparent} convergence of the results in functional approaches within a systematic approximation scheme, this finally will lead to a first principle functional approach to QCD.

The interpretation of the comparison of the dressing $1/Z_l(p)$ from functional methods and the lattice is less clear. To begin with, the lattice result from \cite{Bowman:2005vx}  shows a rather steep slope at momenta $p\gtrsim 1$\,GeV. For $p\lesssim 0.5$\, GeV, it shows a rapidly rising statistical error. Clearly, it would be highly desirable to repeat the computation of the quark dressings with more recent sets of configurations, i.e.~with~\cite{Boucaud:2018xup, Zafeiropoulos:2019flq, Aguilar:2019uob}.
		
In turn, all functional studies, \eg~\cite{Williams:2015cvx, Cyrol:2017ewj, Aguilar:2018epe}, consistently show a smooth rise of the dressing  $1/Z_l(p =\mu)= 1$, that is compatible with perturbation theory. For $p\lesssim 1$\, GeV, these studies show a non-monotonic behaviour, which cannot be identified within the statistical accuracy of the lattice data. This calls for more refined studies, as $1/Z_q(p)$ carries significant systematic errors in this regime $p\lesssim 1$\, GeV. It is interesting to note that the existence and strength of this  non-monotonicity depends on the size of the current quark mass, see \fig{fig:QuarkMass} for a comparison of $1/Z_l$ and $1/Z_s$ and \cite{Cyrol:2017ewj} for a study of the $m_l$-dependence in two-flavour QCD. We hope to resolve this situation in a combined functional-lattice study in the near future.

We emphasise that the present approximation includes \textit{analytically} the full two-loop running of the quark gap equation. Hence, $M_q(p)$ and $1/Z_q(p)$ are two-loop consistent, since the quark gap equation contains all tensors of the quark-gluon vertex. The SDE solution of the latter includes all one-loop diagrams, and hence, the numerical solution for the $\lambda_i(p,q)$ encompasses the full one-loop structure analytically. Furthermore, the input gluon data contain at least the full one-loop momentum dependence. Accordingly, \textit{all} ingredients in the  quark-gap equation carry at least their full one-loop momentum dependence, and hence, the solution is \textit{analytically} two-loop consistent.

\begin{figure}[t]
	\hspace{-1cm}
	\includegraphics[width=0.47\textwidth]{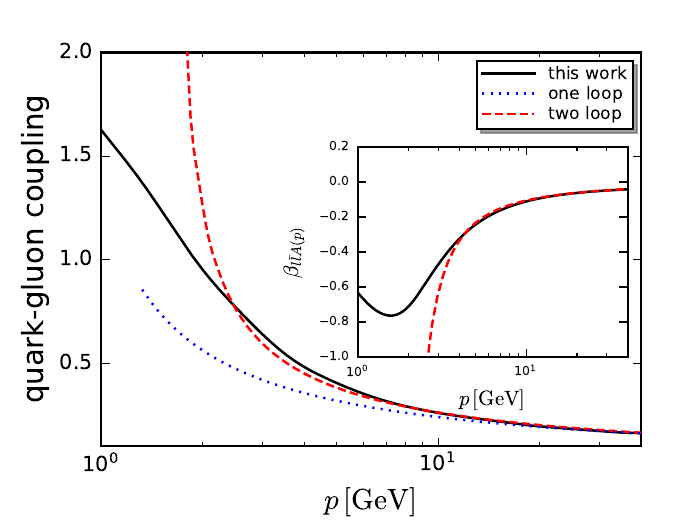}\hspace{1cm}
	\includegraphics[width=0.47\textwidth]{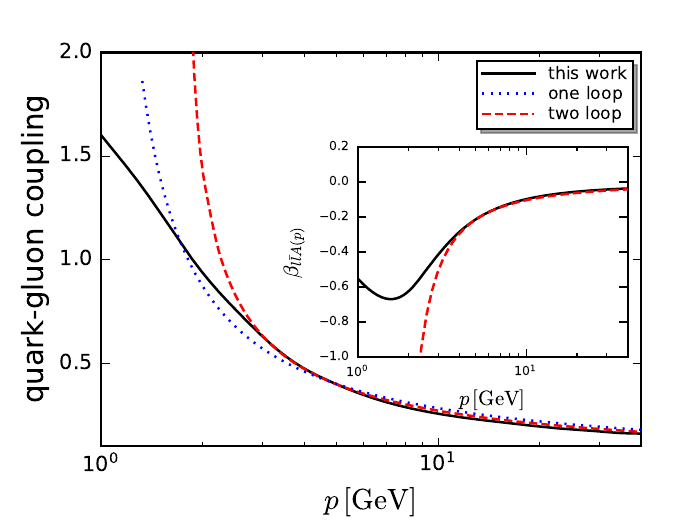}
	\caption{Strong coupling $\alpha_{l \bar l A}(p)$, see \eq{eq:avatars}, in comparison to the one- and two-loop counterparts, adjusted at the RG scale $\mu=40$\,GeV (\emph{left panel}) and $\mu=4.3$\, GeV (\emph{right panel}).  \hspace*{\fill}} \label{fig:alphasComp}
\end{figure}
Of course, as already mentioned in \sec{sec:QuarkGluon}, the
vertex SDE employed (see \fig{fig:SDEVertex})
corresponds to the so-called ``one-loop dressed'' truncation, where
vertices with no classical counterpart, such as the four-quark vertex~\cite{Gao:2020fbl}, have been omitted from the skeleton expansion.
Nonetheless, the contributions of such terms
are very suppressed for perturbative momenta, as we have confirmed in the case of $\alpha_{l\bar l A}(p)= 1/(4 \pi)  \bar\lambda_1^2(p)$. In \fig{fig:alphasComp} we depict our numerical result together with the analytic one- and two-loop strong couplings $\alpha_s^{\textrm{(1loop)}}(p)$ and  $\alpha_s^{\textrm{(2loop)}}(p)$, renormalized at $\mu=40$\,GeV and $\mu=4.3$\,GeV. Then, the respective values for $\Lambda_\textrm{QCD}$ are chosen such that also the $\beta$-functions
\begin{align}\label{eq:betap}
\beta_{l\bar l A}(p)=p\partial_p\, \alpha_{l\bar l A}(p)\,,
\end{align}
match at the renormalization scale $\mu$.
The numerical results in the present work have been obtained with $\mu=40$\,GeV, which lies deep in the perturbative regime. While this choice reduces the systematic error originating from non-perturbative approximations to the SDEs, its successful implementation requires a particularly accurate treatment, in order to reliably connect the wide range of momenta between $\mu$ and the deep infrared.

As can be seen in \fig{fig:alphasComp} (inset left panel), our numerical results for $\beta_{l\bar l A}$ agree quantitatively with the respective two-loop results $ \beta^{\textrm{(2loop)}}_{\alpha_s}$ for momenta $p \gtrsim 5$\,GeV, while the full coupling $\alpha_{l\bar l A}$ and the two-loop coupling $\alpha_s^{\textrm{(2loop)}}$ agree even up to $p\approx 3$\,GeV.

Still, a careful analysis reveals that deviations in the pair $(\alpha_s,\beta_{\alpha_s})$ start to become visible for $p \lesssim 10$\,GeV, for a more detailed discussion see \App{app:MOM2-MOM}; there, it is also shown that the two-loop prediction for  $\Lambda_{\textrm{QCD}}$ is stable for $p\gtrsim 10$\,GeV and as required by RG-consistency. For our RG scale of $\mu=40$\,GeV we find $\Lambda_{\textrm{QCD}}=1.42(2)$\,GeV. We conclude that the current setup is sufficiently accurate to allow for a self-consistent renormalization at a large perturbative $\mu$. However, the $\beta$-function $\beta_{l\bar l A}$, which measures the momentum slope, starts to deviate from the two-loop result $\beta^{\textrm{(2loop)}}_{\alpha_s}$ in the momentum regime $p \in (5-10)$\,GeV, see \fig{fig:alphasComp} (inset left panel). Consequently, in this regime, the required $\mu$-independence of $\Lambda_{\textrm{QCD}}$ in a consistent RG scheme is lost gradually within a two-loop matching, leading to a slightly different $\Lambda_{\textrm{QCD}}=1.49$\,GeV for $\mu=4.3$\,GeV. For more details, see \app{app:MOM2-MOM}, and in particular \fig{fig:Singmu}, where the transition regime between the perturbative and non-perturbative regimes is marked by a red band.

Below $p \approx 3$\,GeV, the strong coupling $\alpha_{l\bar l A}$ rapidly departs from the perturbative two-loop coupling, signalling the onset of non-perturbative physics. The lack of RG-consistency with a two-loop matching for small RG scales is even more apparent within a one-loop matching. There, an adjustment of the one-loop coupling and its $\beta$-function at $\mu=40$ GeV leads to $\Lambda_\textrm{QCD}=0.59$\,GeV, while at $\mu=4.3$\,GeV we are led to  $\Lambda_\textrm{QCD}=0.86$\,GeV. The lack of RG-consistency at one loop is even more manifest in the fact that a respective truncation clearly cannot bridge the wide momentum range between $p=40$\,GeV and the non-perturbative infrared regime with $p\lesssim 5$\,GeV, see \fig{fig:alphasComp} (left panel).

Accordingly, our lower renormalization scale, $\mu=4.3$\,GeV is at the boundary between the perturbative and non-perturbative regimes. The two-loop coupling and the full  $\alpha_{l\bar l A}$ agree well for momenta  $p\gtrsim 3$\,GeV, but, contrary to the case with $\mu=40$\,GeV, the $\beta$-function reveals deviations already in the perturbative regime, see \fig{fig:alphasComp} (inset right panel).

This comparison carries an important message for phenomenological applications: potential inaccuracies of the present method that thwart the reliable bridging of disparate momentum scales can be compensated by choosing a relatively small renormalization scale. Nonetheless, such a choice is limited by a minimal RG scale, $\mu\geq \mu_\textrm{min}$, with $\mu_\textrm{min} \approx 3$\,GeV; smaller RG scales push renormalization clearly into the non-perturbative regime, where the arguments invoked in \sec{sec:MultRenorm} for setting $Z_i=1$ do not apply.

Finally, we note that a fully two-loop consistent analysis would require the omitted diagrams in the quark-gluon SDE, as well as a two-loop consistent gluon input. Both tasks lie within the technical grasp of functional approaches; for a discussion concerning the gluon propagator, see~\cite{Cyrol:2017qkl, Corell:2018yil}.

Also, the results presented here in the MOM${}^2$ scheme can be readily mapped to respective ones in the standard MOM scheme. This is discussed in detail in \app{app:MOM2-MOM}, where it is shown that the present results are in quantitative agreement with the MOM scheme results in the literature. In particular, we provide a comparison of the respective couplings in \fig{fig:Couplingcomp}.

In summary the agreement with the lattice as well as other functional methods is rather impressive, especially
since no phenomenological infrared parameter is involved: the results presented here are obtained within a first-principle setup to QCD, the only input being the fundamental parameters of QCD.

\section{Stability of the numerical results}\label{sec:Stability}

As we will see in detail in this section, the results obtained from our SDE analysis are particularly stable
under variations of the UV cutoff that regulates the loop integrals, the choice of functional or lattice gluon inputs,
and a vast change in the value of the RG scale.

\subsection{Varying the UV cutoff}\label{sec:Lambda-Indep}

\begin{figure}[t]
	\centering
	\begin{subfigure}[t]{0.48\textwidth}
		\includegraphics[width=1\textwidth]{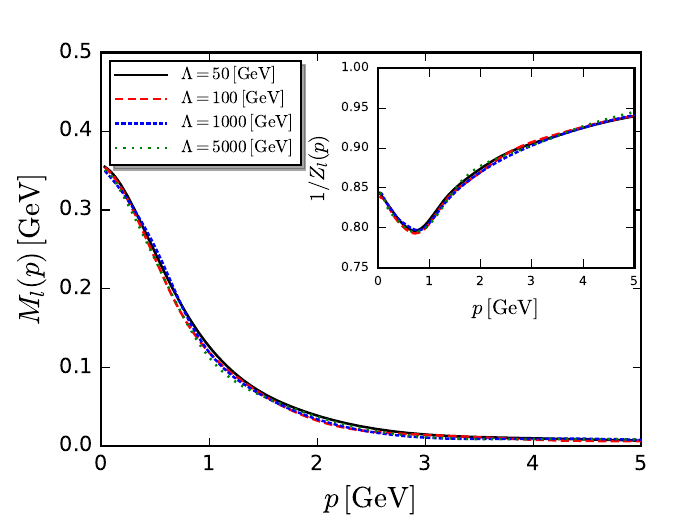}
		\caption{Light quark mass function $M_l(p)$  and  dressing $1/Z_l(p)$ (inset) for different UV cutoffs.\hspace*{\fill}  }
		\label{fig:QuarkMassLambda}			
	\end{subfigure}
	\hfill
	\begin{subfigure}[t]{0.48\textwidth}
		\includegraphics[width=1\textwidth]{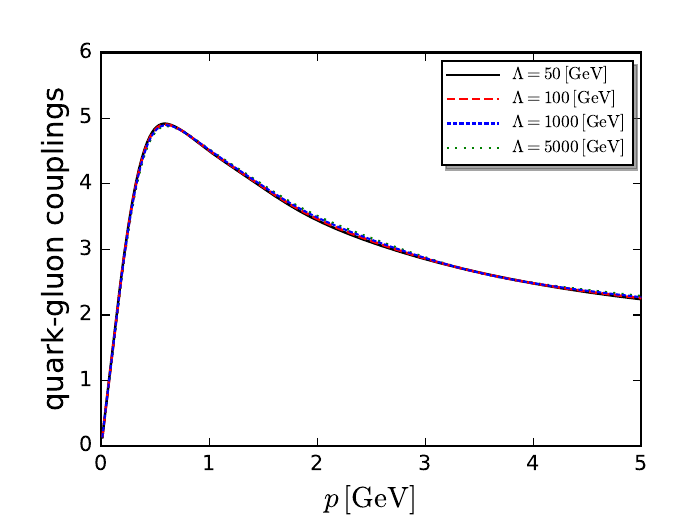}
		\caption{Quark-gluon coupling $\bar\lambda_{1}(p)$ of the classical tensor structure  for different UV cutoffs.\hspace*{\fill} }		
		\label{fig:VertexDressingsLambda}
	\end{subfigure}
	\caption{Numerical results for the coupled system of SDEs for different UV cutoffs \mbox{$\Lambda=50, 100, 1000, 5000$}\,GeV, based on the gluon input data \cite{Gao:2020qsj, Gao:2020fbl} (fRG-DSE in \fig{fig:Ginput}).\hspace*{\fill}}
	\label{fig:ResultsLambda}
\end{figure}

We have verified explicitly that
our results are practically insensitive to variations momentum cutoff $\Lambda$ within the range $\Lambda=50$\,GeV to $\Lambda=5000$\,GeV. As mentioned before, in the MOM${}^2$ scheme this check only tests our numerics implementation, and not the cutoff-independence of the renormalization scheme, as in the MOM scheme. Note that, while $\lambda_1$ displays a marginal (logarithmic) momentum dependence, all remaining  $\lambda_i$ are not subject to renormalization. Moreover, in the Landau gauge, the logarithmic running of $Z_l(p)$ vanishes at one loop, and only $M_l(p)$ shows a one-loop logarithmic running. Accordingly, in \fig{fig:ResultsLambda} we show the absence of cutoff-dependence in $M_l(p)$, $1/Z_l(p)$, and $\bar\lambda_1(p)$. Our results are especially stable, and, in particular, no $\log\Lambda$-dependence may be discerned.

%

\begin{figure}[t]
	\centering
	\begin{subfigure}[t]{0.48\textwidth}
		\includegraphics[width=1\textwidth]{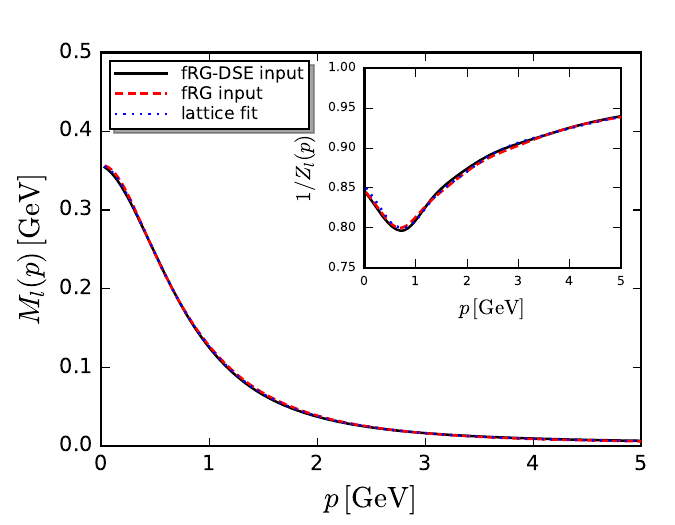}
		\caption{Light quark mass function $M_l(p)$  and  dressing $1/Z_l(p)$ (inset) for different gluon input.\hspace*{\fill}  }
		\label{fig:QuarkPropInput}			
	\end{subfigure}
	\hfill
	\begin{subfigure}[t]{0.48\textwidth}
		\includegraphics[width=1\textwidth]{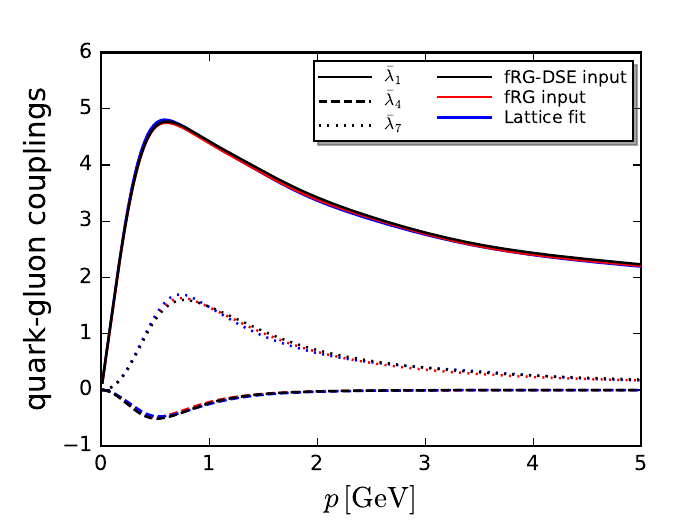}
		\caption{Dominant quark-gluon couplings $\bar\lambda_{1,4,7}(p)$ for different gluon input.		 \hspace*{\fill}  }
		\label{fig:VertexDressingsInput}
	\end{subfigure}
	\caption{Numerical results for the coupled system of SDEs with $\mu=40$\,GeV for different gluon input data, \fig{fig:Ginput} in \sec{sec:InputGluon}:  \cite{Gao:2020qsj, Gao:2020fbl} (fRG-DSE), \cite{Fu:2019hdw} (fRG),  \cite{Boucaud:2018xup, Zafeiropoulos:2019flq, Aguilar:2019uob} ({lattice fit}). \hspace*{\fill}}
	\label{fig:ResultsInput}
\end{figure}
We emphasise that the detection of such a logarithmic dependence in the present system is very difficult, due to its
(Landau gauge) suppression in $Z_l$, and the decay of $M_l(p)$ for large momenta.
This leaves us with $\bar\lambda_1(p)$, whose perturbative momentum-dependence is fixed by the self-consistent determination described in \sec{sec:SC-alphasDet}. Note that these properties, even though they complicate the detection of residual cutoff-dependences,
are a welcome feature rather than a liability: the present setup reduces the sensitivity of the SDE system with respect to the subtleties of a non-perturbative numerical renormalization.

\subsection{Stability with respect to the gluon input data}\label{sec:Gluon-Insensitive}

We proceed with the insensitivity with respect to the gluon input data, described
in \sec{sec:InputGluon} and depicted in \fig{fig:Ginput}. In  \fig{fig:ResultsInput} we compare
the results obtained using as inputs: {\it (a)}
the data from the fRG-DSE
computation~\cite{Gao:2020qsj, Gao:2020fbl} (``{\it fRG-DSE''}\,input);  {\it (b)}
the gluon propagator obtained with the fRG computation of~\cite{Fu:2019hdw} (``{\it fRG''}\,input); and
{\it (c)} the fit to the lattice data of~\cite{Boucaud:2018xup, Zafeiropoulos:2019flq, Aguilar:2019uob},
including an RG-consistent UV extrapolation (``{\it lattice fit}\,'').
The respective results for $M_l(p)$ and $1/Z_l(p)$ are shown in \fig{fig:QuarkPropInput}, while those for $\bar\lambda_{1,4,7}$
in \fig{fig:VertexDressingsInput}.

All results show an impressive quantitative agreement within the statistical and systematic errors. In particular, the difference in the
infrared behavior between the lattice and the functional
data used here [see \fig{fig:Ginput})] does not leave any significant trace on $M_l(p)$ and $1/Z_l(p)$,
as can be seen in \fig{fig:QuarkPropInput}. Accordingly, they do not influence our benchmark prediction for the chiral condensate, given in \eq{eq:DeltalmuRes}.
Moreover, the same independence is seen at the level of the $\bar\lambda_{1,4,7}(p)$, displayed in \fig{fig:VertexDressingsInput}.
This lack of sensitivity to the infrared details of the input gluon propagators stems from
the fact that the latter enter into four-dimensional momentum integrals, whose radial dependence, $p^3$,
suppresses the deep infrared very effectively.

\subsection{Varying the RG scale $\mu$}\label{sec:mu-Insensitive}

Finally we test the response of our results to changes in the RG scale  $\mu$.
In particular, we compare the results obtained when all relevant quantities have been
renormalized at the two vastly different scales $\mu=40$\,GeV and $\mu=4.3$\,GeV;
in both cases we employ the fRG-DSE input.

\begin{figure}[t]
	\centering
	\begin{subfigure}[t]{0.48\textwidth}
		\includegraphics[width=1\textwidth]{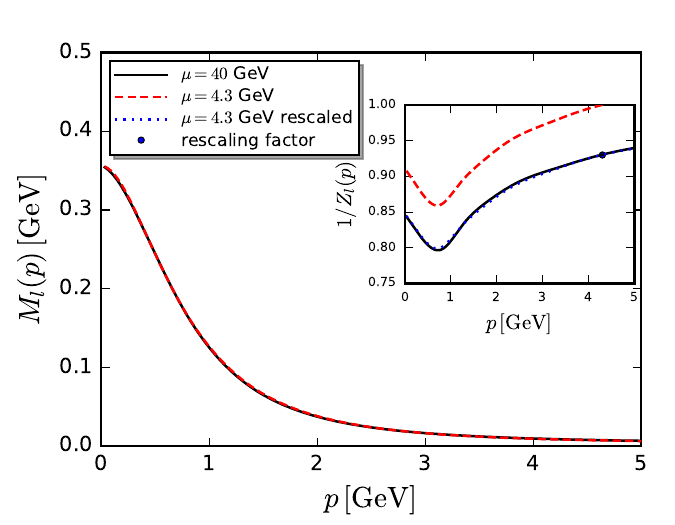}
		\caption{$M_l(p)$ and $Z_l(p)$ using the fRG-DSE input, for $\mu=40$\,GeV (black-solid)
			and $\mu=4.3$\,GeV (red-dashed and  blue-dashed (rescaled)). The blue dot indicates the rescaling factor.\hspace*{\fill}}
		\label{fig:ResultsmudepLatDSE}
	\end{subfigure}
	\hfill
	\begin{subfigure}[t]{0.48\textwidth}
		\includegraphics[width=1\textwidth]{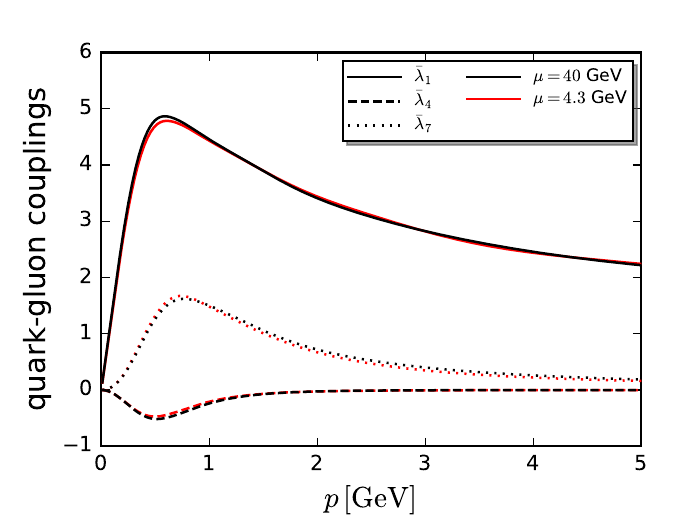}
		\caption{Results for the dominant quark-gluon couplings $\bar\lambda_{1,4,7}$, obtained with the fRG-DSE input,
			and renormalized at $\mu=40$\,GeV and $\mu=4.3$\,GeV. \hspace*{\fill}  }
		\label{fig:Resultsmudep40-4.3}
	\end{subfigure}
	\caption{The $\mu$-independence of $M_l(p)$ and $\bar\lambda_{1,4,7}(p)$, and multiplicative renormalizability of $Z_l(p)$. The RG scales differ by an order of magnitude: $\mu=40$\,GeV and $\mu = 4.3$\,GeV.\hspace*{\fill}}
	\label{fig:Resultsmudep}
\end{figure}

Evidently, $M_l(p)$ and $\bar\lambda_i(p)$ are formally RG-invariant quantities,
and, ideally, they should be $\mu$-independent; in practice,
the amount of residual $\mu$-dependence displayed is an indication of the veracity of the approximations employed.
The results shown in \fig{fig:Resultsmudep} demonstrate clearly that the $\mu$-dependence of these quantities
lies well within the estimated error bars; in particular, the largest visible discrepancy, located at the
peak of $\bar\lambda_1(p)$, is only 3.4\%.

On the other hand, the quantity $Z_l(p)$ is not RG-invariant, depending explicitly on $\mu$,
as can be seen in the inset of \fig{fig:ResultsmudepLatDSE}.
However, multiplicative renormalization, when properly implemented,
dictates that the curves renormalized at two different values of
$\mu$, say $\mu_1$ and $\mu_2$, must be related by
\begin{align}
Z_l^{-1}(\mu_2,\mu_1) Z_l^{-1}(p,\mu_2) = Z_l^{-1}(p,\mu_1)  \qquad\textrm{with}\qquad \mu_2 < \mu_1\,.
\label{eq:multrenZ}
\end{align}
The operation described in \eq{eq:multrenZ} rescales the ``red-dashed'' curve to the ``blue-dotted'' one
in the aforementioned inset.
Note that the rescaling factor is marked on the ``black-solid'' curve with a blue dot; its numerical value is 0.93.
Plainly, the coincidence achieved between original and rescaled curves is excellent, indicating that
multiplicative renormalizability has been adequately implemented at the level of our dynamical equations.

\begin{figure}[t]
	\centering
	\begin{subfigure}[t]{0.48\textwidth}
		\includegraphics[width=1\textwidth]{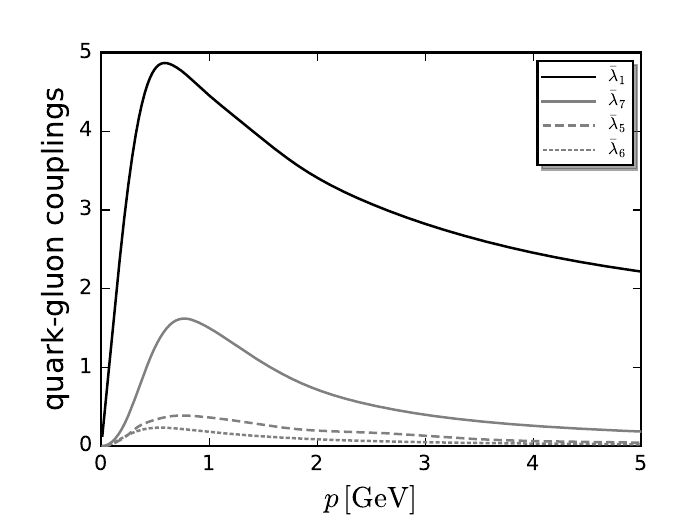}
		\caption{Quark-gluon couplings from chirally-sym\-metric tensor structures ${\cal T}_{1,5,6,7}$. \hspace*{\fill}  }
		\label{fig:VDressingL1}
	\end{subfigure}
	\hfill
	\begin{subfigure}[t]{0.48\textwidth}
		\includegraphics[width=1\textwidth]{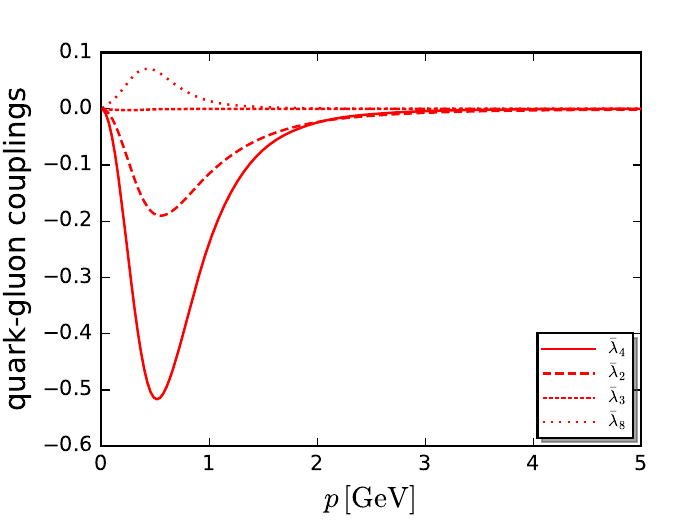}
		\caption{Quark-gluon couplings from of the chiral-symmetry--breaking tensor structures ${\cal T}_{2,3,4,8}$.  \hspace*{\fill} }
		\label{fig:VDressingL2}
	\end{subfigure}
	\caption{Quark-gluon couplings $\bar\lambda_i(p)$, defined at the symmetric point, see \eq{eq:barlambda}. The ordering in the legends is reflecting their strength. \hspace*{\fill} }
	\label{fig:VDressing}
\end{figure}
%

\section{Reliable low-cost approximations}\label{sec:QuantApprox}

The numerical cost of the present work is rather modest: a full simulation with given gluon input data requires about 20 core minutes on a  intel i7 chip. However, if the system is extended by the gluon SDE in order to obtain a fully self-consistent description,
the numerical costs rises significantly. Moreover, for applications to  hadron resonances, see \eg~\cite{Eichmann:2016yit},
the SDE system has to be augmented by BSE, Faddeev equations, and four-body equations, depending on the resonances of interest. Finally, in the study of the QCD  phase structure at finite temperature and density,
a rest frame is singled out, leading to a further proliferation of tensorial structures.
For all the above reasons, any approach that reduces the
computational cost without compromising the veracity of the results, is potentially useful for the above applications.

In what follows we discuss simplified approximations of the treatment of the quark-gluon vertex, that still lead to quantitatively reliable results. To that end, we analyse the numerical impact that the vertex dressings $\bar\lambda_i(p)$ have on the results of the quark dressings $M_l(p)$ and $1/Z_l(p)$. To better appreciate this discussion, we have replotted the results for the  $\bar\lambda_i(p)$, already shown in  \fig{fig:VertexDressings}: in \fig{fig:VDressing} we concentrate on the $\bar\lambda_i(p)$ in the low energy regime, \ie for $p\lesssim 5$\,GeV. The $\bar\lambda_i(p)$ are separated in two groups, those with chiral symmetry preserving tensor structures, \fig{fig:VDressingL1}, and those with chiral symmetry breaking ones, \fig{fig:VDressingL2}.

\begin{figure}[t]
	\centering
	\begin{subfigure}[t]{0.48\textwidth}
		\includegraphics[width=1\textwidth]{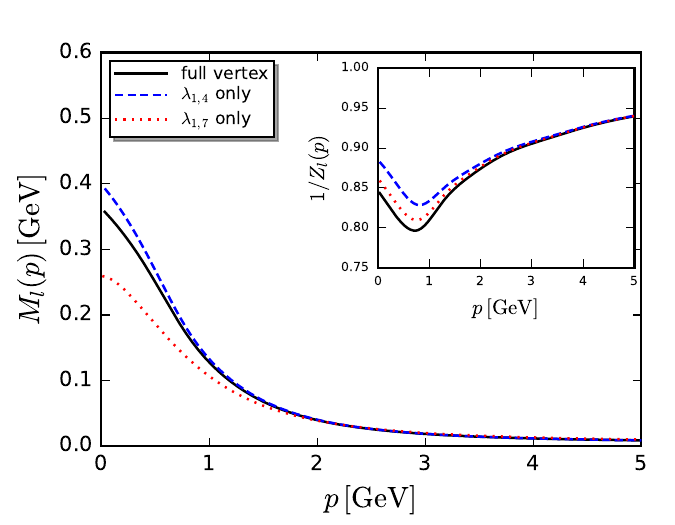}
		\caption{Quark dressings $M_l(p)$ and $1/Z_l(p)$ without  $\bar\lambda_4$ (dashed, red), and without $\bar\lambda_7$ (dotted, blue) in the vertex SDEs in comparison to the full solution (full, black). \hspace*{\fill}  }
		\label{fig:ResultsDrop47qg}
	\end{subfigure}
	\hfill
	\begin{subfigure}[t]{0.48\textwidth}
		\includegraphics[width=1\textwidth]{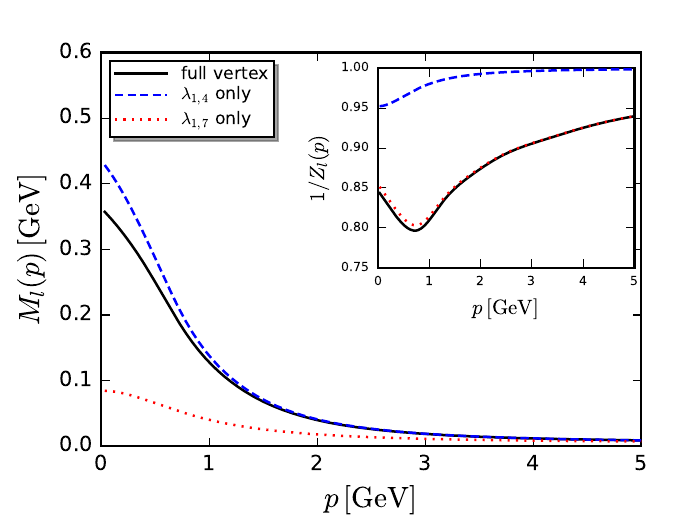}
		\caption{Quark dressings $M_l(p)$ and $1/Z_l(p)$ without  $\bar\lambda_4$ (dashed, red), and without $\bar\lambda_7$ (dotted, blue) in the quark gap equation in comparison to the full solution (full, black).  \hspace*{\fill} }
		\label{fig:ResultsDrop47quark}
	\end{subfigure}
	\caption{Lack of quantitative reliability without quark-gluon couplings $\bar\lambda_{4,7}$ on the right-hand sides of the quark-gluon SDEs and the quark gap equation.   \hspace*{\fill}}
	\label{fig:ResultsDrop47}
\end{figure}

The main outcome of these considerations
may be summarised by stating that  {\it (i)} the inclusions of $\lambda_{1,4,7}(p)$ is {\it necessary and sufficient}
for approximating accurately the results of the full analysis, and
{\it (ii)} $\lambda_{1}(p)$ may be reliably obtained from STI-based constructions, while $\lambda_{4,7}(p)$
from the scaling relations put forth in~\cite{Cyrol:2017ewj, Gao:2020qsj, Gao:2020fbl}.

Point {\it (i)} has been established by considering the relevant SDEs approximations
for the quark-gluon vertex that include $\lambda_1$ (which, obviously, cannot be omitted)
and various subsets of $\{\lambda_{{i}_1},..., \lambda_{{i}_n}\}$.
It is evident from \fig{fig:ResultsDrop47} that the omission of
either $\lambda_4$ or $\lambda_7$ (while keeping the rest)
leads to sizable deviations from our best results for $M_q(p)$ and $1/Z_q(p)$.
Similarly, retaining only the special combination $\lambda_{1,4,7}(p)$ reproduces very accurately
our best results for $M_q(p)$ and $1/Z_q(p)$, as shown in \fig{fig:ResultsDrop47}.

Note also, that the hierarchy of form factors established in {\it (i)}  is compatible with that of the corresponding couplings $\bar\lambda_i(p)$, whose relative size is shown in detail in \fig{fig:VDressing}. As we can see there, $\bar\lambda_{1,4,7}(p)$ are indeed the largest contributions; at the corresponding peaks, $\bar\lambda_{1} > \bar\lambda_{7} >|\bar\lambda_{4}|$.  In fact, the subleading form factors are even less relevant than suggested by the suppression of the couplings.

Importantly, the above study implies that the sole use of STI-derived vertices (which, by construction, do not include $\lambda_{4,7}$) in either the quark-gluon SDE or the gap equation leads to loss of quantitative precision. In particular, we have checked that the inclusion of the BC tensor structures alone in the gap equation reduces dramatically the amount of chiral symmetry breaking, yielding $M_l(0)< 50$\, MeV.

\begin{figure}[t]
	\centering
	\begin{subfigure}[t]{0.48\textwidth}
		\includegraphics[width=1\textwidth]{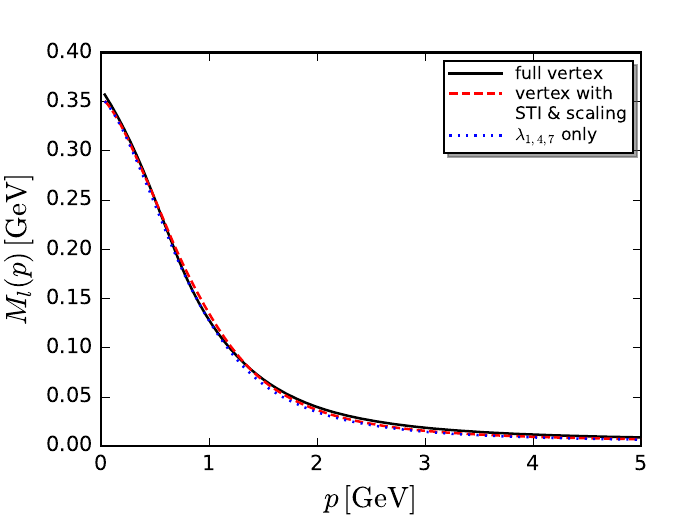}
		\caption{Quark mass function $M_l(p)$:  {full solution} (full black), {Vertex with STI-scaling} (dashed red),  {$\lambda_{1,4,7}$ only} (dotted blue). \hspace*{\fill}  }
		\label{fig:ResultsReduceda}
	\end{subfigure}
	\hfill
	\begin{subfigure}[t]{0.48\textwidth}
		\includegraphics[width=1\textwidth]{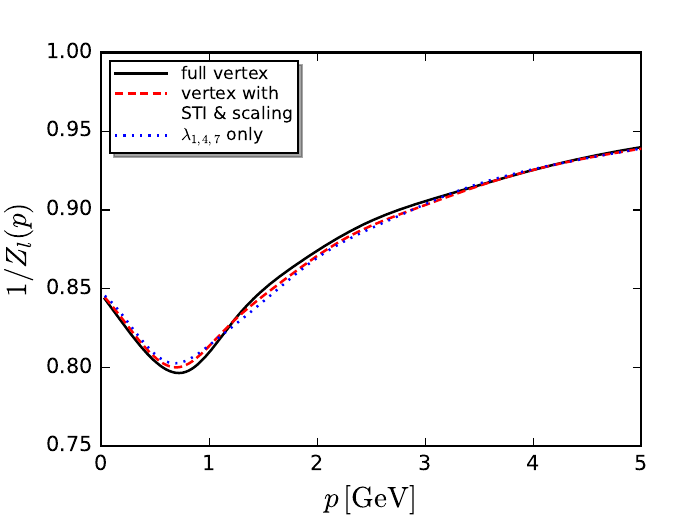}
		\caption{Quark dressing function $1/Z_l(p)$:  {full solution} (full black), {Vertex with STI-scaling} (dashed red),  {$\lambda_{1,4,7}$ only} (dotted blue). \hspace*{\fill}}
		\label{fig:ResultsReducedb}
	\end{subfigure}
	\caption{Quantitative reliability of approximations: {(a)} Quark-gluon couplings $\lambda_{1,2,6}$ from STI, and $\lambda_{4,5,6,7}$ from scaling relations,  see text and \cite{Cyrol:2017ewj, Gao:2020qsj}. {(b)}  $\lambda_{1,4,7}$.    \hspace*{\fill}}
	\label{fig:ResultsReduced}
\end{figure}

For the evaluation of point {\it (ii)}, we have computed $M_q(p)$ and $1/Z_q(p)$ with a dressing $\lambda_{1}(p)$ obtained from the STI construction, while for $\lambda_{4,7}(p)$ we resort to scaling relations suggested by the underlying gauge-invariant tensor structures~\cite{Cyrol:2017ewj, Gao:2020qsj, Gao:2020fbl}. The results obtained are in excellent agreement with those of the full computation, as can be seen in \fig{fig:ResultsReduced}.

The above analysis supports the appealing possibility of implementing relatively simple but quantitatively reliable approximations for hadron resonance computations or the phase structure of QCD, see also \cite{Williams:2015cvx}; such a setup is currently under investigation.

%

\section{Summary}\label{sec:summary}

In this work we have considered the full set of SDEs describing the quark sector of 2+1--flavour QCD. In particular, we have coupled the gap equation of the quark propagator with the one-loop dressed SDE of the quark-gluon vertex, and solved the resulting system of integral equations iteratively. The sole external ingredient used in this analysis is the gluon propagator, which has been taken from the lattice simulations of~\cite{Boucaud:2018xup,Zafeiropoulos:2019flq,Aguilar:2019uob} and results obtained from previous functional treatments~\cite{Gao:2020fbl,Gao:2020qsj,Fu:2019hdw}. Note, in particular, that the gauge coupling has been determined self-consistently, capturing correctly the analytic two-loop running.

The results of our analysis agree quantitatively with those of $N_f = 2+1$ lattice simulations~\cite{Bowman:2005vx}, and the combined (fRG and SDE) functional approach of~\cite{Gao:2020qsj,Gao:2020fbl}. In fact, our agreement with these latter approaches constitutes an important consistency  check within functional methods: SDEs and fRG represent similar but distinct non-perturbative frameworks, and the coincidence of the respective results is highly non-trivial. Moreover, the value for the chiral condensate, our benchmark observable, has been compared to recent lattice predictions compiled in the FLAG review~\cite{Aoki:2019cca}, showing excellent agreement, see \sec{sec:ChiralCond}.

Finally, we have  established that the form factors $\lambda_{1,4,7}(p)$ provide the dominant numerical contribution to the  chiral infrared dynamics, as already suggested by previous studies~\cite{Cyrol:2017ewj, Gao:2020qsj, Gao:2020fbl}. This allows us to devise simplified but quantitatively reliable approximations, which may reduce the numerical costs in the study of systems governed by a large number of intertwined dynamical equations.

In summary, the results of the current comprehensive SDE approach provide physics results without the need of phenomenological infrared parameters that are commonly used explicitly or implicitly. Moreover, we have shown how to self-consistently incorporate
in our analysis general external inputs.

In our opinion, the present comprehensive SDE approach, and in particular the combined use of functional relations for correlation functions, is essential for a successful quantitative investigation of many open physics problems in QCD, ranging from the hadron bound-state properties to the chiral phase structure and critical end point. \\[3ex]

\noindent{\bf Acknowledgements}\\[-1ex]

We thank A.C.~Aguilar, G.~Eichmann, C.F.~Fischer, M.Q.~Huber, and B.-J. Schaefer for discussions. F.~Gao is supported by the Alexander von Humboldt foundation. This work is supported by EMMI and the BMBF grant 05P18VHFCA, by the  Spanish Ministry of Economy and Competitiveness (MINECO) under grant FPA2017-84543-P, and the  grant  Prometeo/2019/087 of the Generalitat Valenciana. It is part of and supported by the DFG Collaborative Research Centre SFB 1225 (ISOQUANT) and the DFG under Germany's Excellence Strategy EXC - 2181/1 - 390900948 (the Heidelberg Excellence Cluster STRUCTURES).

\appendix

\section{Renormalization scheme}\label{app:RG-schemes}

In this Appendix we present additional details related to the MOM type renormalization scheme adapted in the present work, the MOM${}^2$ scheme. We emphasise that the scheme itself is not novel, but is the standard one used explicitly and implicitly in most fRG computations, and in particular in all applications to QCD, \ie \cite{Braun:2009gm, Mitter:2014wpa, Braun:2014ata, Rennecke:2015eba, Cyrol:2016tym, Cyrol:2017ewj, Cyrol:2017qkl, Corell:2018yil, Fu:2019hdw, Braun:2020ada}. It is simply the implementation of the MOM RG-condition at the initial cutoff scale (fRG-MOM${}^2$). It underlies the present work via the fRG gluon input from \cite{Cyrol:2017ewj}, and has been already used in the fRG-assisted SDE computations in \cite{Gao:2020fbl, Gao:2020qsj}. Still, due to its rare use in SDE computations, we would like to provide explicitly its relation to the standard MOM scheme as well as details concerning its numerical implementation and derivation in the present context.

In \app{app:RG-Maps} we first briefly describe the general setup of renormalization conditions in the fRG, in particular putting in perspective the different notion of correlation functions as well as bare and renormalized fields. This should allow the appreciation of the general setting without having to go through the more technical derivations deferred to \app{app:MOM2}. In \app{app:MOM2-MOM} we put all these relations to work. First we fit our numerical results to one- and two-loop formula. This allows us to determine the momentum range of validity of the respective perturbative approximations. Moreover, we use the generalization of the analytic one- and two-loop formula introduced by the present MOM${}^2$ scheme to discuss and quantify its relation to the standard MOM scheme.

A quantitative discussion, including the derivations, is provided in \app{app:MOM2}. There, the present scheme is derived within a Wilsonian approach to the path integral, leading to finite SDEs. Furthermore, we show how this setup can be recast in a standard momentum-subtraction scheme.

\subsection{Mapping RG schemes}\label{app:RG-Maps}
In the fRG approach, the scale dependent effective action $\Gamma_k$ includes all quantum fluctuations (loop corrections) with momentum scales $p^2\gtrsim k^2$. Hence, for $k\to\infty$, all quantum fluctuations are suppressed, and $\Gamma_k$ tends towards the bare action of QCD.  This also entails that the momentum-dependence of the dressings of correlation functions $\Gamma^{(n)}_k$ is suppressed for $p^2/k^2\lesssim 1$: the dressings tend towards the renormalization factors of the bare action, within a momentum-cutoff renormalization. Accordingly, for large IR cutoff scales $k$, the $k$-dependence of the effective action in the fRG approach translates to a dependence on the UV cutoff, $\Lambda$, in the present SDE approach. Moreover, the dependence on the renormalization scale $\mu$ is the same. In particular, logarithmically divergent RG factors run with $\log \Lambda^2/k_\textrm{ref}^2$, where $k_\textrm{ref}$ denotes  some reference scale, that typically is chosen large. This logarithmic $\Lambda$-dependence precisely cancels that produced by the integrated flows, where the cutoff integration runs from $k=\Lambda$ to $k=0$. This integrated flow agrees with the regularised SDE diagrams (again with  UV cutoff $\Lambda$).

Naturally, there is a very specific choice of $k_\textrm{ref}$, namely $k_\textrm{ref} =\Lambda$ (or proportional to $\Lambda$), that minimises the logarithmic corrections. This is very similar to minimising large log contributions in perturbation theory. For such a  choice, the terms proportional to $\log \Lambda^2/k_\textrm{ref}^2$ vanish and we can put all $Z_{\phi_i,k=\Lambda}=1$, $M_{q,\Lambda}=m_q$, and $\lambda_{A^3,\Lambda}=\lambda_{c \bar c A,\Lambda}=\lambda_{q\bar q A,\Lambda}=g_s$, $\lambda_{A^4,\Lambda}=g_s^2$. In practice, this has to be accompanied with setting the RG scale to $\mu^2=\Lambda^2$, and finally removing the cutoff scale $k$. In turn, for $k\to 0$, the fRG dressings are simply the finite dressings of the full, renormalized theory, and no dependence on the cutoff scale is left. This property is called {\it ``RG-consistency''}, see~\cite{Pawlowski:2005xe,Pawlowski:2015mlf,Braun:2018svj}.

We call this scheme the MOM${}^2$-scheme. With the above information we now translate this (typically implicit) RG scheme within the fRG to an explicit RG scheme in the SDEs. In short, this can be achieved by absorbing the wave function renormalizations in the fields at the RG scale $\mu$. As a result, correlation functions, while being finite, carry the RG properties of correlation functions of \textit{finite} bare fields, where the term ``bare'' only reflects their RG properties: they are invariant under RG-rescalings\footnote{Note that this invariance only reflects the property that the dispersion of these fields is classical at the RG point. In turn, at a fixed momentum $p$, the dispersion changes when changing $\mu$.}. Moreover, the MOM condition is defined for these correlation functions and not for the standard renormalized ones. Consequently, this is a MOM-type scheme whose
wave function renormalizations also equal to unity, hence the name MOM${}^2$. Note that, if translated to a standard  RG condition for renormalized $n$-point correlation functions, MOM${}^2$ is not MOM, but the differences are proportional to $c_n (\alpha_s(\mu))^{j_n}$, where $c_n$ are a $\mu$-independent constants and $j_n=1,3/2,...$ is some positive power. Accordingly, these differences vanish for $\mu\to\infty$. In summary, the current scheme has the advantage of a natural RG condition without fine-tuning for neither the correlation functions nor their renormalization constants. The price to pay is the necessity
of mapping certain quantities, \eg the resulting running couplings, to their standard MOM counterparts.
Note, however, that this necessity does not appear at the level of the actual computation, which is now freed from the
delicate adjustments of the renormalizations constants that are typically required.

\subsection{Running couplings $\alpha_s(p)$ and $\Lambda_{\textrm{QCD}}$: From MOM${}^2$ to MOM}\label{app:MOM2-MOM}

In \sec{sec:results}, the full quark-gluon coupling has been compared with its one- and two-loop counterparts, see in particular \Fig{fig:alphasComp}. It is here, where the difference between the current MOM${}^2$ scheme, described in \sec{sec:MultRenorm} and \app{app:RG-Maps}, and the standard MOM scheme becomes most apparent. Its explicit derivation is discussed in detail in the next Appendix, \app{app:MOM2}. In the present section we map the MOM${}^2$ couplings to MOM couplings, our findings in comparison to MOM results in the literature are summarised in \fig{fig:Couplingcomp}.

It is evident from \app{app:RG-Maps}, that the MOM${}^2$ scheme allows for rescalings of the couplings with ratios of renormalization functions. Hence, we define the strong coupling through a slight generalization of the standard one, \eg~\cite{Prosperi:2006hx} with a multiplicative factor, $z_s$. Accordingly, for $\mu\to\infty$, the $z_s$ is the difference between the MOM the MOM${}^2$ couplings. This leads us to
\begin{align}\label{eq:alpha-1loopMOM2}
\alpha_s^{\textrm{(1loop)}}(p)=\frac{z_s}{ \beta_0\,\textrm{ln}(p^2/\Lambda^2_\textrm{QCD})}\,,
\end{align}
with $\beta_0=\frac{1}{4\pi}(11-\frac{2}{3}N_f)$. The two parameters  $z_s$ and $\Lambda_\textrm{QCD}$ are then determined by a best
$\chi^2$ fit of \eq{eq:alpha-1loopMOM2} in a momentum regime about $\kappa=p$. Naturally, for a given RG scale $\mu$, one may identify $\mu=\kappa$. However, the variation of $\kappa$ also provides some information about the range of validity of the one-loop approximation. \eq{eq:alpha-1loopMOM2}. In the present work we use the interval,
\begin{align}
	\label{eq:FitInterval}
	p\in(\kappa, \kappa+\Delta\kappa)\,,
\end{align}
and the size of the regime $\Delta\kappa\approx 10 - 60$\,GeV is minimised under the constraint that it includes enough data points from our solution of the SDEs for achieving accurate results. It hence increases with $\kappa$.

To begin with, we first neglect the factor $z_s$, setting $z_s\to 1$. At one loop this is tantamount to a rescaling of $\Lambda_\textrm{QCD}$. As is clear from \eq{eq:alpha-1loopMOM2}, $\Lambda_{\textrm{QCD}}$ provides the position of the singularity of $\alpha_s^{\textrm{(1loop)}}(p)$ in momentum space, $p_\textrm{sing}:=\Lambda_{\textrm{QCD}}$. This definition is also used below for the two-loop coupling, $\alpha_s^{\textrm{(2loop)}}(p)$. We also emphasise that, natural as it may be, this  is not the only possible definition of $\Lambda_{\textrm{QCD}}$ beyond one loop, see \eg \cite{Deur:2016tte}.

Even though $\Lambda_\textrm{QCD}$ is, in principle, $\kappa$- and hence $\mu$-independent, this property is not exhibited at the level of the one-loop formula. Specifically, we obtain $\Lambda_\textrm{QCD}=0.59$\,GeV for $\mu=40$\,GeV, and  \mbox{$\Lambda_\textrm{QCD}=0.86$\,GeV} for $\mu=4.3$ GeV. The large difference between the respective $\Lambda_\textrm{QCD}$ is yet another manifestation of the limitations of the one-loop approximation. These limitations prevent us from using large RG scales even in the context of the standard MOM scheme, if the approximation does not at least leads to two-loop consistent results. Moreover, the analysis also entails that $\Lambda_{\textrm{QCD}}$ in the current MOM${}^2$ scheme is significantly different from the respective value in the standard MOM scheme. This has been already discussed in detail in~\cite{Ellwanger:1997tp}, where a comparison of the present scheme with the $\overline{\textrm{MS}}$ scheme was carried out within Yang-Mills theory. This analysis extends straightforwardly to a comparison with MOM. The difference with the standard MOM scheme is also clearly seen in the comparison of the Yang-Mills data of~\cite{Cyrol:2016tym} (fRG, present scheme) with those of~\cite{Huber:2020keu} (SDE, MOM scheme), both featuring correlation functions in quantitative agreement with the respective lattice results.

\begin{figure}[t]
	\centering
	\includegraphics[width=.5\textwidth]{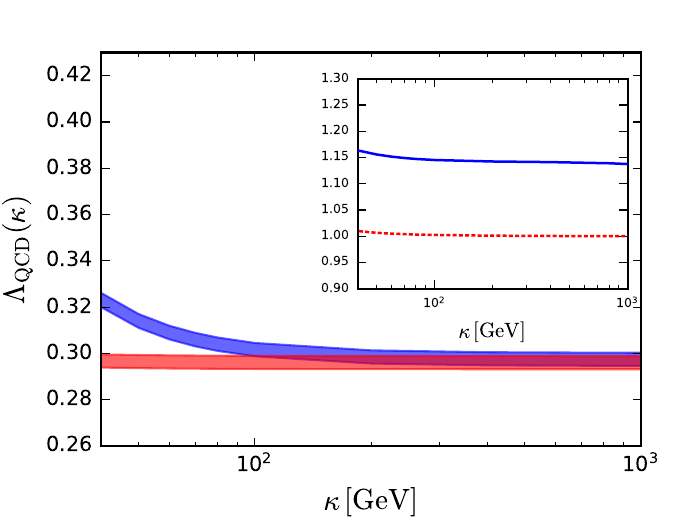}
	\caption{Comparison of $\Lambda^\textrm{(1loop)}_{\textrm{QCD}}(\kappa)$ derived from \eq{eq:alpha-1loopMOM2}  (blue) and the quark mass function, \eq{eq:UVlimDeltal} (red) from a fit in the regime \eq{eq:FitInterval} with $\Delta\kappa=60$\,GeV. The inset shows $\Delta_{l,\chi}(\kappa)$ (red dashed) normalized by $\Delta_{l,\chi}(\kappa=10^3 \rm GeV)=(244(6)\,\textrm{MeV})^3$ as well as $z_s(\kappa)$ (blue solid).  \hspace*{\fill}  }
	\label{fig:LambdaQCD1loop}
\end{figure}

Now we proceed to the situation with a non-trivial $z_s$. Then the pair $(z_s, \Lambda_\textrm{QCD})$ is determined by fitting the asymptotic momentum behaviour of the running coupling to \eq{eq:alpha-1loopMOM2} in the interval \eq{eq:FitInterval}. This allows us to deduce the standard MOM coupling $\alpha_{s,\textrm{MOM}}$ from that in \eq{eq:alpha-1loopMOM2}, henceforth called $\alpha_{s,\textrm{MOM}^2}(p)$: For $\mu\to\infty$, the present MOM${}^2$ RG condition for the rescaled fields also implies the MOM condition for the standard renormalized fields, up to rescalings. For the coupling, these rescalings are carried by $z_s$, and we obtain
\begin{align}\label{eq:MOM2toMOM}
		\alpha_{s,\textrm{MOM}}(p)= \lim_{\mu\to\infty} \frac{1}{z_s}\alpha_{s,\textrm{MOM}{}^2}(p)\,,
\end{align}
which holds quantitatively at our large RG scale $\mu=40$\,GeV. In \fig{fig:LambdaQCD1loop} we depict $\Lambda_\textrm{QCD}^{\textrm{(1loop)}}$ derived from \eq{eq:alpha-1loopMOM2} (blue) in comparison to that derived from the asymptotic momentum behaviour of the quark mass function (red), see \eq{eq:UVlimDeltal} and the discussion below. We conclude that, for $p\gtrsim 10^2 $\, GeV, the necessary condition for one-loop compatibility is satisfied. This is in line with respective considerations in perturbation theory, see \eg \cite{Deur:2016tte}.  Moreover, the inset in \fig{fig:LambdaQCD1loop} also shows the stability of the chiral condensate $\Delta_{l,\chi}$ for $\kappa\gtrsim 10^2$\,GeV. In the inset we also show $z_s(\kappa)$. The respective asymptotic chiral limit condensate $\Delta_{l,\chi}$, as well as  $\Lambda_\textrm{QCD}$ in the MOM${}^2$ scheme, are given by,
\begin{figure}[t]
	\centering
	\includegraphics[width=.5\textwidth]{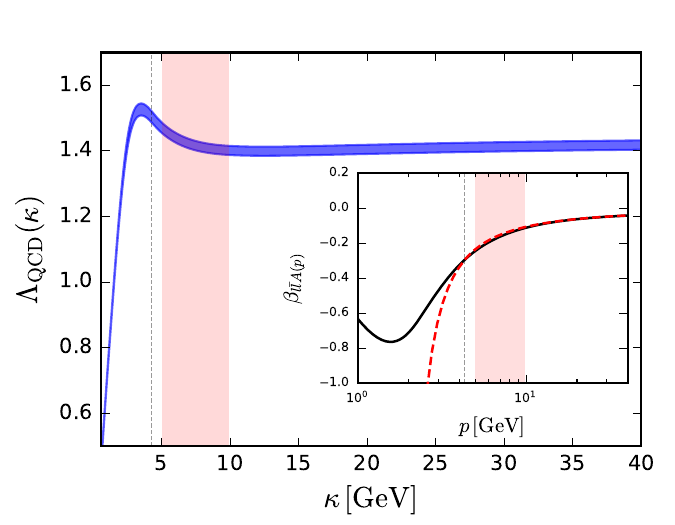}
	\caption{$\Lambda_{\textrm{QCD}}(\kappa)$, obtained by the fit of \eq{eq:alphas2loop} with $z_s=1$ to the numerical $\alpha_s$-data in the regime \eq{eq:FitInterval} for $\kappa\in (1, 40)$\,GeV. The RG scale $\kappa=4.3$\,GeV is indicated with a dashed vertical line. The inset shows the full $\beta$-function, $\beta_{l\bar l A}(\alpha_s)$, in comparison to its two-loop counterpart. The red vertical band indicates the transition momentum regime, in which the full $\beta$-function starts to deviate from the perturbative two-loop $\beta$-function. \hspace*{\fill}  }
	\label{fig:Singmu}
\end{figure}
\begin{align}\label{eq:LQCDMOM21loop}
\Delta_{l,\chi}=  (244(6)\,\textrm{MeV})^3\,, \qquad \qquad 	\Lambda_\textrm{QCD}^{\textrm{(1loop)}} =295(18)\,\textrm{MeV}\,.
\end{align}
For the evaluation at two loop we use the standard parameterization given, \eg~in~\cite{Prosperi:2006hx, Deur:2016tte}, together with the rescaling factor $z_s$ from renormalized to bare correlation functions. This leads us to,
\begin{align}\label{eq:alphas2loop}
\alpha_s^{\textrm{(2loop)}}(p)=-\frac{\beta_0}{\beta_1}\frac{z_s}{1+W_{\!-1}(y)}\,,
\qquad y=-\frac{\beta^2_0}{e\beta_1}\left(\frac{\Lambda^2_\textrm{QCD}}{p^2}\right)^{\!\!\beta_0^2/\beta_1} \,,
\end{align}
with $\beta_1=\frac{1}{(4\pi)^2}(102-\frac{38}{3}N_f)$ and $\beta_0=\frac{1}{4\pi}(11-\frac{2}{3}N_f)$, where $W_{\!-1}(y)$ denotes the ``physical'' branch of the real valued Lambert function. We also note in passing that the approximate formula~\cite{Pich:2020gzz},
\begin{align}
\alpha_s^{\textrm{(2loop)}}(p)=\frac{ z_s \,\alpha(\mu) }{1+  \beta_0\,\alpha(\mu)\left[1+\alpha(\mu) \frac{\beta_1}{\beta_0}\right]\ln(p^2/\mu^2)}\,,
\end{align}
fits the full coupling $\alpha_{l \bar l A}(p)$ even better than \eq{eq:alphas2loop} for momenta $p\gtrsim 10$\,GeV (in terms of $\chi^2$).This may be interpreted as an indication of the effectiveness of the resummation scheme underlying our approximation in the perturbative regime. Note that this statement holds true for $z_s=1$. Note also that in the present scheme with $z_s\neq 1$ we have $\alpha_s^{\textrm{(2loop)}}(\mu)= z_s \alpha(\mu)$, the latter being the MOM value, $\alpha(\mu)=\alpha_{s,\textrm{MOM}}(\mu)$. We emphasise that the use of the one- and two-loop fit formula introduced above does not change any results or the renormalization procedure, they are simply introduced for elucidating the scheme used here.

Again we first discuss the two-loop fits with $z_s=1$: as is clear from \fig{fig:Singmu},  $\Lambda_{\textrm{QCD}}$ of $\alpha_s^{\textrm{(2loop)}}(p)$ is stable  under changes of $\mu$, for $\mu\gtrsim 10$\,GeV, and is still compatible in a transition regime with $p\in(5-10)$\,GeV. This is in clear contradistinction to its one-loop counterpart, where such an RG-consistency does not hold. In particular, we find that $\Lambda_{\textrm{QCD}}=1.42$\,GeV for $\mu=40$\,GeV, and $\Lambda_{\textrm{QCD}}=1.49$\,GeV for our  low RG scale of $\mu=4.3$\,GeV. The latter RG scale is at the limit or slightly below the lower bound of the two-loop consistent regime. Note that this low RG scale has been chosen because it represents the largest momentum accessible by the lattice data.

\begin{figure}[t]
	\centering
	\includegraphics[width=.45\textwidth]{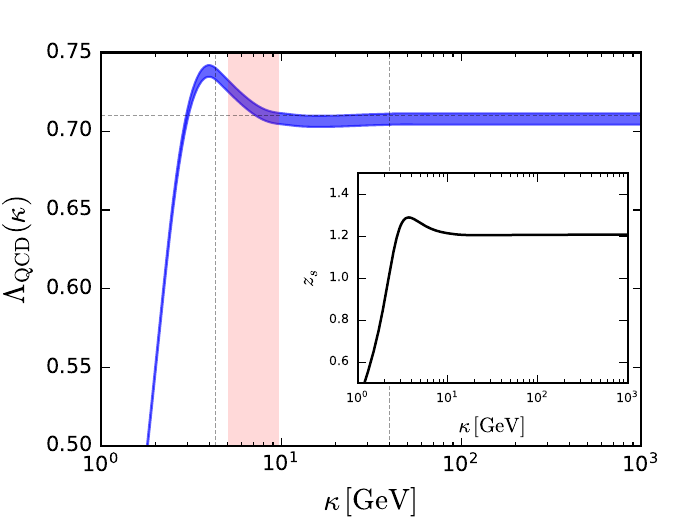}\hspace{1cm}
\includegraphics[width=.45\textwidth]{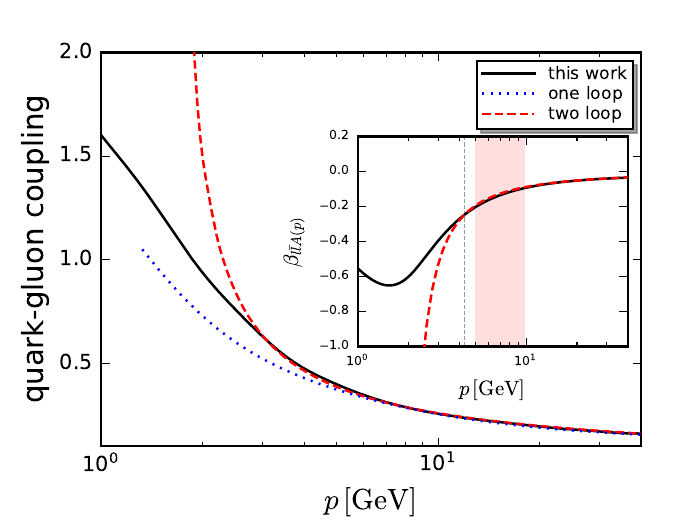}
	\caption{\emph{Left panel}: $\Lambda_{\textrm{QCD}}(\kappa)$, obtained by the fit of \eq{eq:alphas2loop} to the numerical $\alpha_s$-data in the regime \eq{eq:FitInterval} for $\kappa\in (1, 40)$\,GeV. The RG scales $\kappa=4.3, 40$\,GeV is indicated with a dashed vertical lines. The inset shows the rescaling factor $z_s$ defined in \eq{eq:alphas2loop}. \emph{Right panel}:
		Full quark-gluon coupling $\alpha_{q \bar q A}$ in comparison to the one-loop, \eq{eq:alpha-1loopMOM2}, and two-loop counterpart, \eq{eq:alphas2loop}, with $\mu=40$\,GeV. The inset shows the full $\beta$-function, $\beta_{l\bar l A}(\alpha_s)$, in comparison to its two-loop counterpart. The red vertical band indicates the transition momentum regime, in which the full $\beta$-function starts to deviate from the perturbative two-loop $\beta$-function. \hspace*{\fill}  }
	\label{fig:Singmuzs}
\end{figure}
Now we proceed to the fits with $(z_s, \Lambda_{\textrm{QCD}})$. The respective results are depicted in \fig{fig:Singmuzs}. As for the fit with $z_s=1$, we see no $\kappa$-dependence of $\Lambda_{\textrm{QCD}}$ in the regime $\mu\gtrsim 10$\,GeV. This $\kappa$-independence also extends to the rescaling $z_s$, and with $\kappa\gg 10$\, GeV we arrive at the asymptotic values,
\begin{align}\label{eq:MOMLambda+zs}
\Lambda_{\textrm{QCD}} = 708(3)\, \textrm{MeV}\,, \qquad \qquad z_s= 1.19(5) \,.
\end{align}
Both the stability of our two-loop determination for $\kappa\gtrsim 10$\,GeV as well as its $\kappa$-dependence for $\kappa\lesssim 10$\,GeV, are in perfect agreement with respective considerations in perturbation theory: there, two-loop and three- or four-loop running strong couplings begin to deviate at about $p\approx 10$\,GeV.

With the definition of the MOM${}^2$ couplings in terms of the vertex and propagator dressing in \eq{eq:avatars} and the relation between MOM and MOM${}^2$ couplings in \eq{eq:MOM2toMOM} we are led to
\begin{align}
	\label{eq:avatarsMOM}
\alpha_{c\bar c A}(\bar p) =\frac{1}{z_s}\frac{1}{4 \pi} \frac{[\lambda^{(1)}_{c\bar c A}(\bar p)]^2}{Z_A(\bar p)Z_c^2(\bar p) }\,,	\qquad\qquad	\alpha_{q\bar q A}(\bar p) = \frac{1}{z_s}\frac{1}{4 \pi} \frac{[\lambda^{(1)}_{q\bar q A}(\bar p)]^2}{Z_A(\bar p)Z_q^2(\bar p)} \,,
\end{align}
where $\bar p$ signals the symmetric point \eq{eq:SymPoint}.  The relation \eq{eq:MOM2toMOM}  can be used for asymptotically large RG scales, and we have shown above that $\mu=40$\,GeV is indeed sufficiently large, and hence we will use \eq{eq:MOMLambda+zs}. The respective MOM couplings are depicted in \fig{fig:Couplingcomp}. The values of the quark-gluon coupling in the MOM-scheme at selected momenta are given by
\begin{align}
	\alpha_{q\bar q A}(4.3\,\textrm{GeV}) = 0.365\,,  \qquad \quad \alpha_{q\bar q A}(40\,\textrm{GeV}) =  0.140\,,\qquad  \quad \alpha_{q\bar q A}(M_Z) =  0.119\,,
	\label{eq:alphaMOMp}
\end{align}
which is quantitatively compatible with the literature values for $2+1$-flavour QCD, see, \eg \cite{Deur:2016tte}. In \fig{fig:Couplingcomp} we also compare our prediction for the Taylor coupling, $\alpha_T(p)$, with infrared ($p\lesssim 4$\,GeV) lattice results in~\cite{Zafeiropoulos:2019flq}. The Taylor coupling is defined from the ghost and gluon dressings as
\begin{align}\label{eq:Taylor}
	\alpha_T(p) = \frac{1}{z_s} \frac{1}{4 \pi} \frac{[\lambda^{(1)}_{c\bar c A}(\bar p=\mu)]^2}{Z_A(p)Z_c^2(p) }\,,
\end{align}
with $\bar p$ defined in \eq{eq:SymPoint}. In the present scheme we have $\lambda^{(1)}_{c\bar c A}(\bar p=\mu)=1$.

The Taylor coupling $\alpha_T(p)$ in \eq{eq:Taylor} is closely related to the ghost-gluon vertex coupling defined in \eq{eq:avatarsMOM}, the difference being the momentum dependence of the ghost-gluon vertex dressing present in the latter: In the Landau gauge, the latter carries no renormalization scale dependence and hence the  $\beta$ functions agree up to two-loop. However, the respective momentum dependence differs beyond one loop.
\begin{figure}[t]
	\centering
	\includegraphics[width=.48\textwidth]{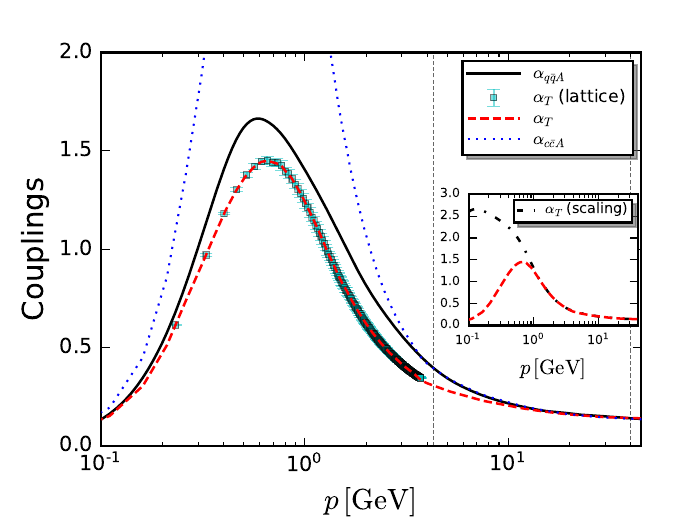}\hspace{.6cm}\includegraphics[width=.48\textwidth]{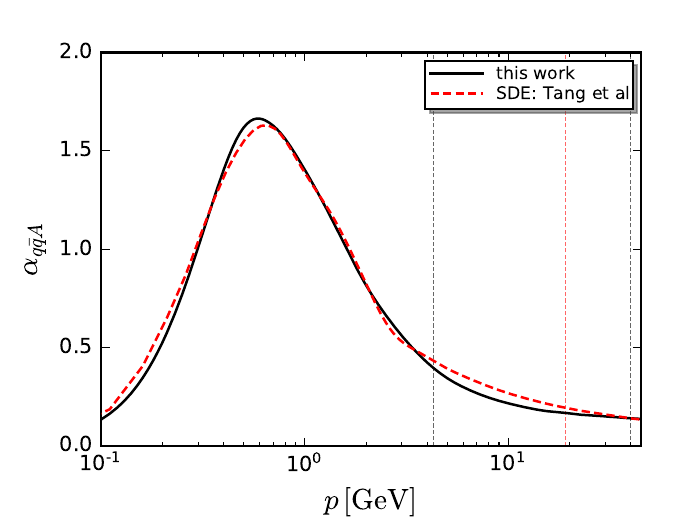}
	\caption{MOM scheme couplings defined in \eq{eq:avatarsMOM} and \eq{eq:Taylor}. The grey dashed vertical lines indicate the RG scales used in the present work. \emph{Left panel}: Quark-gluon coupling $\alpha_{q\bar q A}$ ({black solid line}), ghost-gluon coupling $\alpha_{c\bar c A}$  ({blue dotted line}), and Taylor coupling $\alpha_T$ ({red dashed line}) in comparison to the lattice Taylor coupling~\cite{Zafeiropoulos:2019flq}. The inset shows $\alpha_T$ with the lattice infrared ghost ({red dashed line}) and scaling ghost dressing ({black dashed-dotted line}). \emph{Right panel}: $\alpha_{q\bar q A}$ from the present work ({black solid line}) in comparison to that constructed from the vertex dressing $\lambda^{(1)}_{q\bar q A}$, and the quark dressing $Z_q$ from \cite{Tang:2019zbk} as well as the $Z_A$ from the present work ({red dashed line}). The vertical red dashed line indicates the RG scale used in \cite{Tang:2019zbk}. \hspace*{\fill}  }
	\label{fig:Couplingcomp}
\end{figure}
The Taylor coupling determined in the present work depends on the input gluon data for $Z_A$ in 2+1 flavour QCD from functional methods, \cite{Fu:2019hdw, Gao:2020qsj, Gao:2020fbl}, based on the two-flavour results in \cite{Cyrol:2017ewj}. The ghost is a combination of the infrared tail from ~\cite{Zafeiropoulos:2019flq} for $p\lesssim 4$\, GeV (lattice) and the UV tail from \cite{Cyrol:2017ewj} (2-flavour QCD). The ghost propagator in the latter work is of the scaling-type, the respective Taylor coupling $\alpha_s$(scaling) is shown in the inset in \fig{fig:Couplingcomp}, and agrees up to $p\approx 1$\,GeV with the Taylor coupling with the lattice (decoupling) IR tail, $\alpha_s$(lattice). The two-flavour approximation for the functional ghost propagator is based on the fact that the ghost is nearly insensitive to the $s$-quark, and the functional data in \cite{Cyrol:2017ewj} also underly the gluon propagator. We emphasise that the quantitative agreement of this result with the lattice result solely tests the reliability of our determination of the MOM coupling at $p=40$\,GeV, as well as the accuracy of our input data in the regime $p\in (4,40)$\,GeV, where no lattice data is available. Hence, the quantitative agreement in \eq{fig:Couplingcomp} shows the reliability of the present determination of correlation functions and the input data used on the 5\% level, well within our conservative systematic error estimate.

Finally, the quantitative agreement of the quark-gluon coupling $\alpha_{q \bar q A}(p)$, computed in the present work, and the ghost-gluon coupling $\alpha_{c\bar c A}$ for momenta $p\gtrsim 4$\,GeV entails that, in the present framework the STIs are accurately fulfilled, being the cornerstone of any quantitatively reliable approximation scheme. Note that this STI consistency of all avatars \eq{eq:avatars} of the strong coupling is already present in \cite{Cyrol:2017ewj}, and hence underlies \cite{Fu:2019hdw, Gao:2020qsj, Gao:2020fbl} and the present computations. However, importantly, our determination of the running quark-gluon coupling discussed in \sec{sec:SC-alphas} re-enforces this property. Both procedures discussed there, \textit{(i)} and \textit{(ii)}, have this property, and in \textit{(ii)}, being  based on STIs, this is most apparent. Accordingly, while rooted in the input data, its persistence in the current approach is highly non-trivial, and an asset of the present approach: apart from stabilising the computations, it enforces gauge-consistency and self-consistency.

In the right panel of \fig{fig:Couplingcomp} we compare the quark-gluon coupling $\alpha_{q \bar q A}(p)$ with SDE results from \cite{Tang:2019zbk}, also based on the full tensor basis for the quark-gluon coupling. Further comparisons with the fRG-DSE results from \cite{Cyrol:2017ewj, Gao:2020fbl, Gao:2020qsj} can be found in \fig{fig:ResultsCompareLitqg}.

In \cite{Tang:2019zbk} a gluon model was used. For the sake of the comparison we hence have augmented the vertex dressing $\lambda_1(\bar p)$  and quark dressing $Z_q(p)$ in \cite{Tang:2019zbk} with the current 2+1 flavour gluon dressing $Z_A(p)$ from the current work for defining the respective $\alpha_{q \bar q A}(p)$, \eq{eq:avatars} with MOM dressings. Below approximately 3\,GeV the coupling from \cite{Tang:2019zbk}  agrees well with the present one, as required for quantitative results for the quark dressings. Indeed, in \cite{Tang:2019zbk} the size of the coupling at the RG-scale $\mu=19$\,GeV was adjusted with the size of the quark mass function in the infrared. In turn, the coupling and its momentum running differs significantly for momenta larger than $3$\,GeV, where the present coupling is two-loop compatible as discussed above.

In summary, we have shown that the present results for correlation functions and couplings in the MOM${}^2$ scheme are fully consistent with quantitative results in the MOM scheme. A final remark concerns possible choices of the RG scale $\mu$: as already discussed in the main text, it is best taken asymptotically large. However, when using lattice input data, we are restricted to relatively small $\mu\lesssim 5$\,GeV. From \fig{fig:Couplingcomp} we can infer yet again that, for $\mu\lesssim 4$\,GeV, the standard ``perturbative'' renormalization procedure based on the equality of vertex couplings, $\alpha_i(\mu)=\alpha_s(\mu)$, is bound to fail. Loosely speaking, while apparently in line with the STIs, in reality such a choice violates them, as the vertex couplings cease to agree in the infrared. We emphasise that the latter feature is \textit{not} an artefact of a particular approximation, but rather, an intrinsic property of the system.

\subsection{Derivation of the MOM${}^2$ scheme for the SDE}\label{app:MOM2}

We close this section with a brief technical derivation of the above statements and relations. Its understanding requires a minimal
familiarity the fRG-approach, since a full self-contained derivation is beyond the scopes of the present work. For more details on fRG renormalization schemes in the context of QCD we refer to the reviews \cite{Pawlowski:2005xe, Gies:2006wv, Rosten:2010vm, Braun:2011pp, Dupuis:2020fhh}.

\subsubsection{General remarks}
Functional relations for QCD, such as SDE, fRG flow equations, and nPI (n-particle irreducible) hierarchies, are different but equivalent loop equations for one-particle irreducible (1PI) correlation functions in full and dressed vertices and propagators.

While the first two hierarchies are \textit{closed}, being one- and two-loop exact functional relations for 1PI correlation functions, the latter nPI hierarchies are not: They are closed in terms of nPI correlation functions, but are infinite loop order equations for 1PI correlation functions. Accordingly, we can present concise closed functional 1PI relations for the full hierarchies of fRG and SDEs, but not for the nPI ones. We now discuss the derivation of these concise relations in the presence of an infrared cutoff term for the effective action $\Gamma_k[\phi]$ for both, the fRG and the SDE approaches. The cutoff term is introduced by changing the (scalar part of the) dispersion $p^2$ for gauge fields and ghosts in the bare action with $p^2 \to p^2 +R_k(p)$, the latter being a momentum dependent mass, that decays rapidly for momenta $p^2/k^2 \to \infty$.  This leads to propagators $G_k$ and vertices $\Gamma^{(n)}_k$ that depend on the cutoff $k$, or rather, the regulator function $R_k$.

\subsubsection{Functional identities and finite SDEs}

Finally, both hierarchies of equations are used to compute the (1PI) correlation functions, which are related to the moments of the path integral/generation function
\begin{align}
	Z[J]\propto\int d\varphi\,e^{-S[\varphi]+\int J\cdot \varphi} \,,
	\label{eq:ZJ}
\end{align}
where, for the sake of simplicity, we have dropped the normalization. \Eq{eq:ZJ} has to be UV-regularised and renormalized. Assume now that this has been done and we have already integrated out the UV momentum modes $\varphi_+= \varphi(p^2\geq \Lambda )$. Then, only the integration over $\varphi_-= \varphi(p^2\leq \Lambda )$ is left, and we are led to,
\begin{align}
	Z[J]\propto \int  d\varphi_-\,e^{-S_{\textrm{eff},\Lambda}[\varphi_-, J_+]+\int J \cdot \varphi_-}\,, \quad\textrm{with}\quad
	e^{-S_{\textrm{eff},\Lambda}[\varphi_-,J_+]} =  \int d\varphi_+\,e^{-S[\varphi_-+\varphi_+]+\int J \cdot \varphi_+}\,,
	\label{eq:ZJksharp}
\end{align}
where we have introduced the Wilsonian effective action $S_{\textrm{eff},\Lambda}$. Furthermore, we have used, that $\int J\cdot \varphi_\pm = \int J_\pm \cdot \varphi= \int J_\pm \cdot \varphi_\pm$. The constrained functional integral for the Wilsonian effective action in \eq{eq:ZJksharp} can be represented in terms of an unconstrained functional integral with an infrared regularization of the classical dispersion in the exponent of the path integral, to wit,
\begin{subequations}\label{eq:WilsonZ}
\begin{align}
	e^{-S_{\textrm{eff},\Lambda}[\phi_-, J_+]}  =&\, \int d\varphi \, e^{-S[\varphi+\phi_-] -\frac12 \int_p \varphi(-p)
	R_{\Lambda}(p) \varphi(p) +\int J \cdot \varphi}  \,,
\label{eq:SWilson}\end{align}
with
\begin{align}\label{eq:Rsharp}
	R_{k,\textrm{sharp}}(p) = \left\{\begin{array}{lcl}
		\infty & \qquad \qquad &p^2 \leq k^2\,,\\[1ex]
		0 & & p^2 \geq k^2 \,,
	\end{array}\right.  \qquad  \textrm{and}\qquad R_k(p) = p^2 \left(\frac{1}{\theta\left(\frac{p^2}{k^2}-1\right)}-1\right)\,,
\end{align}
\end{subequations}
for a generic infrared cutoff scale $k$, and $\theta(x)$ is the Heaviside step function. Note that the current analysis straightforwardly extends to smooth regulators, as typically used in advanced numerical applications. For the performance of numerical momentum integrations, the non-analyticities introduced by the sharp cutoff lead to a significant slowing down. Moreover, it can be shown that sharp cutoffs are not optimised, \cite{Litim:2000ci, Litim:2001up, Pawlowski:2005xe, Pawlowski:2015mlf}, since their use typically increases the truncation artefacts. For smooth regulators, the distinction between infrared and ultraviolet modes $\phi_\pm$ and their integrations requires more care. Given that these practical considerations do not add to the structural understanding of the MOM${}^2$ renormalization scheme, in what follows we concentrate on the simple sharp cutoff case.

The sharp cutoff regulator \eq{eq:Rsharp} leads to propagators with
\begin{align}
	\label{eq:G=0}
	G_k(p^2\leq k^2 ) = 0\,,
\end{align}
and all momentum loops are cut-off in the infrared. As a result, the UV regularization and renormalization are implicit, since finally we will work only with finite functionals and functional relations.

For the integration over $\varphi_-$ we now utilise, that the cutoff procedure applies to the situation with a generic infrared cutoff $k$.  This cutoff is eventually taken to zero. Within this setup we write,
\begin{align}
	Z_k[J]\propto  \int  d\varphi \,e^{-S[\varphi] -\frac12 \int_p \varphi(-p)
		R_{k}(p) \varphi(p)+\int J\cdot \varphi} = \int   d\varphi_-\, e^{-S_{\textrm{eff},\Lambda}[\varphi_-,J_+] -\frac12 \int_p \varphi_-(-p) R_k(p) \varphi_-(p)+\int J\cdot \varphi_-} \,,
	\label{eq:ZJk}
\end{align}
where we have used \eq{eq:WilsonZ} for the introduction of a Wilsonian UV action. Note that $S_{\textrm{eff},\Lambda}$ does not depends on $k$, as $k<\Lambda$. For $k\to 0$,  the generating functional $Z_k$ in \eq{eq:ZJk} approaches the full path integral $Z=Z_0$. In turn, for $k\to \Lambda$,
it approaches $\exp\{-S_{\textrm{eff},\Lambda}[\varphi] \}$. For large $\Lambda$, the Wilsonian action $S_{\textrm{eff}}$ tends towards the (classical) bare action, $S_{\textrm{eff},\Lambda}\to S_\textrm{bare}$. Note that the latter necessarily depends on $\Lambda$. Evidently, we also have
\begin{align}\label{eq:ZRG}
	\Lambda \frac{ Z_k[J]} {d 	\Lambda } =0\,;
\end{align}
the path integral $Z_k$. Hence, also the full one, $Z=Z_0$, is independent on the ``intermediate'' UV-cutoff $\Lambda$. The reasoning behind \eq{eq:ZRG} can also be applied to the (implicit) UV regularization and renormalization with a UV cutoff (or regularization scale) $\Lambda_{\textrm{QCD}}\textrm{UV}$. Accordingly, there is also no $\Lambda_\textrm{UV}$-dependence. It is for this reason that we can safely ignore the (implicit) UV renormalization behind using a \textrm{finite} Wilsonian action $S_{\textrm{eff},\Lambda}$.

In summary, finiteness of $S_{\textrm{eff},\Lambda}$, together with \eq{eq:ZRG}, encodes the regularization and renormalization of the path integral. Indeed, the dependence of $S_{\textrm{eff},\Lambda}$ on $\Lambda$, or that of $Z_k$ on $k$, encode the full RG-equations of the theory. This will become even more evident later.

We also remark, that the propagators in any (explicit) loop expression derived from \eq{eq:ZJk} are not only cut off in the infrared, but also in the UV,
which restricts us to $\phi(p^2 \leq \Lambda^2)$. This leads us to
\begin{align}
	Z_\Lambda[J]\propto e^{- S_{\textrm{eff},\Lambda}[0 ,J_+]} \,.
\end{align}

Taking the first derivative of the $k$-dependent $Z_k$ with respect to $t=\log k/k_\textrm{ref}$,
or using the translation invariance of the path integral measure under $\phi(p) \to \phi(p) + c(p)$, leads us to functional flow equation and functional SDEs, respectively, see e.g.~\cite{Pawlowski:2005xe, Dupuis:2020fhh, Alkofer:2000wg,  Fischer:2006ub},
\begin{align}\label{eq:fRG-DSE}
	\partial_t \Gamma_k[\phi]= \frac12 \Tr \,G_k[\phi]\,\partial_t R_k\,,\qquad \qquad
	\frac{\delta\Gamma_k[\phi]}{\delta\phi} = 	\frac{\delta S_{\textrm{eff},\Lambda}[\varphi_-,J_+]}{\delta\varphi}\left[\varphi=G_k \frac{\delta}{\delta\phi}+\phi\right]\,,
\end{align}
where $t=\log k/k_\textrm{ref}$ is the (negative) RG-time, $J_+$ is a functional of the mean field, and the field $\phi$ comprises all fields, $\phi=(A_\mu, c,\bar c, q,\bar q)$. Note that the effective action is defined with
\begin{align}
	\Gamma_k[\phi]=\int J\cdot \phi - \log Z_k[J]  - \frac12  \int_p \phi(-p) R_k(p)\, \phi(p)\,,
\end{align}
which is the modified Legendre transform of $\log Z_k$. This implies that
\begin{align}\label{eq:GLambda}
	\Gamma_\Lambda[\phi]= \int J_+ \phi_++   S_{\textrm{eff},\Lambda}[\phi_-, J_+] \,.
\end{align}
Evidently, it follows from \eq{eq:GLambda}, that
\begin{align}
	 \frac{\delta  S_{\textrm{eff},\Lambda}[\phi_-, J_+] }{\delta \phi_- } =  \frac{\delta  \Gamma_{\Lambda}[\phi] }{\delta \phi_- } \,.
	 \end{align}
Accordingly, if we restrict ourselves to momenta smaller than the intermediate cutoff scale, $p^2<\Lambda^2$, we arrive at
\begin{align}\label{eq:fRG-DSE-}
\frac{\delta\Gamma_k[\phi]}{\delta\phi_-} = 	\frac{\delta \Gamma_{\Lambda}[\varphi]}{\delta\varphi_-}\left[\varphi=G_k \frac{\delta}{\delta\phi}+\phi\right]\,.
\end{align}
This concludes our derivation of functional flow equations and finite functional SDEs.

\subsubsection{Some Consequences from the functional SDEs \& fRG}

While the standard functional SDEs for QCD are two-loop exact functional relations that  depend on the full correlation functions and the bare action, the fRG master equation for the effective action in \eq{eq:fRG-DSE} is one-loop exact, and depends only on fully dressed correlation functions. The latter has the advantage that the \textit{self-consistency} of the renormalization scheme is trivially guaranteed, see \cite{Pawlowski:2005xe, Rosten:2010vm, Pawlowski:2015mlf, Braun:2018svj}. Indeed, the flow equation interpolates between the fRG UV action (which is the bare action in the given RG scheme) and the full effective action. The price to pay for both, the trivial RG-consistency and the one-loop closure, is the representation of loop hierarchies in terms of integrals over the infrared cutoff scale $k$, or the (negative) RG-time. In terms of numerical cost, and formulated in terms of SDE integral equations, this can be interpreted as another momentum dependence in the loop integrals (mapping a $d$-dimensional theory onto a $(d+1)$-dimensional one).

In the last section we have combined the advantages of both: we have derived a finite SDE in terms of  $\Gamma_k[\phi]$ (with uniform scaling properties) that smoothly interpolates between the UV effective action (bare action) and the full effective action. This has the advantage that we can invoke the more general RG schemes present in the fRG approach also in the context of SDEs. In general, this facilitates the computation, as it guarantees RG-consistency by construction. However, this advantage comes with a price, since the interpretation of running couplings, such as $\alpha_s$, and of correlation functions is more intricate.

The fRG hierarchy can be understood as a differential SDE in the presence of an infrared cutoff.
It can be shown formally, \cite{Pawlowski:2005xe}, that
\begin{align}\label{eq:fRG-diffDSE}
\frac{\delta}{\delta\phi}\left(	\frac12 \Tr \,G_k[\phi]\,\partial_t R_k \right)= \partial_t \left(\frac{\delta S_{\textrm{eff},\Lambda}[\varphi]}{\delta\varphi}\left[\varphi=G_k \frac{\delta}{\delta\phi}+\phi\right]\right)\,,
\end{align}
where we have used $S_{\textrm{eff},\Lambda}[\varphi]$ in a slight abuse of notation. \Eq{eq:fRG-diffDSE} is nothing but the integrability condition
\begin{align}\label{eq:IntCond}
	\frac{\delta}{\delta\phi}\partial_t  \Gamma_k[\phi] = \partial_t 	\frac{\delta}{\delta\phi} \Gamma_k[\phi] \,.
\end{align}
\Eq{eq:fRG-diffDSE} offers the exciting practical possibility to use either the SDE or fRG relation for one or several of the correlation functions in a consistent way in the hierarchy of the others. This promising option has been pursued, \eg in \cite{Fischer:2008uz, Gao:2020fbl, Gao:2020qsj}. In particular, in \cite{Fischer:2008uz}, the right hand side of \eq{eq:fRG-diffDSE} has been used for the flow of the ghost propagator, as the SDE of the latter only depends on the full ghost-gluon vertex (apart from ghost and gluon propagator), while the respective fRG (left hand side of \eq{eq:fRG-diffDSE}) also depends on the ghost ($c \bar c c\bar c $) and ghost-gluon ($c\bar c A A$) scattering vertices. Moreover, the ghost-gluon vertex is protected by Taylor's non-renormalization theorem, displaying a relatively mild momentum dependence. In turn, in \cite{Gao:2020fbl, Gao:2020qsj}, the quantitative two-flavour vacuum results from the fRG computation of \cite{Cyrol:2017ewj} have been used as input, constituting the most advanced functional QCD computation of the full system to date. Then, the difference SDEs have been solved for thermal, density, and $s$-quark fluctuations. This builds on the fact that functional relations for additional fluctuations are more stable and less sensitive to approximations in the difference system.

\subsubsection{Consequences and computational setup}

We next concentrate on mapping RG schemes in the presence of a sharp cutoff with the properties \eq{eq:Rsharp}, leading to \eq{eq:G=0}. We integrate the SDE representation of the fRG on the right hand side of \eq{eq:fRG-diffDSE} from $k=\Lambda$ to $k=0$. We also make allowances for formal inaccuracies in the treatment of $\phi_+$, and use \eq{eq:fRG-DSE-} also for $\phi_+$ without further proof. With $\Gamma[\phi]=\Gamma_{k=0}[\phi]$ and $\Gamma_\Lambda\stackrel{!}{=} S_{\textrm{eff},\Lambda}=:S_\textrm{bare}$ we arrive at,
\begin{subequations}\label{eq:finiteDSEfull}
\begin{align}\label{eq:IntegratedFlow}
	\frac{\delta\Gamma[\phi]}{\delta \phi} = \frac{\delta S_{\textrm{bare}}[\phi]}{\delta\varphi}\left[\varphi=G \frac{\delta}{\delta\phi}+\phi\right]-\frac{\delta S_{\textrm{bare}}[\phi]}{\delta\varphi}\left[\varphi=G_\Lambda \frac{\delta}{\delta\phi}+\phi\right] + \frac{\delta S_\textrm{bare}[\phi]}{\delta \phi}\,.
\end{align}
Note that the identification of $S_\textrm{bare}=\Gamma_{\Lambda}$ implies that the SDEs include $n$-loop terms, with $n>2$, as $S_{\textrm{bare},\Lambda}[\varphi]$ contains terms with $\phi^{n+2}$. However, all such contributions
are subleading, and are dropped consequently.

The difference of SDEs does not contain the bare classical terms, which cancel identically. Trivially, these terms are fully contained in the last term on the right-hand side of \eq{eq:IntegratedFlow}.

Now we concentrate on diagrammatic terms, which are schematically written as
\begin{align}\nonumber
& \frac{\delta S_{\textrm{bare}}[\phi]}{\delta\varphi}\left[\varphi=G \frac{\delta}{\delta\phi}+\phi\right]-\frac{\delta S_{\textrm{bare}}[\phi]}{\delta\varphi}\left[\varphi=G_\Lambda \frac{\delta}{\delta\phi}+\phi\right]\\[3ex]
 =
& 	\frac{\delta S_{\textrm{bare}}[\phi]}{\delta\varphi}\left[\varphi=\hat  G_\Lambda \frac{\delta}{\delta\phi}+\phi\right]_{\hat G^n = G^n-G_\Lambda^n} - \frac{\delta S_{\textrm{bare}}[\phi]}{\delta\phi}\,,
\label{eq:UVcutoffDSE}
\end{align}
This entails that the SDEs for correlation functions simply are given by the standard diagrams with full propagators and vertices, subtracted by the diagrams with the propagators $G_\Lambda$ and bare vertices, as $\Gamma_\Lambda^{(n)} = S_\textrm{bare}^{(n)}$. The propagators $G_\Lambda$ vanish for momenta $p^2\leq \Lambda^2$.
Accordingly, the difference of propagators (and vertices) in loop integrals have the property
\begin{align}\label{eq:diff}
	\prod_{i=1}^n G(l+p_i)-\prod_{i=1}^n G_\Lambda(l+p_i)^n =  \left\{\begin{array}{lcl}
		G^n(p) & \qquad \qquad &(l+p_i)^2 < \Lambda^2\,,\\[1ex]
		\approx 0  & & (l+p_i)^2 \geq \Lambda^2 \,,
	\end{array}\right.
\end{align}
\end{subequations}
where, for the sake of simplicity,
we have restricted ourselves to the one-loop SDE diagrams. Note that, while the product of propagators is not vanishing for $(l+p_i)^2\geq \Lambda^2$, it is rapidly decaying as the full propagator and full vertices approach the bare ones for these momenta.

\subsubsection{MOM${}^{\,2}$ scheme}

Having concluded our derivation of the finite SDE, we readily can use it in its form \eq{eq:finiteDSEfull}: the flows of propagators and vertices are computed by taking the respective field derivatives of \eq{eq:IntegratedFlow}. Note that we have not chosen a renormalization point yet. Naturally, the present setup suggests to take $\mu=\Lambda$, as the bare UV action $S_{\textrm{bare}}=\Gamma_\Lambda$ is fixed there. For large \mbox{$\mu=\Lambda$}, the latter UV effective action tends towards the classical bare action with $\Lambda$-dependent vertex factors and $\Lambda$-dependent wave function renormalizations. The latter are absorbed in the definition of the fields, leaving us with a classical dispersion, ignoring the subleading terms in $\Gamma_{\Lambda}$. Schematically this reads
\begin{align}
	Z^{1/2} \phi \to \phi\,.
\end{align}
In terms of standard RG theory we effectively reformulate the effective action in terms of bare fields. Then, with the rescaled field $\phi$ we are led to $Z_3, \tilde Z_3, Z_2=1$ (see \eq{eq:Bare-Ren}), effectively setting the wave function renormalizations to unity by absorbing them in the fields. Consequently, the vertex factors have the RG-properties of the respective powers of the strong coupling. Moreover, they satisfy the STIs for the coupling, and reduce to $z_s g_s$ (ghost-gluon, quark-gluon, three-gluon vertex) and $z_s^2 g^2_s$ (four-gluon vertex). Here, $z_s$ is the \textit{unique} rescaling factor for all diagrams that arises from the rescaling of all fields. The uniqueness is a direct consequence of the STIs. We now define
\begin{align}\label{eq:z_sDef}
z_s g_s \to g_s\,,
\end{align}
and write the UV effective or bare action $\Gamma_\Lambda$ in terms of the field $\phi$ and $g$. Within this parameterization, $\Gamma_\Lambda$ reduces to the classical action up to subleading UV-irrelevant terms. In terms of standard renormalization conditions this leads us to
\begin{align}\label{eq:MOMfull}
	Z_3, \tilde Z_3, Z_2, Z_g=1\,.
\end{align}
We emphasise that this re-parametrization leaves the form of the SDEs unchanged. Moreover, it implicitly fixes our RG conditions. For their determination, or rather, their estimate, we assess the value of the correlation functions at the renormalization point $\bar p^2=\Lambda^2=\mu^2$, from the finite SDE, \eq{eq:finiteDSEfull}. Here, $\bar p^2$ indicates a symmetric point configuration in the respective vertices. Schematically, these corrections have the form
\begin{align}\label{eq:RG-condsub}
\Gamma^{(n)}(\bar p) = \Gamma_\Lambda^{(n)}(\bar p) + \left[\textrm{Diagrams}(G) - \textrm{Diagrams}(G_\Lambda)\right](\bar p)\,,
\end{align}
with $\bar p^2 =\mu^2=\Lambda^2$, and $\Gamma_\Lambda^{(n)}(p)$ is only non-vanishing (up to \, terms) for the primitively divergent vertices (including the propagators). Within our rescaled form, we have $\Gamma_\Lambda^{(n)}(\bar p) =S^{(n)}(\bar p)$, see \eq{eq:Bare-Ren}. Thus, the remaining task is to determine the loop corrections.

To that end, first we remark that, for a large RG scale, deep in the perturbative regime, a one loop analysis is sufficient. Evidently, for large momentum scales, the logarithmic running of the loops is proportional to $\alpha^n_s(\bar p^2)\log \bar p^2/\Lambda^2$ or $\alpha_s^n(\bar p^2)\log (\bar p^2 + c \Lambda^2)/\Lambda^2$, where $c$ is some constant (the physical mass scales drop out). Depending on the vertex discussed, $n$ is given by $n=1, 3/2, 2, ...$  At $\bar p^2=\Lambda^2$, the logarithmic running of the loops drops out,as there we have  $\alpha_s(p^2)^n \log \bar p^2/\Lambda^2=0$, and more generally
\begin{align}
	\label{eq:asympt0}
	\lim_{\Lambda/\Lambda_{\textrm{QCD}}\to\infty} \left[\alpha_s(\bar p)^n \log (\bar p^2 + c \Lambda^2)/\Lambda^2\right]_{\bar p^2=\Lambda^2}=
	\lim_{\Lambda/\Lambda_{\textrm{QCD}}\to\infty}\alpha_s(\Lambda)^n \log (1+c) = 0\,,
\end{align}
where we have used asymptotic freedom: $\alpha_s(\Lambda\to\infty)^n =0$. In conclusion, for a sufficiently large renormalization scale $\Lambda /\Lambda_{\textrm{QCD}}$ this provides us with a 'doubly'-MOM scheme, both the bare $Z$'s and the full dressings are unity. For this reason we have called it MOM${}^2$ scheme.

Indeed, we can now slightly modify our procedure: we insist on the standard MOM condition \eq{eq:RG-conditions} and tune the bare $Z$'s accordingly. However, in the present scheme with \eq{eq:asympt0} this leads us to $Z_i\to 1$ for sufficiently large $\Lambda$. We conclude this analysis with the remark that, in the present work, the loop corrections are of the order of $10^{-2}$, owing to the small value of $\alpha_s$ at $\mu= 40$\,GeV.

\subsubsection{MOM${}^{\,2}$ scheme in the fRG approach}

As mentioned before, most fRG computations use the MOM${}^{\,2}$ scheme. In particular it is implemented in the fRG approach to QCD in \cite{Mitter:2014wpa, Cyrol:2016tym, Cyrol:2017ewj, Cyrol:2017qkl}, and hence underlies our fRG input. In these works one uses $\Gamma_\Lambda=S_\textrm{cl}$ with a large cutoff scale $\Lambda/\Lambda_{\textrm{QCD}}\to\infty$. Here, $S_\textrm{cl}$ is the standard classical action (no $Z's$). This translates into a standard RG scheme with $\mu=\Lambda$ and $S_\textrm{bare}$ with bare renormalization constants $Z_i=1$.

In conclusion, as already discussed for the finite SDEs, for a sufficiently large renormalization scale $\Lambda \gg \Lambda_{\textrm{QCD}}$, this provides us with the MOM${}^{\,2}$ scheme, since both the bare $Z$'s and the full dressings are unity. Indeed, we can now slightly modify our procedure: we insist on the standard MOM condition \eq{eq:RG-conditions} and tune the bare $Z$'s accordingly. However, in the present scheme with \eq{eq:asympt0} this leads us to $Z_i\to 1$ for sufficiently large $\Lambda$. This leads us to a coinciding RG scheme for both, fRG and finite SDE approaches, and allows for the efficient and formally consistent incorporation of results obtained within either of the approaches into the other.

\subsection{Practical implementation in the present work}

While the finite SDE \eq{eq:finiteDSEfull} can be implemented straightforwardly, it is convenient to recast it into a form, which readily allows for the use of standard SDE numerics. In the following we reformulate the finite SDE in terms of a BPHZ scheme, which allows for a more direct interpretation of the subtraction procedure. The effects of the respective approximations can be estimated along the same lines as the difference between the bare renormalization constants and the full dressings, and lead to further terms of the order \eq{eq:asympt0}, hence vanishing for large renormalization scales.

To begin with, a good approximation for the leading order contribution is
\begin{align}
	\hat G^n_\Lambda[\phi] \approx \bar G_\Lambda^n[\phi]\,\quad \textrm{with}\quad \bar G_\Lambda[\phi] = G[\phi] -G_\Lambda[\phi]\,,
\end{align}
which implies for the momentum dependence of the propagators $\bar G(p^2) = \bar G[\phi=0](p^2)$,
\begin{align}
	\bar G_\Lambda(p^2)  \approx G(p^2)\theta(\Lambda^2-p^2)\,.
\end{align}
Note, however, that there is no technical obstruction in simply using the full expressions. In other words, \eq{eq:IntegratedFlow} reduces to
\begin{align}\label{eq:finiteDSE}
	\frac{\delta\Gamma[\phi]}{\delta \phi} = \frac{\delta S_{\textrm{bare}}[\varphi]}{\delta\varphi}\left[\varphi=\bar G_\Lambda \frac{\delta}{\delta\phi}+\phi\right]\,.
\end{align}
\Eq{eq:finiteDSE} is a \textit{finite} SDE with a UV momentum cutoff in each line. Formally the finite SDE derived in the previous Sections is cutoff-independent, see \eq{eq:ZRG},
\begin{align}\label{eq:G-indepL}
	\Lambda\partial_\Lambda  	\frac{\delta\Gamma[\phi]}{\delta \phi} =0\,.
\end{align}
However, the approximations used for arriving at the convenient form \eq{eq:finiteDSE} lead to some residual $\Lambda$-dependence that has to be checked.

The appearance of $\bar G_\Lambda(p^2)$, or rather $\hat G_\Lambda^n$, in the loops is similar to the standard momentum cutoff in loops. However, it has the advantage that this \textit{consistent} procedure maintains translational symmetry in momenta, which is broken
in the presence of a hard momentum cutoff. Note that this symmetry is important, \eg for the anomalous triangle diagrams that contribute to $\pi\to\gamma\gamma$ scattering, being related to the pion decay constant in the chiral limit.

\subsubsection{MOM${}^{\,2}$ scheme with BPHZ implementation}

This setup entails that we can translate the SDEs \eq{eq:finiteDSE} into a more standard representation for large $\Lambda$: we yet again use, that the contributions of the UV-part of the loops at $\bar p^2$ tends towards zero and substitute the cutoff procedure in \eq{eq:finiteDSE} with a subtraction at $\bar p^2=\mu^2=\Lambda^2$. This is the standard MOM procedure, and amounts to substituting the difference of diagrams in \eq{eq:RG-condsub} with
\begin{align}\label{eq:RG-condBPHZ}
\textrm{Diagrams}(G)(p) - \textrm{Diagrams}(G_\Lambda)(p) \to&\, \textrm{Diagrams}(G)(p) \\[1ex] \nonumber
&\hspace{-2cm}- \textrm{Diagrams}(G)(\bar p) - (p^2-\bar p^2)\left[  \partial_{p^2} \textrm{Diagrams}(G)(p) \right]_{p=\bar p}\,,
\end{align}
where the linear expansion term is only used for the propagators, and renormalizes the propagator dressing.  \Eq{eq:RG-condBPHZ} directly implements the MOM RG conditions, as it leads to
\begin{align}\label{eq:MOM2}
	Z_{\phi}(\bar p^2)=1\,, \qquad\alpha_i(\bar p^2)= \alpha_s\,,\quad\textrm{with}\quad\phi=(A,c,\bar c, q, \bar q)\,,\quad  i= c\bar c A\,,\, q\bar q A\,,\,  A^3\,,\,  A^4\,.
\end{align}
and $\bar p^2=\mu^2$ is the momentum squared in two-point functions and a symmetric point configuration in three- and four-point functions. Consequently, the $Z$'s in the ``bare'' action $\Gamma_{\Lambda}$ deviate from unity by corrections of the order \eq{eq:asympt0}, that vanish proportional to $\alpha_s(\mu^2)\to 0$ for $\mu/\Lambda_\textrm{QCD}\to \infty$. Accordingly, \eq{eq:MOMfull} is only satisfied approximately, but to a very good degree for large RG scales. Moreover, the relation \eq{eq:RG-condBPHZ} is valid up to corrections of the order \eq{eq:asympt0}.

In summary, we have arrived at a SDE implementation of the {MOM${}^2$ scheme, that allows to make use of standard SDE numerics: we use \eq{eq:MOMfull} and \eq{eq:MOM2} for asymptotically large RG scales. Note that the absence of non-trivial renormalization constants can be traced back to the use of bare fields and couplings in terms of standard RG schemes. Such a rescaling comes naturally in the present Wilsonian approach with its underlying effective action $\Gamma_k$, which interpolates between the bare (but finite) UV effective action $\Gamma_{\Lambda}$ and the full renormalized effective action $\Gamma=\Gamma_{k=0}$.

This finally sets up the numerical procedure used in the present work. We close this section with a final comment on the momentum-dependence of general non-perturbative renormalization procedures: the identification of $\Gamma_\Lambda$ with the bare action in \eq{eq:finiteDSE} entails that the latter also contains subleading momentum dependences, which are again of the order \eq{eq:asympt0} for momenta $p^2\lesssim \Lambda^2$. In the present work they are dropped (as well as in the fRG works \cite{Mitter:2014wpa, Cyrol:2016tym, Cyrol:2017ewj, Cyrol:2017qkl}), and we can investigate the reliability of this procedure as follows:
\begin{itemize}
\item[(i)] We disentangle the RG scale $\mu=\Lambda$ from the cutoff scale $\Lambda_\textrm{UV}$ in the loop integrals of the SDEs within the MOM formulation (subtraction of the loops at $\bar p^2=\mu^2$). Note that this leads to the standard MOM renormalization conditions, see \eq{eq:RG2} in the present work. This is defined by
the combination of \eq{eq:MOMfull} and \eq{eq:MOM2} for $\alpha_s(\bar p)\to 0$.

\item[(i)] Then we extend the momentum range of the loop integrals to large UV cutoff scales: $\Lambda_\textrm{UV}/\Lambda\to\infty$. This gives access to the momentum dependence of $\Gamma_\Lambda$ for $p^2\lesssim \Lambda^2$. We explicitly checked in the present work that this does not change our results, hence providing a further self-consistency check of our procedure.
\end{itemize}
With (i) \& (ii) we have converted the present scheme into a standard BHPZ scheme, the mathematically soundest standard regularization and renormalization scheme used for most proofs of perturbative renormalizability and beyond. Indeed, one of the first uses of functional renormalization group investigations was the simplification of such proofs with the Polchinski equation \cite{Polchinski:1983gv}, evoking the standard fRG renormalization procedure that generalises BHPZ.

\subsubsection{MOM${}^{\,2}$ versus MOM}

We close this Appendix with a brief evaluation of similarities and differences between the MOM${}^{2}$ and the standard MOM scheme. First of all, the RG conditions for propagators and primitively divergent vertices are identical, and given by \eq{eq:RG-conditions}. Evidently, the conditions for the bare $Z's$ are different, as the standard MOM scheme implies non-trivial and cutoff-dependent values for them, while in the present scheme we have $Z_i\to 1$ for sufficiently large RG scales. The difference comes from an inherent generalization of the RG schemes in Wilsonian approaches. This property is explicitly built in in the fRG approach: the flow interpolates between a given bare action $\Gamma_\Lambda$ and the full effective action. Trivially, this allows for more general rescalings of fields and couplings that are not taken into account in standard RG-schemes as they are cutoff and RG scale dependent. In standard RG schemes this may cause consistency problems, whereas within the fRG, RG-consistency is built in.

This entails that the MOM${}^{2}$ scheme necessarily implies a rescaled running coupling $\alpha_{s,\textrm{MOM}^2}$ with
\begin{align} \label{eq:MOM2-MOM}
	\alpha_{s,\textrm{MOM}^2}(p^2) = z_s\, \alpha_{s,\textrm{MOM}}(p^2)\,.
\end{align}
This property has been seen in \cite{Cyrol:2016tym} for Yang-Mills theory. Indeed, in comparison with the recent quantitative evaluation of the Yang-Mills system with SDEs and the standard MOM scheme in \cite{Huber:2020keu}, one finds \eq{eq:MOM2-MOM} with approximately $\alpha_{s,\textrm{MOM}^2}(p^2)\approx 4/3 \,\alpha_{s,\textrm{MOM}}(p^2)$. Note that the prefactor depends on the RG scale chosen within the MOM${}^2$.

In QCD, the fRG results in \cite{Cyrol:2017ewj} obtained for the full system also show a larger coupling, roughly given by $\alpha_{s,\textrm{MOM}^2}(p^2)\approx  1.2 \,\alpha_{s,\textrm{MOM}}(p^2)$ for an RG scale $\mu=40$\,GeV. The present SDE computation within the MOM${}^2$ scheme in 2+1 flavour QCD confirms this result with $z_s=1.19$, see \eq{eq:MOMLambda+zs}.

\section{Kernels of the quark-gluon vertex SDE }\label{app:cof}

The kernels
$K_{ijk}(p,q,k)$ and $\widetilde{K}_{ijk}(p,q,k)$ appearing in \eq{eq:aandb} have the general form
\begin{align}
K_{ijk}(p,q,k) = \sum_{\alpha=1}^{2} C^{\alpha}_{ijk}(p,q,k)\sigma_{\alpha}(k)\,,
\qquad \widetilde{K}_{ijk}(p,q,k) = \sum_{\alpha=1}^{4} \widetilde{C}^\alpha_{ijk}(p,q,k) \widetilde{\sigma}_\alpha(p,q,k)\,,
\end{align}
with
\begin{align}
\sigma_{1}(k) = \frac{1}{Z_q(k) [k^2+M^2_q(k)]} \,,\qquad\qquad
\sigma_{2}(k) = M_q(k)\,\sigma_{1}(k)\,,
\label{eq:sigma}
\end{align}
and
\begin{align}
\widetilde{\sigma}_{1}(p,q,k) =&\,\sigma_{1}(p+k)  \sigma_{1}(q+k)\,,\qquad  \widetilde{\sigma}_{2}(p,q,k) = M_q(q+k) \,\widetilde{\sigma}_{1}(p,q,k)\,,\nonumber\\[2ex]
 \widetilde{\sigma}_{3}(p,q,k) =& \, M_q(p+k)\,\widetilde{\sigma}_{1}(p,q,k)\,,\qquad
\widetilde{\sigma}_{4}(p,q,k) = M_q(p+k)\,  M_q(q+k)  \,\widetilde{\sigma}_{1}(p,q,k) \,,
\label{eq:tildesigma}
\end{align}
and the closed expressions for the kinematic functions $C^{\alpha}_{ijk}(p,q,k)$ and $\widetilde{C}^\alpha_{ijk}(p,q,k)$ are reported in the github (https://github.com/coupledSDE/FormDerive).


\bibliography{ref-lib}
\end{document}